\documentclass[11pt,preprint]{aastex}

\received{}
\shortauthors {Stanghellini et~al.}
\shorttitle {Compact Galactic Planetary Nebulae} 

\newcommand{\noprint}[1]{}
\newcommand{\figsetstart}{{\bf Fig. Set} }
\newcommand{\figsetend}{}
\newcommand{\figsetgrpstart}{}
\newcommand{\figsetgrpend}{}
\newcommand{\figsetnum}[1]{{\bf #1.}}
\newcommand{\figsettitle}[1]{ {\bf #1} }
\newcommand{\figsetgrpnum}[1]{\noprint{#1}}
\newcommand{\figsetgrptitle}[1]{\noprint{#1}}
\newcommand{\figsetplot}[1]{\noprint{#1}}
\newcommand{\figsetgrpnote}[1]{\noprint{#1}}

\begin{document}

\title {Compact Galactic Planetary Nebulae: A  \textit{HST}/WFC3 Morphological Catalog, and a Study of their Role in the Galaxy}

\author {Letizia Stanghellini\altaffilmark{1}, 
Richard A. Shaw\altaffilmark{1}, 
Eva Villaver\altaffilmark{2}
}
\altaffiltext{1}{National Optical Astronomy Observatory, 950 N.\ Cherry Avenue, Tucson, AZ 85719}
\altaffiltext{2}{Universidad Aut{\'o}noma de Madrid, Departamento de F{\'i}sica Te{\'o}rica C-XI, E-28049 Madrid, Spain}

\begin{abstract} 

We present the images of a \textit{Hubble Space Telescope} (\textit{HST}/WFC3) snapshot program of angularly compact Galactic planetary nebulae (PNe), acquired with the aim of studying their size, evolutionary status, and morphology.  
PNe that are smaller than $\sim4\arcsec$ are underrepresented in most morphological studies, and today they are less well studied than their immediate 
evolutionary predecessors, the pre-planetary nebulae. 
The images have been acquired in the light of [\ion{O}{3}]$\lambda5007$, which is commonly used to classify the PN morphology, in the UV continuum with the aim of detecting the central star unambiguously, and in the $I-$band to detect a cool stellar companion, if present. 
The sample of 51 confirmed PNe exhibits nearly the full range of primary morphological classes, with the distribution more heavily weighted toward bipolar PNe, but with total of aspherical PNe almost identical to that of the general Galactic sample. A large range of microstructures is evident in our sample as well, with many nebulae displaying attached shells, halos, ansae, and internal structure in the form of arcs, rings, and spirals. Various aspherical structures in a few PNe, including detached arcs, suggest an interaction with the ISM. 
We studied the observed sample of compact Galactic PNe in the context of the general Galactic PN population, and explore whether their physical size, spatial distribution, reddening, radial metallicity gradient, and possible progenitors, are peculiar within the population of Galactic PNe. 
We found that these compact Galactic PNe, which have been selected based on apparent dimensions, 
constitute a diverse Galactic PN population that is relatively uniformly distributed across the Galactic disk, including the outskirts of our Galaxy. 
This unique sample will be used in the future to probe the old Galactic disk population.

\end{abstract}


\section {Introduction} 

Planetary nebulae (PNe) are ionized shells of gas that were ejected by Asymptotic Giant Branch (AGB) stars at the end of their evolution. The ejection symmetry, slow expansion rate, rapidly evolving central star (CS) that may feature a fast wind and magnetic fields, presence of interstellar medium materials, dust content, and 
central star multiplicity produce PNe with a variety of shapes that may change during their short, hydrodynamical evolution. In the Galaxy, studies based on PN image catalogs \citep[e.g.,][]{1992A&AS...96...23S, 1996iacm.book.....M} have shown that PN 
projected macro-morphologies can be usefully grouped into a few basic classes, and that round (R), elliptical (E), bipolar core (BC), and bipolar (B) shapes are the most commonly observed. Morphology is best studied with narrow-band imagery, and typically gives a reasonable classification for all PNe that are larger than $\sim4\arcsec$ to obtain a well spatially-resolved PN image as observed with ground-based telescopes. This empirical limitation is due to the minimum number of resolution elements needed to sensibly determine morphology. For this reason, Galactic morphological catalogs cannot classify the angularly small PNe, which are then labelled unresolved or point-source.

With \textit{HST} cameras it has become possible to determine the morphology of spatially compact PNe. In the past decades \textit{HST} has revealed the shape and apparent size of extragalactic PNe such as those in the Large and Small Magellanic Clouds (e.g., Stanghellini et~al. 2000, 2003; Shaw et~al. 2001, 2006), which are almost invariably unresolved from the ground, and whose apparent diameters are typically $\sim0.5\arcsec$. 
On the other hand, compact Galactic PNe, with apparent radius $<4\arcsec$, have not been studied as an independent group to date, but several such objects have corresponding \textit{HST} images \citep[e.g.,][]{2014ApJ...787...25H} that reveal their shapes. 
Compact Galactic PNe are important for a variety of reasons: as a group, they may be predominantly farther away than other Galactic PNe, or they may be younger -- recently ejected -- with respect to the general population. In the former case, this sample gives the chance to peer farther into the Galactic disk when using PNe as probes of Galactic structure and evolution, while in the latter case it gives a chance to study early PN morphology, soon after the ionization onset. Only by studying a sizable sample of compact PNe we would be able to disclose their original stellar population and evolutionary stage.

Compact PNe offer another advantage with respect to the general population: their spectra, both optical and IR, can be acquired with just one pointing to include the whole nebula, which in turns provides  plasma diagnostics and chemical analysis of the PN as a whole. This is important for ground-based optical spectroscopy, and even more so for mid-IR space spectroscopy. For example, their compact shapes allowed the study of dust content with {\it Spitzer}/IRS spectroscopy;  dust analysis of more than 150 compact  Galactic PNe is available to date (Stanghellini et~al. 2012). Dust plays an important role in PN formation, where radiation pressure on dust grains causes stellar winds, and thus must affect PN shaping. By studying Magellanic Cloud PNe, it has been shown that dust type and morphology are correlated \citep{Stang_etal07}. 

In this paper we present a morphological catalog of compact Galactic PNe observed with \textit{HST}/WFC3. This is the largest {bf homogeneous} sample of compact Galactic PNe observed to date. The observations made it possible to determine the PN apparent radii, their morphology, and to study their CSs. 
Here we present the nebular aspect of this sample, while the CS analysis, including that of possible companion stars, will be published in \citet{MVSS16} and elsewhere.

In Section 2 we describe the observations that yielded to 51 PNe images in 3 filters. 
In $\S$ 3 we give the photometric radii, fluxes, and classification of individual morphology, measured from the images, and a discussion on their distances and physical sizes. Section 4 includes a discussion on the relation between morphology and dust types of compact PNe. In $\S$ 5 we compare compact PNe with the general Galactic PN population. This includes comparing physical dimensions, morphology, dust type, Galactic distribution, and radial metallicity gradients. Section 6 includes a discussion of our findings. Finally, $\S7$ gives the conclusions, and some directions for potential future research.

\section {Observations}

The  goal of our \textit{HST} snapshot program (GO 11657; PI: L. Stanghellini) was to obtain high-resolution images of a substantial sample of angularly small PNe (i.e., with published angular diameters $<4\arcsec$) whose ground-based observations are insufficient both to determine nebular morphology with adequate detail, and to measure CS fluxes against the bright nebular emission. 
The target selection for this program includes all known compact Galactic PNe. 
The population of spectroscopically confirmed Galactic PNe is 
described in the Strasbourg-ESO catalog of Galactic PNe \citep{Acker92}, and in the MASH survey \citep{2006MNRAS.373...79P}.
Of the 1143 PNe listed by \citet{Acker92}, 143 are point-sources and 86 have ground-based measured radii $\theta<4\arcsec$. 
The MASH survey adds two compact PNe with $\theta<4\arcsec$, but no point sources. 
From these compact targets we explicitly removed all Galactic Bulge and Halo PNe, to obtain a pure Galactic disk population of compact PNe. 
To assess whether a PN belongs to the Galactic bulge or halo we used the criteria defined by \citet[][hereafter SH10]{2010ApJ...714.1096S}
but more precise population assessment, which is based also on their distance, can be done only a posteriori for compact PNe, once their statistical distances can be reliably calculated from their photometric radii. 

From the target sample we excluded those PNe that have been previously observed with \textit{HST} WFPC2 and ACS in similar imaging modes to avoid duplications. We did not however exclude PNe already observed with pre-COSTAR \textit{HST} imaging, since the PN dimensions obtained from these images are not reliable (Stanghellini et~al. 2000).  By proposing these observations in snapshot mode we further reduced the target list by only including targets that are bright enough the be observed within one orbit. Our approved program consisted of 130 targets. 

Observations of  compact Galactic PNe were acquired with \textit{HST} Wide-Field Camera~3 \citep{Kimble_etal08} through four filters. We used F502N to obtain images of the targets in the light of  [\ion{O}{3}] $\lambda5007$, which is very often among the brightest emission lines in PNe, and it is customary used to classify the morphology and measure the angular dimensions with aperture photometry. 
We also imaged each target with F200LP, which passes essentially all wavelengths to which the detectors are sensitive, and in particular takes advantage of the extraordinary sensitivity of this camera to near-UV light \citep{WFC3_ihb}.  We complemented the NUV exposures with F350LP, which blocks UV light but passes all visible light; the difference between the calibrated flux of the CS in F200LP and that in F350LP yields a UV continuum magnitude. 
Finally, we imaged each target with F814W to measure the stellar the $I-$band continuum. All images except those in F814W were split into sub-exposure pairs to enable cosmic-ray rejection. 
Images in the UV and IR continuum allowed measuring the magnitude and color of the CS
\citep{MVSS16}. 

We planned our exposure durations in F502N to yield an average signal-to-noise ratio (S/N) of at least 16~pixel$^{-1}$, based on the $H\beta$ fluxes of \citet{Acker91}, and the $I(5007)/I(H\beta)$ relative intensities and angular diameters as compiled by \citet{Acker92}. 
For objects with no published [\ion{O}{3}] flux we assumed  $I(5007)=I(H\alpha)$ at zero reddening. 
Our experience with prior \textit{HST} programs indicates that images so exposed yield useful morphological detail. We note that the substantial error in the published angular diameters (which are all taken from ground-based observations) yielded large uncertainties in the estimated exposure times. 
Our estimates of the optimal exposure durations for the continuum filters was more uncertain still, since we had no information on the CS magnitudes. 
The absolute brightness of a CS can vary by 10~mag in $V$ as it evolves at nearly constant luminosity from 30,000~K to well over 100,000~K, before fading to the white dwarf cooling track.
In the end we planned the F200LP and F350LP exposures to yield a S/N of $\sim20$ for a star with apparent magnitude $V=25$ and a temperature of $10^5$~K. 
We added a second exposure in these filters that was usually shorter by a factor of a few to 10. 
This combination of exposures allowed sufficient dynamic range to minimize the risk of saturation for bright, nearby CSs while providing useful S/N for faint, distant stars. 
For F814W we planned a single exposure, modified for interstellar extinction, that would yield useful S/N for a possible companion between types G2V and M5V. 

Minimizing visit duration without compromising science is always a key concern for \textit{HST} snapshot programs. 
Our strategy was to reduce the detector read-out time by using sub-array apertures, which allows all the exposures during a visit to be stored in the instrument memory, rather than incurring the overhead of a buffer dump to external storage. 
For F502N it was important to image the entire nebula (which could have structure larger than the nominal 4\arcsec), and to allow for the possibility of errors in the coordinates, which were taken from \citet{Kerber03}. 
We therefore used the UVIS1-2K4-SUB aperture, which covers a roughly 2K~$\times$~4K~pixel (82\arcsec$\times$164\arcsec) field of view using CCD1. 
The continuum exposures used the UVIS1-C512A-SUB aperture which covers a $512\times512$~pixel ($20\arcsec\times20\arcsec$) sub-array located in the corner of CCD~1, near a read-out amplifier. 
This approach minimizes read-out time and associated overheads, as well as the number of charge transfers during read-out; this aperture includes virtual overscan for robust bias subtraction. 

Only 54 targets of the original 130 in the target list were observed successfully. 
Of these, three objects (PN~G011.7--06.6, PN~G274.1+02.5, and PN~G311.1+03.4) are likely misclassified PNe, and will not be analyzed here.
The exposure for F502N failed for one target, PN~G348.8--09.0.
The observations were taken throughout the course of Cycle 17; the observing log is presented in Table~\ref{ObsLog}, where we give the PN names (PN~G designations and common names), the observing date, the dataset name, the filter used in the observation, the exposure time, and the position angle of each observation.
The images were processed using the standard WFC3 calibration pipeline (CALWFC3 version 2.0, 08-Mar-2010). 
The processing steps in CALWFC3 are described in detail by \citet{WFC3_dhb}. 
In brief, the instrumental signature is removed from the raw data by subtracting the bias level and residual bias structure, removing the overscan regions, scaling and subtracting a master dark frame, normalizing to unit gain, and correcting for photometric uniformity (including pixel-to-pixel sensitivity variations within a CCD, and the focal plane illumination pattern) by dividing out a master flat-field. 
Cosmic rays are rejected when the CR-split images for each filter (except F814W) are co-added. 
The combined images are geometrically rectified and the brightness is normalized to unit exposure time.

\section {Dimensions, Fluxes, and Morphology}

\begin {figure}
\plottwo {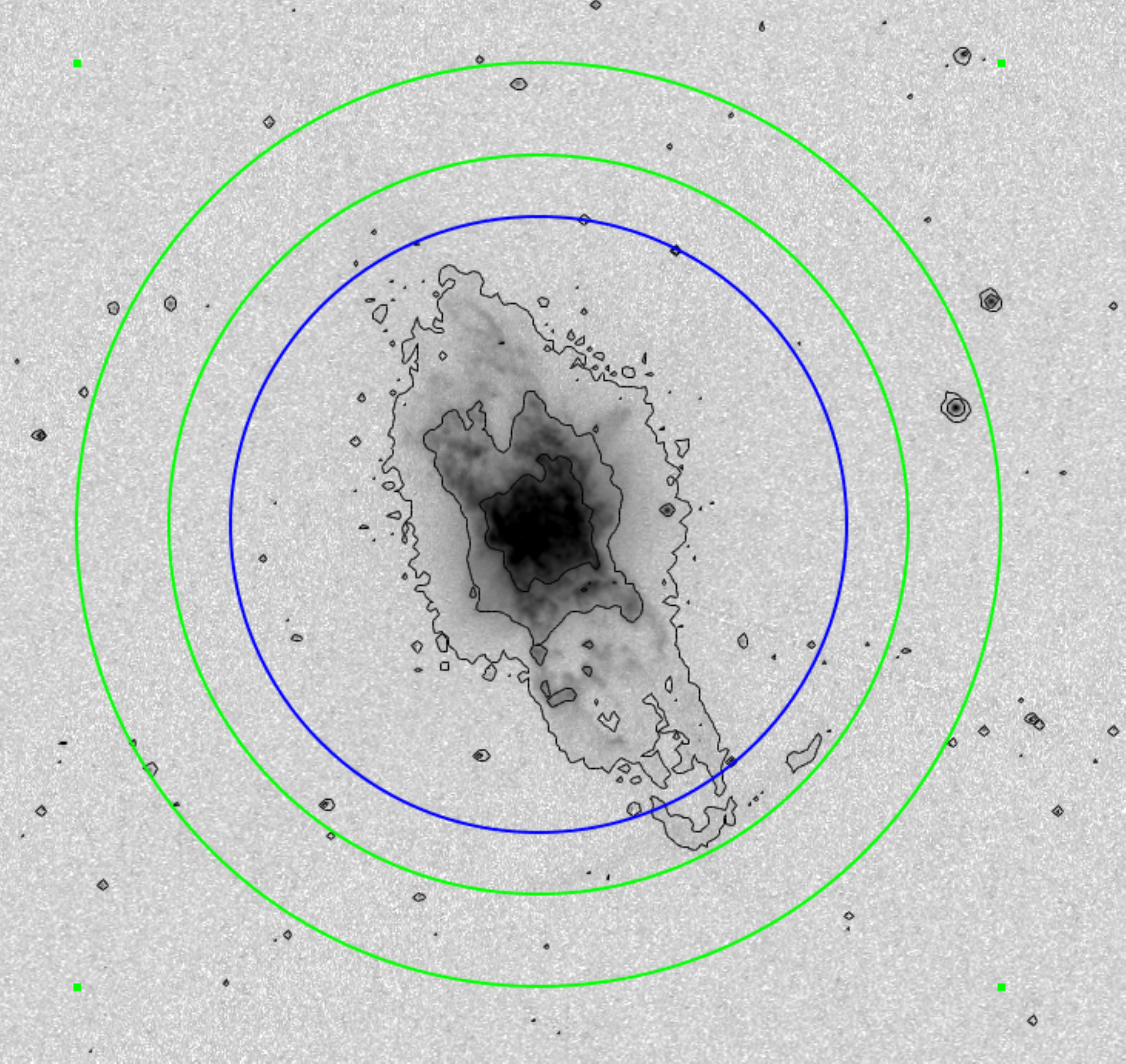}{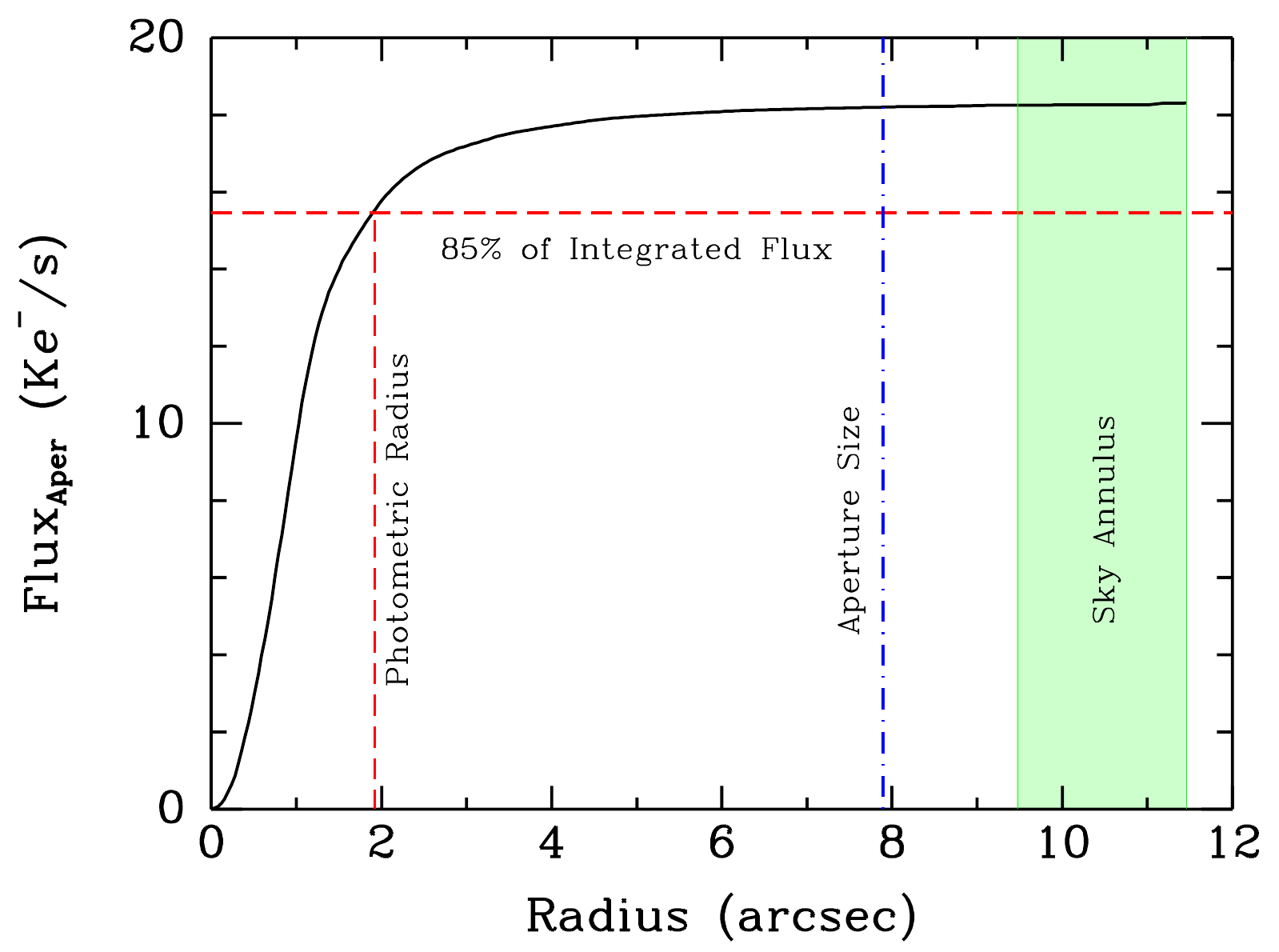}
\figcaption{
\textit{Left:} Image of PN~G285.4+01.5 with contours (at 10\%, 1\%, and 0.1\% of the maximum target intensity) over plotted. Photometric aperture (\textit{blue circle}) and sky annulus (\textit{green circles}) are shown.
\textit{Right:} Curve of integrated flux within the aperture as a function of aperture size, showing the photometric radius (\textit{dashed line}), the size of the photometric aperture used to determine the flux in F502N (\textit{dot-dashed line}), and the bounds of the annulus used to determine the sky background (\textit{shaded band}).
\label{fig:photRad}}
\end {figure}

We present all 51 observed targets in the compact Galactic PN image catalog in Appendix A. 
We determined the dimensions of each nebula by measuring the radius that contains 85\% of the integrated light in [\ion{O}{3}] $\lambda5007$, 
using the \textit{Aperture Photometry Tool} \citep{Laher12} on the F502N images. 
We measured the flux within a virtual circular aperture with a radius significantly larger than the target, and with brightness clipping to minimize the effect of field stars. 
The local background was estimated with a circular annulus, again with brightness clipping, using inner and outer radii that were customized for each nebula to be as large as possible. 
We have used a similar technique systematically in the past \citep[see, e.g.,][]{Shaw_etal01, Stang_etal03, Shaw_etal06} with great success.
The photometric radius provides a consistent measure of nebular size and is insensitive to the nebular morphology, complexity of the stellar background, and the signal-to-noise ratio in the image.
The technique is illustrated in Figure~\ref{fig:photRad} on a highly non-circular PN in our sample, showing the smooth, monotonic rise in the integrated fraction of nebular light with radius.

\begin {figure}
\plotone {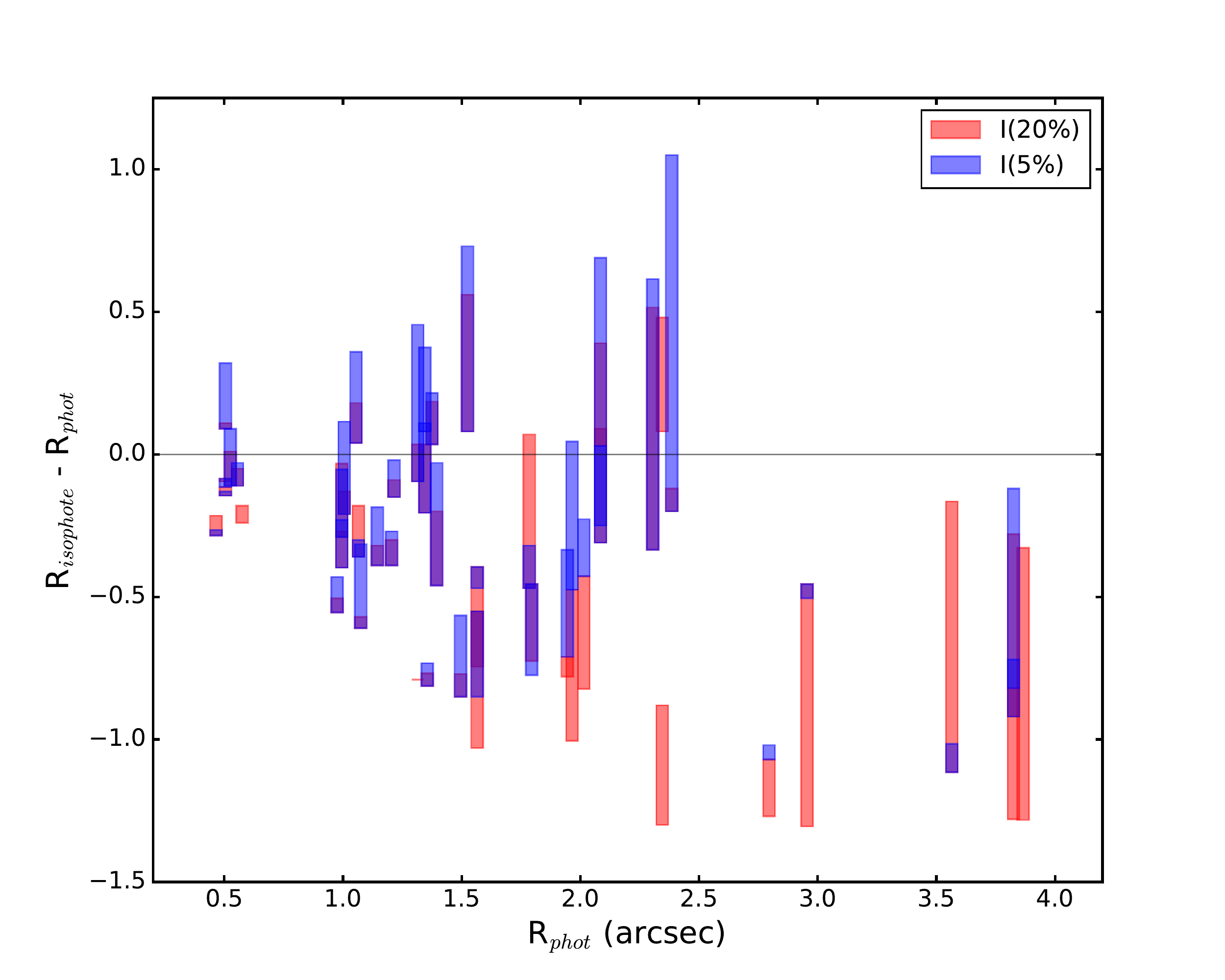}
\figcaption{
Difference between the nebular radius as determined from our photometric technique (85\% of the integrated narrow-band light of [\ion{O}{3}] $\lambda5007$), and that determined from the isophotal contours at 20\% (\textit{red}) and 5\% (\textit{blue}) of the peak intensity in the same band. The bars indicate the range in isophotal radius as determined from the major and minor axis of the nebula.
\label{fig:photTechnique}
}
\end {figure}

We also measured the nebular dimensions using one or more isophotes: at 20\% of the peak intensity (not counting the CS), at $\sim5\%$, and in some cases at fainter isophotes if the S/N was high.
We measured the diameters along the major and minor axes of the isophote in question, usually on the F502N image unless the S/N was poor, in which case we used the F200LP image.
While this technique is the most commonly used (and it can be useful for planning follow-up spectroscopic observations), the results depend strongly upon the S/N ratio in the image and upon the spatial resolution.
There is also a level of subjectivity in these measurements, depending for instance on whether microstructure is considered (e.g., ansae), and in choosing how to weight the results for non-circular nebulae.
Figure~\ref{fig:photTechnique} compares our photometric radii with those determined from isophotal contours in the same emission line at two different intensity levels, and for both the major and minor axes.
The degree of agreement is very poor and highlights the inconsistency of the isophotal technique, particularly for angularly small nebulae. 
We note that the statistical distance scale derived by \citep[][hereafter SSV]{Stang_etal08} was calibrated using photometric radii determined in the way described here, and that distances derived using nebular radii as measured in a different waveband or with a different technique are likely to result in significant systematic errors.
For these reasons, isophotal radii are certainly not appropriate for deriving physical dimensions of the nebulae, or statistical distances. 

We give in successive columns in Table~\ref{tab:Morph} the PN~G target designation (column 1), coordinates as measured from the F814W image (columns 2 and 3), the flux in [\ion{O}{3}] $\lambda5007$ and $H\beta$ (columns 4 and 5), the extinction constant at $H\beta$ (column 6), the photometric radius $R_{\rm phot}$ in arcsec, followed by the isophotal diameter(s) and the reference isophote(s) (columns 7, 8, and 9), the morphological class (column 10), and notes on individual nebulae (column 11). 
Coordinates are given in the ICRS reference system, 
and correspond to the position of the PN central star if detected, otherwise the approximate geometric center. The [\ion{O}{3}] $\lambda5007$ fluxes are measured directly from the \textit{HST} images as described above, the $H\beta$ fluxes have been calculated from the [\ion{O}{3}] $\lambda5007/H\beta$ flux ratio from the \citet{Acker92} catalog. 
All extinction constants are taken from data published elsewhere, as given in the Table references.

We classified the projected morphology of all 51 PNe in the sample based on their appearance in the [\ion{O}{3}] $\lambda5007$ images, 
following the morphological scheme for Galactic PN morphological classification described in Stanghellini et~al. (2006). The major morphological classes are:
Round (R) PNe; elliptical (E) PNe, which are distinguished from R PNe if an axial difference of at least $5\%$ is detected;  bipolar (B) PNe, implying the presence of at least one pair of lobes, and including quadrupolar and other multi-polar PNe; point-symmetric (P) PNe, implying features with central symmetry and no lobes. We also identified elliptical or round PNe with inner bipolar structure or the presence of a ring as bipolar core (BC) PNe, and set them in a class apart. Some PNe display attached or detached shells or haloes, ansae, and other inner and outer features. These additional, secondary morphological features are noted in column 11.  In Figures 3 through 5 we show examples of the principal morphological types.
It is worth noting that morphological classification based on imagery alone may be in some cases misleading, since some round or elliptical shapes may derive from bipolar lobes projected into the plane of the sky, and round PNe in particular could also be elliptical in projection. These hidden morphologies can only be detected with the help of nebular kinematic studies (e.~g., Kwok 2010), and this is beyond the scope of the present paper, but should be addressed in future studies.
Two PNe are not classifiable within the above scheme, and we classify them as Irregular (Irr) in Table~\ref{tab:Morph}. The remaining PNe are morphologically distributed into R PNe (2), E PNe (25), BC PNe (4), B PNe (15), and P PNe (3).

\subsection {Peculiar Morphologies}

While morphological classifications are provided in Table~\ref{tab:Morph} for all observed PNe, several objects have additional features that are worth elaboration. Indeed, macro-morphological details such as the presence of halos indicate that the ionization front has passed completely through the nebula, from which we can infer advanced nebular age and low optical depth to Lyman radiation within the nebula. Other morphological details may suggest shaping by jets, interaction of the nebula with the surrounding ISM, motion of the CS relative to the nebula, and even the presence of toroidal rings which may harbor molecular gas. 
Even though our classifications are based solely on the projected shape, with no supporting velocity information, these details are clues to the youth or maturity of the nebular evolution. 

{\it PN G000.8--07.6} (Fig.~5): The bipolar structure of this PN is evident in the [\ion{O}{3}] $\lambda5007$ image. 
The lobe on the east side of the nebula has been truncated, possibly by interaction with the ISM. 
Bipolar morphology is preserved close to the CS, i.e., in the region internal to the bow-shock, where the shaping mechanism  is not expected to be disrupted (e.g., see Villaver et~al. 2012). 
The bipolar lobe opposite to motion has grown mostly unaltered. 
The morphology is characteristic of a small relative velocity interaction. 
Unfortunately there is no proper motion information available for this object for the comparison with dynamical models.

{\it PN G014.0--05.5} (Fig.~12): The main body of the PN is elliptical, with ansae that could be bipolar lobes seen in projection. 
The shaping mechanism of this object could be similar to that in the ``Cat's Eye" PN, in at least two phases: Early ejection of a symmetrical shell, and later asymmetrical shaping (Balick et al. 2001, and references therein). Note that PN G026.5--03.0 also shows morphological features indicative of the same two-phase shaping process.

{\it PN G025.3--04.6} (Fig.~14): 
This PN is multi-polar when looking at the [\ion{O}{3}] $\lambda5007$ image with an appropriate intensity cut.

{\it PN G044.1+05.8} (Fig.~15): This is a very faint PN with an apparently very bright CS. 
Given the nebular asymmetry and the comparative brightness of the PN and CS, it is 
possible that the star is not the central remnant, but rather is an unresolved companion or a field star. 

{\it PN G053.3+24.0} (Fig.~17): This PN has a bright central ring (probably a torus in projection), with bubble-shaped ansae extending in a direction that could be orthogonal in projection to the plane of the ring. A faint halo extends more than twice the angular size of the ring, with a surface brightness $\sim1000$ times fainter. Radial structure is evident in the halo, similar to that seen in HST image of NGC~6543 (Balick et al. 2001). The central star is visible in the continuum images, at the center of the ring. 

{\it PN G068.7+01.9} (Fig.~18): There is a clear signature of an interaction of this PN with the surrounding ISM, in the form of a rim or a partial ring to the north through west. This feature is easily visible in broad-band continuum, but is quite faint in [\ion{O}{3}] $\lambda$5007. It is worth noting that the PN 
morphology interior to the rim does not appear to be particularly disturbed. 
The large opening angle of the ring segment could be a bow-shock, 
which would indicate a low velocity interaction through a low density ISM \citep{Vmg12}. 

{\it PN G107.4--02.6} (Fig.~21): The shape of this PN is highly unusual, if not unique, with an elliptical shell surrounding an oddly shaped main body. 
Two stars of nearly equal brightness lie within the nebula, separated by 2\arcsec: the brighter star is located near the center of the nearly elliptical interior shell, and the fainter star lies near the center of curvature of an extension to the interior shell. Each star appears to lie within a small cavity of lower emission. There is a significant brightness enhancement at the east rim of the interior shell. It is almost as if this PN surrounds two central stars, each of which contributed to the shaping (if not the formation) of the PN, unlikely as that may be given their wide separation. Yet a chance alignment within the nebular shell of a field star of nearly equal brightness to the CS also seems unlikely. This object is worthy of follow-up spectroscopy. 

{\it PN G205.8--26.7} (Fig.~22): The shape of this nebula, a round shell surrounding a bright interior ring, with an equatorial brightness enhancement,  is characteristic of some early post-AGB stars. Several of the PNe in our survey display this secondary morphology, while none of the non-compact PNe have shown it.

{\it PN G275.3--04.7} (Fig.~24): This PN also has a bright shell surrounding a brighter ring, which is similar to PN G205.8--26.7, 
except that the equatorial brightness is not as prominent.

{\it PN G285.4+01.5} (Fig.~25): This is a spectacular PN, with remarkable microstructure within the bipolar wings, and on the periphery of the pinched waist. The central star is just visible above the surrounding, bright nebular continuum. There is also a faint arc (invisible with this intensity cut) at the end of one lobe, extending from the west to the northwest, some 8.8\arcsec\ removed from the CS, which suggests an interaction of the outer shell with the surrounding ISM. 

{\it PN G295.3--09.3} (Fig.~27): This PN seems to present extremely early evolution morphology. It is 
modestly extended in the \textit{HST} image, and it shows a strong dual ring-like feature.
The contours are rectangular in the interior (consistent with a bipolar waist), and the orientation of the rings is consistent with bipolar lobes.

{\it PN G309.5--02.9} (Fig.~29): There is clearly a structure reminiscent of interaction between the PN and the ISM. The opening angle of the putative bow-shock reveals a likely low velocity ( $\approx20$ Km s$^{-1}$) interaction.

{\it PN G327.8--06.1} (Fig.~30): This nebula appears to be multi-polar, with plenty of microstructure. The streak running north-south is probably a diffraction artifact. 

{\it PN G344.8+03.4} (Fig.~34): This nebula is very strange. The marginally resolved, bright inner core is shaped like a four-leafed clover, but there is very faint structure in the form of blobs and arcs that are removed from the CS by 5\arcsec\ to 8\arcsec, and lie along one axis of symmetry in the NW and SE. The nebulosity in the outer structure is fainter than the core surface brightness by a factor of 1000. We speculate that a jet may be operative in this very young object. 

{\it PN G348.4--04.1} (Fig.~35): The shape of this bright PN looks like a classic pinched waist of a bipolar, except that the lobes seem to be truncated. The CS is quite prominent in the continuum images. 

{\it PN G351.3+07.6} (Fig.~36): A bipolar core is 
distinguishable in this PN, the outer contours of which also square in shape.

{\it PN G356.5+01.5} (Fig.~36): This is a point-symmetric PN with 
a distinct, asymmetrical arc of emission extending over $180\deg$ on the north side of the nebula. 
This is a clear signature of interaction with the ISM.

{\it PN G358.6+07.8} (Fig.~37): This elliptical PN shows an inner core and an attached shell, both with a good deal of microstructure that suggests a mature nebula. The CS is very prominent, even in F502N.

\subsection{Multiple Ejections, Halos, and Interaction with the ISM}

Several PNe show fainter shell structures surrounding the brighter central main nebula. The faint nebula is likely the remnant ejection of mass from the progenitor star as it ascended the asymptotic giant branch (AGB). These structures also represent a clear examples of a shaping mechanism that changes with time, becoming more asymmetrical. PNe showing these features in the sample of compact objects presented here are PN G014.0-05.5, PN G026.5-03.0, PN G052.9-02.7, PN G053.3+24.0, PN G068.7+01.9, PN G079.9+06.4, PN G184.0-02.1, PN G275.3-04.7, PN G309.5-02.9, PN G336.9+08.3, PN G356.5+01.5, and PN G358.6+07.8.

The stellar motion is capable of influencing the structure and dynamics of the ejected AGB envelope, as the stellar mass-loss interacts with the local ISM.  As the AGB envelope becomes ionized, marking the birth of the PN, the asymmetries developed from the interaction with the ISM are in many cases major morphological features (Villaver et~al. 2012).  About $10\%$ of the compact objects under study show morphological features that clearly reveal the interaction between the stellar ejecta and the ISM. This fraction is similar in the compact and the general Galactic PN samples. Note that when deeper images are taken, reveling the external layers, this fraction tends to increase \citep{Corradi03}. 

\subsection{Comparison to other classification schemes}

Sahai et al (2011, hereafter SMV) studied morphology of pre- and young PNe, and built a morphological classification scheme especially suitable for these  early post-AGB ejecta. They analyzed in depth a sample of such targets, determining several physical parameters, and classified them based on their new scheme, including statistical analysis of the resulting distribution among morphological classes. 

As a corollary, they shown that their pre- and young-PN morphological scheme is applicable to any given sample of PNe. They used some of the WFC3/{\it HST} images publicly available from our snapshot program  to prove their point. Their publication occurred when our  snapshot survey was still ongoing, and images of only 44 PNe were available (hereafter, the {\it preliminary sample}). Their study of the preliminary sample is limited to morphological classification, presented in their Table 5. They did not study the physical parameters, evolution, or distribution amongst morphological classes of the preliminary sample PNe, thus our comparison with their work will be limited to the analysis of the similarities and differences in the classification schemes, and we will also analyze in detail the PNe for which the classification across the two schemes differ.

The morphological class R (round), defined by us for elliptical PNe whose axial difference is $<5\%$, broadly correspond to the R {\it round} class in SMV, where {\it round} PNe have maximum diameter measuring less than 1.1 times their average diameter. There are two PNe in the preliminary sample that we classify as R, which have been also classified as R by SMV.  Our definition is more stringent than that of SMV, thus we expect a few of our elliptical PNe to be classified as round in their scheme.  

Our E {\it elliptical} PNe class does not have a direct correspondence in SMV's scheme; nonetheless, they define E {\it elongated} PNe as those with elongation along one axis, and we expect most of our E PNe to be classified as elongated by SMV. We indeed found that 13 of the 20 E PNe in the preliminary sample have been classified as E (elongated) by SMV. Three of our E PNe are R in SMV's scheme;  these PNe are those whose ellipticity is below 10 $\%$, and their different classification  in the two schemes are compatible within the class definition. The E PN PN~G048.5+04.2 is classified as B (bipolar) by SMV. By looking at Figure 16 (left panel) we see why it could not have been defined elongated. We still prefer the E rather than B classification for this target, but we recognize that B is also a possibility. PN~G285.4+02.2, which is E in our classification, is irregular in SMV's. While this target is definitely elliptical, it may not be exactly elongated, thus the respective classifications are agreement within the individual schemes. PN~G309.5-02.9 is classified as L by SMV, a class typically used only for pre-PNe; both studies detected the same structures (see Fig. 29, left panel), which we note in $\S$3.1. Finally, the E (or possibly P) PN~G344.8+03.4 is classified as S (or, with spiral arms) by SMV, which is compatible with our P class.

Our BC class does not have a correspondence in SMV's scheme. Following the geometrical definitions of the two schemes, BC PNe should all be classified as E (elongated) by SMV. In fact, all 4 BC PNe in the preliminary sample are classified as E (elongated).

Both our and SMV's schemes consider B {\it bipolar} PNe as a fundamental class. We do not distinguish between bipolar and multipolar PNe, while SMV do (M stands for multipolar in their scheme), but we do include a note to multipolarity in our classification table. There are 12 B PNe in the preliminary sample, and 11 of them are classified as B or M by SMV. In addition, the B PN PN~G042.9-06.9 (Fig. 15, left panel) is L in SMV classification, which is compatible with B in our scheme. 

We use the separate class P to indicate point-symmetry, while a similar main class is not included in SMV's scheme. Of the 3 P PNe of the preliminary sample, two (PN~G025.3-04.6 , Fig. 14, left panel; and PN~G356.5+01.4, Fig. 36, right panel) are classified as {\it elongated}, which is how we would have classified if we were working with SMV's scheme. The third one, PN~G296.3-03.0 (Fig. 28, left panel) is classified as B by SMV. While the PN has a pinched waist, we do not see the axial-symmetry that is expected in a bipolar PN, while we do see the point-symmetry. This may be the target where the respective classifications definitely diverge.

We classify 2  PNe in the preliminary sample as irregular: PN~G309.0+00.8 (Fig. 28, right panel) and PN~G98.2+04.9 (Fig. 20, right panel), respectively {\it bipolar} and {\it elongated} in SMV's paper. The former PN may be bipolar (or multipolar) with evolved morphological pattern, although we prefer to classify it as irregular. The latter PN is definitely elongated, but not elliptical, thus both classifications agree within the respective schemes.

To summarize this comparison, there are three PNe (PN~G48.5+04.2, PN~G309.0+00.8, and PN~G296.3-03.0) whose morphology we believe may be different from SMV's classification, once taken into account the different schemes. Misclassification represent a very small fraction of the preliminary sample, thus a comparable analysis as we present in this paper but using SMV's morphological scheme would have very similar outcome.

\subsection {Distances and Physical Sizes}
 
We did not know a priori whether to attribute the small sizes of the target PNe to young dynamical age or large distance. 
PNe would be dynamically younger than 2000~yr if they are angularly smaller than $\sim4\arcsec$ and closer than 6 kpc \citep{Vgm02}. Only by measuring statistical distances  could we establish whether a compact PN is young or merely distant. 
But statistical distances are based on 
apparent radii, which has been impossible for compact PNe before the \textit{HST} images became available. 
It is worth noting that apparent PN radii smaller than 4\arcsec~are always extremely uncertain if measured from ground-based images, thus not sufficient to soundly determine whether the nebula is dynamically young or just located at larger distance. 

Once we measured the photometric radii, the distances to the PNe were determined via the standard statistical techniques, as described by SSV. 
In order to determine the distances we follow Eqs. 8ab of SSV, which give the distances to the PNe if the angular radius, the 5 GHz flux, and the optical thickness parameter $\tau$ are available. 
For our calculation, we used the measured photometric radii ($R_{\rm phot}$) listed in Table~\ref{tab:Morph}. 
We used the 5 GHz fluxes from Cahn et~al. (1992, hereafter CKS). 
In the cases where the 5 GHz flux was not available in CKS, we utilized the $F_{\rm \nu}$(5 GHz) to $F_{\rm H\beta}$ relation (see CKS, Eq. 6), using for input the $H\beta$ flux and extinction constant  from Table~\ref{tab:Morph}.  

The resulting distances, and physical sizes, of the PNe are given in Table~\ref{tab:Dist}, where we give the 
 $F_{\rm \nu}$(5 GHz), the flux adopted for the distance calculation in column (2), its reference in column (3), which also indicated the cases where the 5 GHz flux had been inferred from the $H\beta$ flux. 
Column (4) gives the calculated optical thickness parameter $\tau$, which determines the formula used to calculate the distance (see SSV). 
Columns (5) through (8) give the physical parameters dependent on the distance scale, respectively, the physical radius $R$, the PN distance from Earth $D$, the PN distance from the Galactic center $R_{\rm G}$, and the absolute scale-height from the Galactic plane, $|z|$. 
For PNe whose morphology is markedly bipolar we use the distance scale based on the relation between surface brightness and size, as explained in SSV.  

In addition to those deselected at proposal time, few additional compact PNe may be members of the Galactic halo or bulge population, based on our new distances, and following the prescription by SH10. The possible halo PNe in our sample are PNG~53.3+24.0, PN~G264.4-12.7, and PN~G275.3-04.7; possible bulge PNe are  PN~G000.8-07.6, PN~G356.5+01.5, and PN~358.6+07.8.

\section{The Correlation of Nebular Morphology and Dust Type}

Most (45) of our targets have simultaneous \textit{HST} images and  {\it Spitzer}/IRS observations, thus we can relate the dust type to other characteristics for the Galactic PN sample. We use the dust type classes as described by Stanghellini et~al. (2012). These classes are based on the $4-40~ \mu$m spectra of the PNe, displaying either carbon dust features, which may include either or both aromatic and aliphatic dust emission (carbon-rich dust, or CRD); oxygen-dust features, which may include either or both crystalline and silicates (oxygen-rich dust types, or ORD); or both dust types (mixed-chemistry dust, or MCD). In some cases, there are no visible features above the dust continuum (featureless, or F).  
If we separate the PNe into two morphological groups, namely, ``spherical" (R and E: 26 PNe) and ``aspherical" (B, BC, and P: 19 PNe), we can see evident differences in the dust type distributions of the two groups. In fact, most ($\sim 60\%$) spherical PNe are CRD, while only $\sim 17\%$ of them are ORD. The situation is reversed for aspherical PNe, where 
 a large fraction of PNe  ($\sim 47\%$) are ORD, and only $25\%$ are CRD. MCD PNe are respectively $8$ and $16 \%$ of the spherical and aspherical samples. Featureless PNe are frequent in the spherical ($25\%$)  sample, but not among the aspherical ones ($5\%$). Stanghellini et~al. (2007, 2012) have shown that lack of molecular features above the continuum in IRS spectra are linked to more evolved PNe, both in the Galactic than in the Magellanic Cloud samples. It seems clear that overall the aspherical sample has more un-evolved targets, both in the compact and general Galactic samples.
 
\section{Compact Galactic PNe Compared to the General Galactic Population}

In order to acquire a deeper knowledge of our ensemble of compact Galactic PNe  (hereafter, the ``compact sample") one should study them in the context of the general Galactic population. 
We use the general sample of spectroscopically-confirmed PNe from the \citet{Acker92} catalog. 
Physical parameters for all but the compact sample are taken directly from SH10, where all Galactic PNe in the \citet{Acker92} catalog had been searched for reliable published physical parameters. 
In other words, the database used here is the same as in SH10, with the exception of morphologies, physical sizes, and distances of compact sample, which are from this paper.

There are 938 Galactic PNe whose radii and statistical distances are known to date. The compact sample consists of all 51 PNe listed in Table~\ref{tab:Dist}. In all the following comparisons we include the compact sample in the general Galactic sample, unless otherwise noted. Also, the general and the compact samples have been scaled by number for easy comparison in the plots.

In Figure 6 we show the distribution of physical sizes of the compact sample (shaded histogram) compared to the general sample (thick histogram). It is evident from this figure that selecting apparently compact PNe excludes the most extended nebulae, i.e., those with radii larger than $\sim0.2$ pc when calculated based on the SSV distance scale. But in order to determine what fraction of compact PNe are dynamically young we need to examine their optical thickness, and distance, distributions. 

In order to discriminate between optically-thin and optically-thick PNe, we use the optical thickness parameter $\tau$, given for the compact sample in Table 3.  Optically-thick (or radiation-bounded) PNe have $\tau<2.1$, and their ionized mass depends on the progression of the ionization front toward the outer edge of the nebula. Optically-thin (or density-bounded) PNe have $\tau>2.1$,  their nebular ionized mass is assumed constant in this approximation, and the nebular gas is completely ionized within the nebula. This is based on a very simple PN model, yet it can be used as an additional tool to determine the evolutionary stage of compact Galactic PNe. We found that $<\tau>=3.65\pm1.21$ for the general sample. On the other hand, $<\tau>=2.7\pm0.85$ when calculated for the compact sample exclusively, indicating that a larger fraction of PNe in the compact sample are optically-thick, compared to the general sample. Yet the majority of the compact sample PNe are optically thin, as is the majority of the general sample. 
In summary, dynamically young, optically-thick PNe are more abundant amongst the compact sample than the general sample; nonetheless, they do not represent the majority of the former sample, and most of the compact sample PNe are evolved and optically-thin.

In Figure 7 we show the distance distribution of compact sample PNe (shaded histogram), compared to that of the general sample. All distances have been determined with the statistical scheme by SSV. By selecting angularly compact PNe, as done to build the compact sample, we probe distances that cluster around 10--12 kpc, and also probe the farther reaches of the Galaxy, while the general sample PNe resides within $\sim10$ kpc, and peaks around 5 kpc.  Only a few PNe in the compact sample are closer than 6 kpc, thus dynamically young.

In Figure 8 we show the Galactic distribution of the compact sample  (large dots) with respect to the general Galactic sample (black dots). Here we plot the 
$X_{\rm Gal}=D cos(b) cos(l)$ and $Y_{\rm Gal}=D cos(b) sin(l)$, where $l$ and $b$ and the target's longitude and latitude.  It is important to note that the compact sample PNe generally populate the Galaxy periphery, rather than its center. Generally speaking, PNe in the compact sample are seen at farther distances than the general sample, as also seen in Figure 7. It is worth recalling that  {\it HST} observations did not privileged any particular sector of the Galaxy when observing a subsample of the original target list, thus selection effects are unlikely.  All Galactic PN with known distances, including bulge and halo PNe, have been included in Figure 8; disallowing possible bulge and halo PNe would not change the plot significantly.

More insight can be obtained by looking at the distribution of the compact sample in the $z-R_{\rm G}$ plane, as in Figure 9, where $z$ is the distance form the Galactic plane, and $R_{\rm G}$ is the radial distance from the Galactic center, calculated as in SH10. 
We found that the compact sample PNe (large dots) are not concentrating in any particular location of the Galaxy with respect to the general sample, showing that they represent the general PN population both in scale height and galactocentric radius. The plot shows that the compact sample is not tightly confined to the proximity of the Galactic plane. In fact, the average absolute distance of the compact sample PNe from the plane is $<|z|>=1.42\pm1.57$ kpc, compared to  $<|z|>0.66\pm1.22$ kpc of the general sample. 
This finding agrees with the population of the compact sample being just a subsample of the general population, i.e., these compact PNe are merely, on average, farther away from us than the general population.  Distances from the Galactic center of compact PNe cover the same domain than the general population, as seen in Figure 9, although the average galactocentric distance of compact PNe is $<R_{\rm G}>=10.10\pm5.01$ kpc, against the complete sample with average $6.56\pm4.14$ kpc, showing that compact sample PNe probe the farther reaches of the Galaxy. 

In Figure 10 we show the radial oxygen abundance distribution of compact sample (large dots) and that of the general sample. Here we excluded bulge and halo PNe. 
Oxygen abundances are from SH10, except that the SH10 database has been updated with the abundances by \citet{2014A&A...567A..12G}, from which we chose only those PNe observed directly by the Authors, and  by \citet{2015ApJ...803...23D}. 
By plotting metallicity vs. galactocentric radius, we have a sample of 206 disk PNe whose oxygen abundance is known from direct abundance analysis. 
The radial oxygen gradient of all these PNe is shallow, with slope of $\Delta{\rm log}(O/H)/ \Delta R_{\rm G}=-0.026$ dex kpc$^{-1}$, and intercept
${\rm log}(O/H)+12=8.8$. 
By selecting only compact sample PNe we obtain a metallicity gradient gradient slope of $\Delta{\rm log}(O/H)/ \Delta R_{\rm G}=-0.013$ (dex kpc$^{-1}$) and intercept ${\rm log}(O/H)+12=8.63$. 
The difference in gradient slopes is not significant, given the large scatter of the distributions. 
The average oxygen abundance of the general sample is slightly higher than that of the compact sample, but the number of PNe in the compact sample with known oxygen abundance is still too small to draw evolutionary conclusions based on these averages. 
The fact that compact PNe probe farther Galactic regions implies that this sample has more potential for improving the estimates of the O/H gradient in the Galaxy, and should be followed up with more spectroscopic analysis in the future. 

There are other ways to assess the distribution of our compact sample within the Galaxy. For example,
by comparing the optical extinction distribution of Galactic PNe for which extinction at $H\beta$ is reliable (see SH10), we obtain an average $c_{\beta}=1.25\pm0.87$, which is very similar to what obtained for the compact sample, $c_{\beta}=1.12\pm0.72$. 
This indicates that PNe of the compact sample do not reside at particularly high extinction patch of the Galaxy.

We have compared the frequency of morphological distribution of the compact sample to those of the general sample. 
We could assign 579 Galactic PNe to a definite morphological class, including the compact PNe with known morphology studied in this paper. 
About half of both the general ($48\%$) and the compact ($51\%$) sample are elliptical . The other half is divided between R (approximately $14$ and $4 \%$ of the general and compact samples respectively), BC ($\sim15$ and $8 \%$), B ($\sim19$ and $31\%$), and P ($\sim6$ and $3\%$). 
If we group B and BC PNe into one class, then the percentages of these PNe are very similar for the general ($35\%$) and the compact ($39\%$) samples, which may indicate that the different BC and B fraction of the two groups may be due to higher definition of the \textit{HST} images. 
In fact, most of the morphological class assignment of the extended PNe is based on ground-based images. Similarly, the higher fraction of R and P PNe in the general group may also be due to an image definition issue. 
We are inclined to believe that the morphological distribution of the compact sample is very similar to that of the general sample. 
This strengthens the conclusion that the compact sample is representative of the general PN population in the Galaxy. 

\section {Discussion}

The main goal of this project was to present a morphological catalog of compact, Galactic PNe, through narrow-band images (Appendix A), and to explore the detailed properties of this understudied class of PN. 
The observed targets were selected based on their compact size. 
The 130 original targets represent  a large fraction of all spectroscopically-confirmed Galactic PNe whose ground-based measured radius is smaller than 4\arcsec. 
The further selection down to the 51 PNe presented here is due mostly to random sky position available for the snapshot program to be activated, thus is similarly distributed to the original selection, although it could slightly favor brighter PNe, which are observable in shorter time slots. Our compact sample is representative of all compact PNe in the Galaxy. 

We did not know a priori that this sample would result in a representative Galactic PN sample, which is slightly skewed to more distant PNe, and with a higher-than-average optically-thick, radiation-bounded PN fraction. One of the original goals of this survey was to study the early onset of morphology, assuming that the majority of compact PNe would be also young, i.e., recently ejected, and with a short dynamical age. 
This does not appear to be the case after studying the sample against a homogeneous Galactic sample of both compact and extended PNe. 
Most of the compact sample PNe are not in an earlier dynamical evolutionary stage than the general Galactic PN population, although the fraction of optically-thick PNe is higher in the compact sample ($\sim40\%$) than in the general sample ($\sim 11\%$ ).

We perform morphological classification of the nebulae based on projected shapes. The morphological distribution of PNe in the Galaxy is not very different if the PNe are compact or extended, as it appears, if we consider the broad morphological groups of spherical (R and E) and aspherical (B, BC, and P) PNe. 
On the other hand, there are notably more lobed B PNe among the compact than the general sample, with BC/B number ratio being $90\%$  in the general group and only $30\%$ in the compact sample.  An important distinction between the compact and general samples is that morphology of compact PNe are all from \textit{HST} images, thus the definition of certain details may be superior to those of ground-based images which constitute the majority of all PN images available to date for Galactic PNe. This could augment the relative number of B PNe relative to the BC PNe that one would observe from the ground at a lesser resolution. A direct comparison of our morphological classification with that of SMV's shows differences in only 3 targets, making both classification schemes reliable and non subjective.

To reinforce the conclusion that compact sample belongs to a general population is their extinction and spatial distributions. They are distributed through the Galaxy, with relative predominance toward the larger galactocentric radii, making them ideal probes to determine Galactic metallicity gradients. Compact sample PNe may be on average brighter than the general population. 
It is beyond the scope of this paper to perform a detailed study of the luminosities of the two populations, but we can calculate the [\ion{O}{3}] $\lambda5007$ equivalent magnitudes for all PNe discussed here that have a flux determination, and determine their absolute magnitude distribution based on the statistical distances. 

In Figure 11 we show the absolute equivalent magnitude distribution in the light of  [\ion{O}{3}] $\lambda5007$ for the compact sample (shaded histogram) and the general sample (thick histogram). 
The  [\ion{O}{3}] $\lambda5007$  equivalent absolute magnitude has been calculated from the standard formulation:  $M_{\rm 5007} = -2.5 {\rm log}(I_{\rm 5007}) - 13.74$, where $I_{\rm 5007}$ is the absolute intensity in the line at 10 pc. 
There are 778 PNe, 45 of which compact, whose absolute  [\ion{O}{3}] $\lambda5007$ magnitude is known, and that have been included in Fig. 11. 
The compact PN distribution peaks at lower absolute magnitudes than the general sample, and their distribution also lacks the lower luminosity tail that is seen in the general sample. This strengthen the finding that compact Galactic PNe, while similar in population than the general Galactic PN sample, are a brighter subset of the general Galactic sample, and can be found at larger distances.

\section {Conclusions and Future Work}
We have present the catalog of compact Galactic PNe, based on \textit{HST}/WFC3 images. 
We have determined photometric radii, distances, morphological types, and other physical characteristics of this nebular sample, which has been homogeneously and unbiasedly selected form the general Galactic PN sample, based on apparent sized being smaller than 4\arcsec. 

The compact sample presented here has unique attributes, in the simultaneous availability of their \textit{HST} images---thus photometric and morphological properties presented here, and the central stars studies presented elsewhere---and Spitzer IRS spectra, ideally suited to disclose their dust and molecular properties. Once high quality optical spectra become available the abundances will be highly accurate because the collisionally excited IR lines in our IRS spectra offer more diagnostics and ionic species not seen in the optical. An optical spectroscopic program is already underway.

Morphological analysis of the PNe distributes them into the commonly used macro- morphological classes, with similar spherical to non-spherical number ratios than the general population, except the compact sample includes more bipolar members than the general sample. 

We determined physical radii and distances of the observed PNe, examined their distribution within the Galaxy, and compared it with that of the general Galactic sample. We found that not all PNe in the compact sample are dynamically young, and the majority of them represents a typical Galactic PN population, with their distribution skewed toward more distant, and brighter, PNe, allowing to peer into the far reaches of the Galaxy.
The studied sample is typical of the larger, general population except that it excludes physically large PNe, and has a higher proportion of quite small PNe.
 
It is worth noting that another sample of compact Galactic PNe would not necessarily have the same properties of the one presented here: let us recall that we did not add any selection bias a priori, and that we confirmed a posteriori that the sample is representative of the Galactic PN population. 

Our conclusions have a broader impact. In particular, in the next decade or so astronomers will finally be able to determine absolute physical properties of Galactic PNe based on GAIA distances, which are expected to be much more accurate than the statistical distances. We will be able to use directly the GAIA distances and all the related physical properties presented here and elsewhere of this particular Galactic PN sample as the most complete to date. 

\appendix

\section{The Catalog of compact Galactic PNe}

In this Appendix we show the images of all compact Galactic PNe observed. Figure Set 12 shows the compact Galactic PN images through the F502N filter, from our program, except for PN~G348.8-09.0, where we show the F200LP image (its F502N image is not available). Typically, images are $15\arcsec \times15\arcsec$, with the exception of the few very large PNe, with a log intensity stretch. The complete catalog with images in all filters will be published in the near future in a dedicated webpage within the MAST HLSP archive.

\acknowledgements 

Partial support for this work was provided by NASA through grant GO-11657 from Space Telescope Science Institute, which is operated by the Association of Universities for Research in Astronomy, Incorporated, under NASA contract NAS5--26555. E.~V. acknowledges support from the Spanish Ministerio de Econom'a y Competitividad under grant AYA2014-55840P.

\clearpage


\begin{figure}
\plottwo{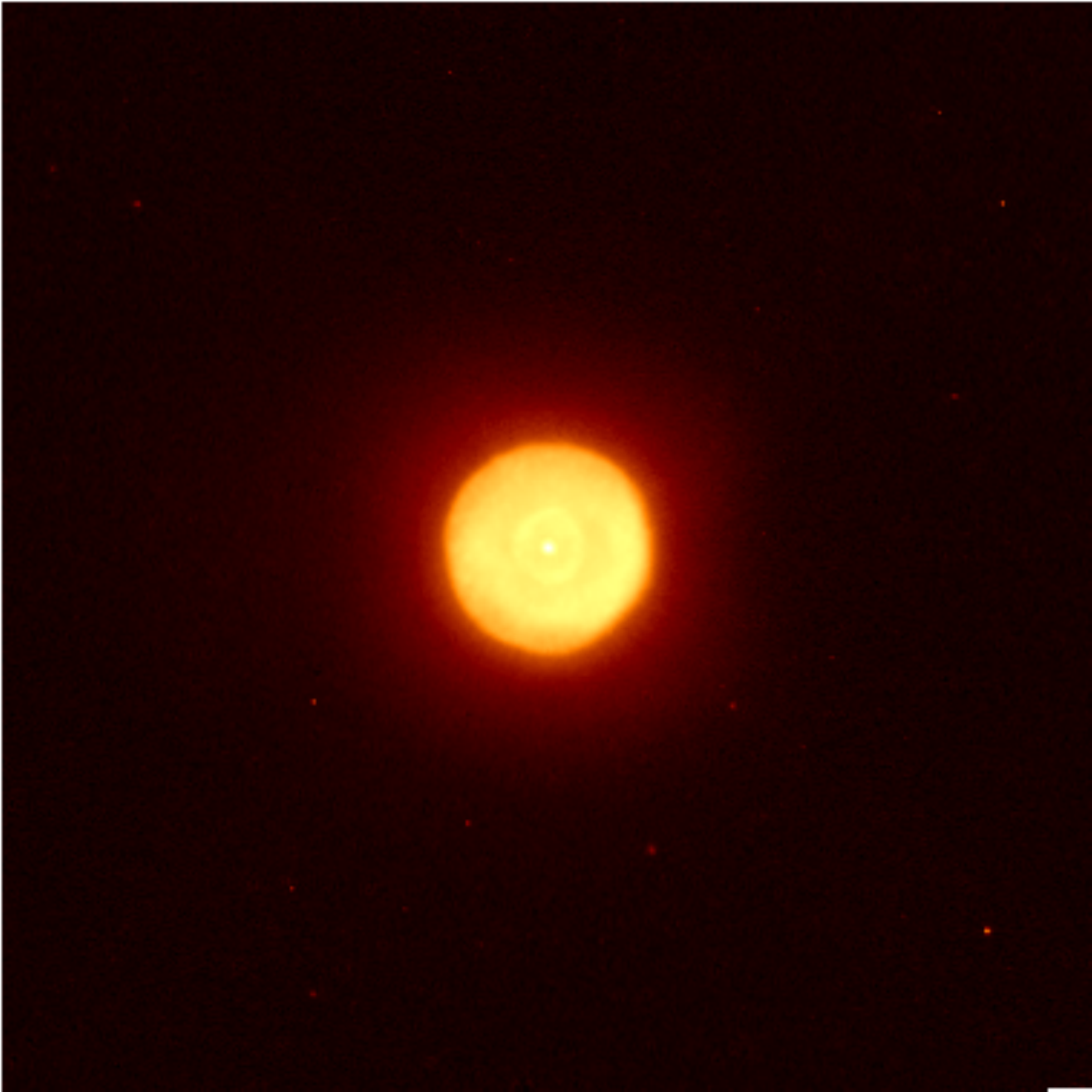}{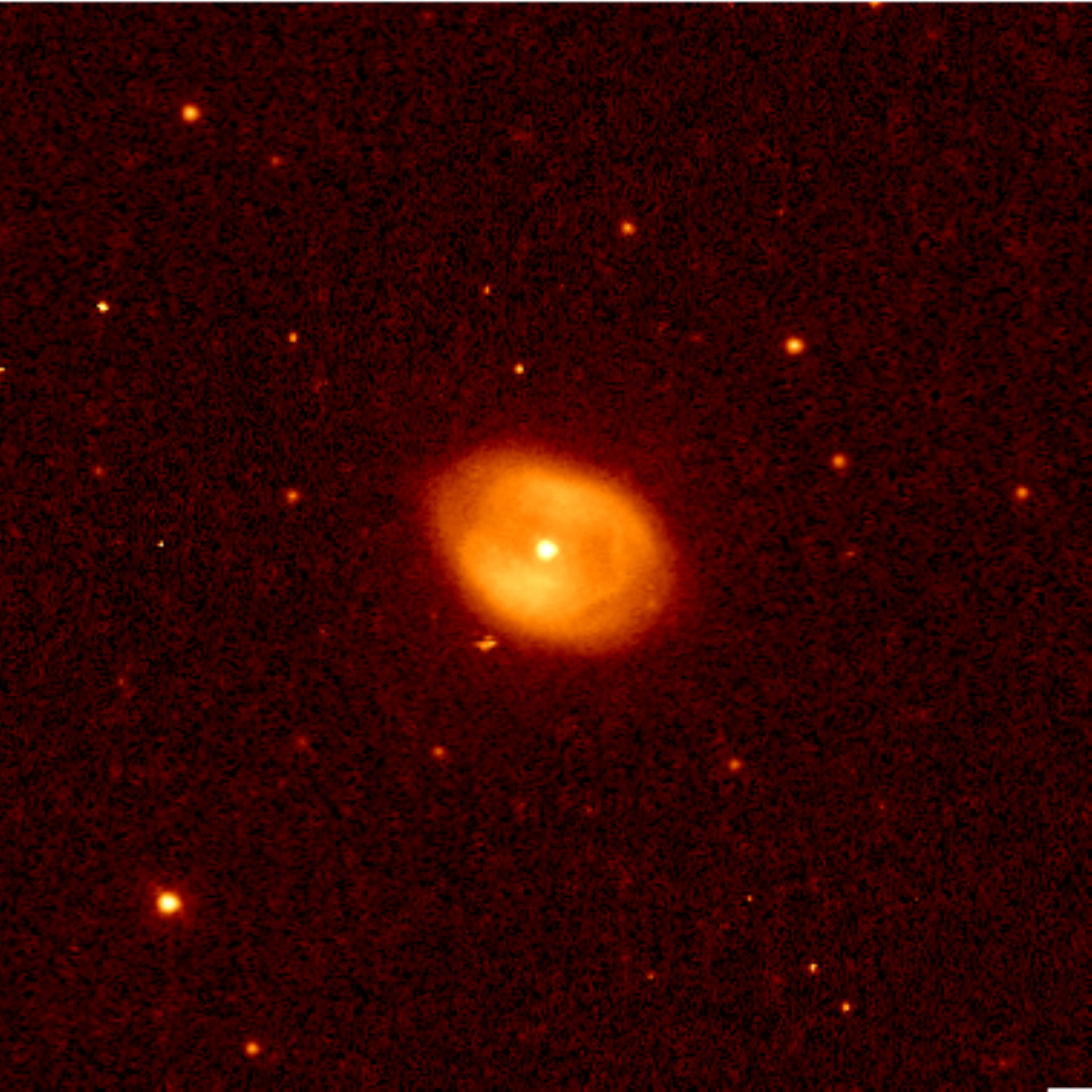}
\figcaption{
False-color images of PN~G264.4--12.7 (\textit{left}) and PN~G327.1--01.8 (\textit{right}) in the F502N filter, illustrating the R and E morphological types, respectively. Note the faint halo around PN~G264.4--12.7. Images are $15\arcsec \times15\arcsec$, with a log intensity stretch. 
\label{fig:RE}
}
\end{figure}

\begin{figure}
\plottwo{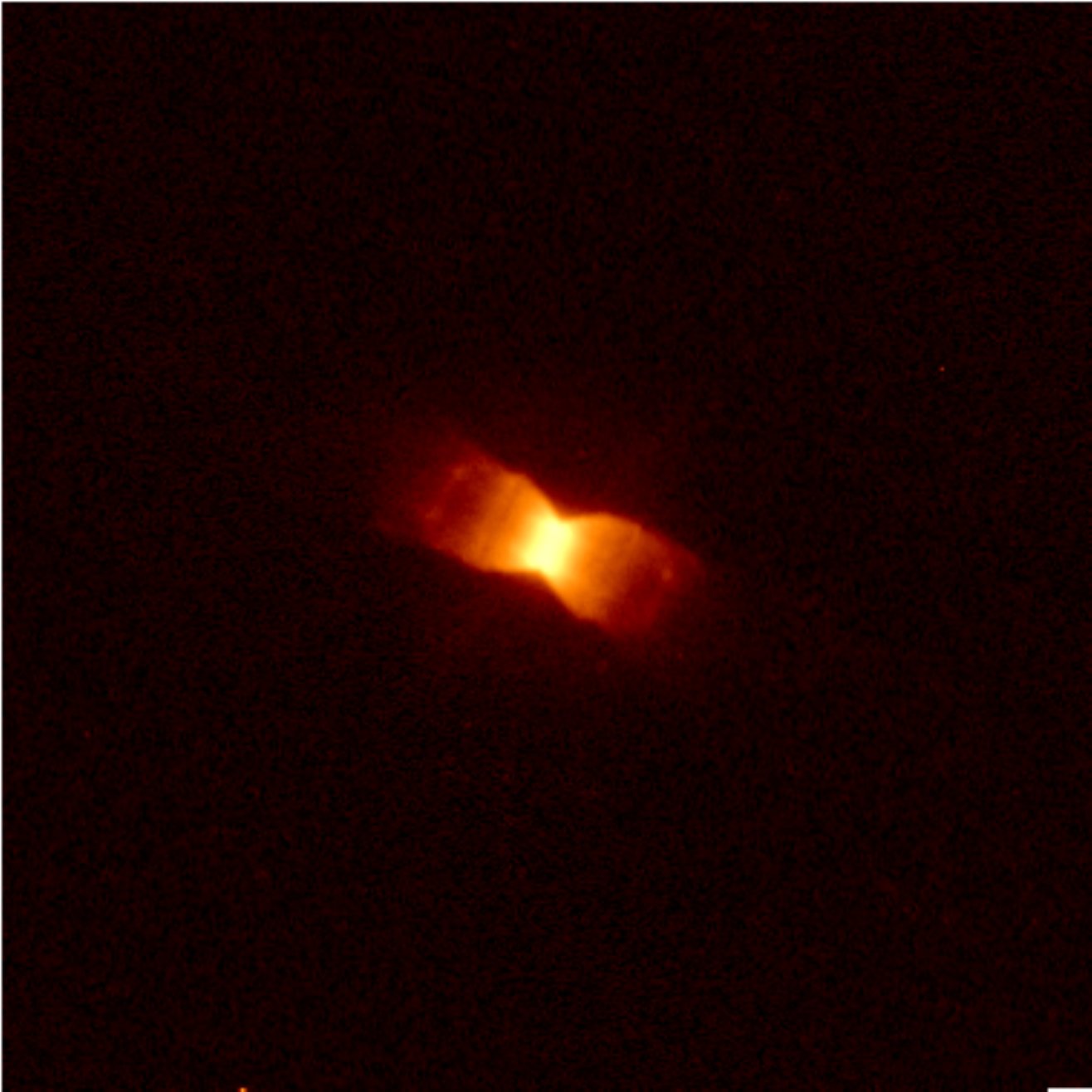}{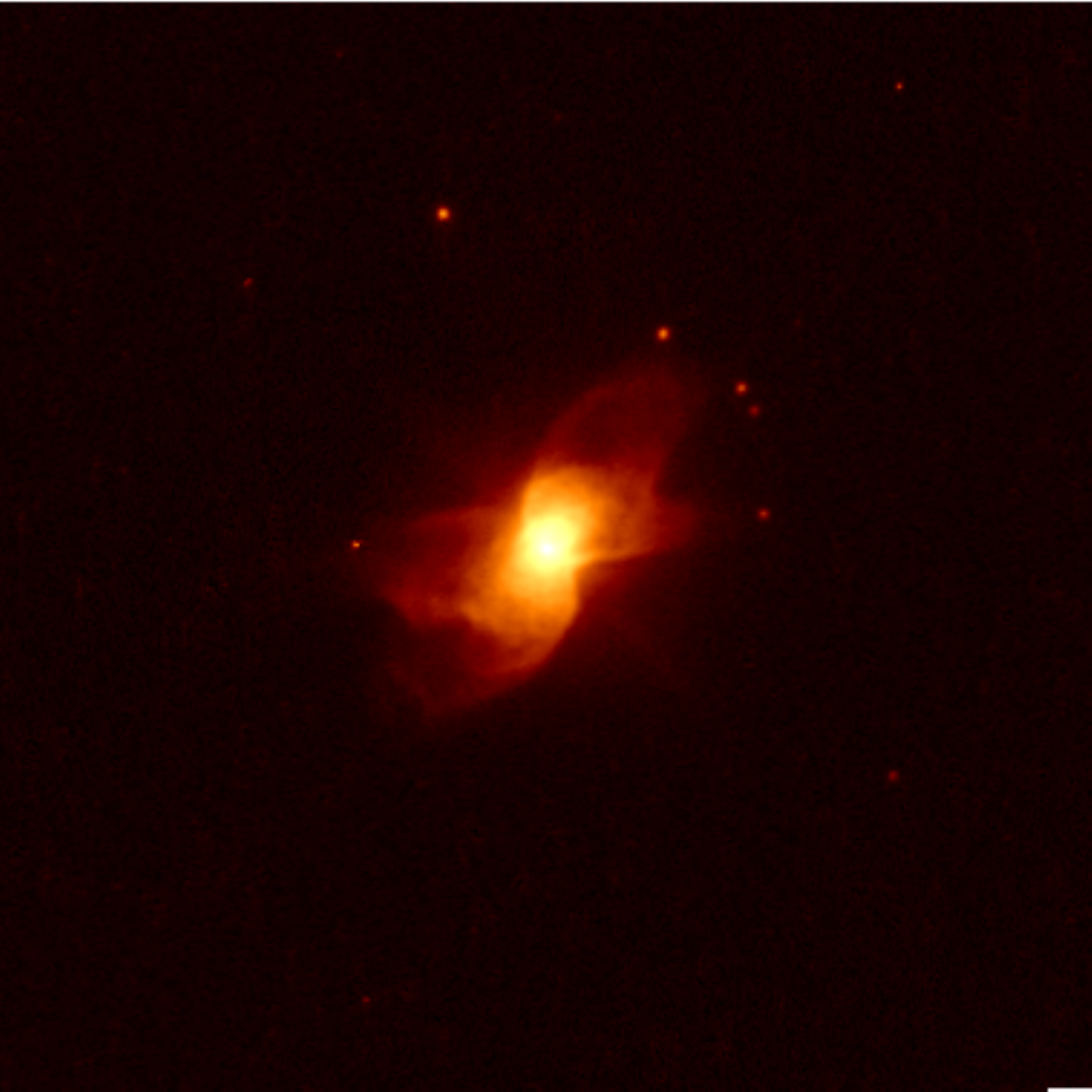}
\figcaption{
Same as Figure~\ref{fig:RE} for PN~G286.0--06.5 (\textit{left}) and PN~G327.8--06.1 (\textit{right}), illustrating the B and multi-polar morphological types, respectively.
}
\end{figure}

\begin{figure}
\plottwo{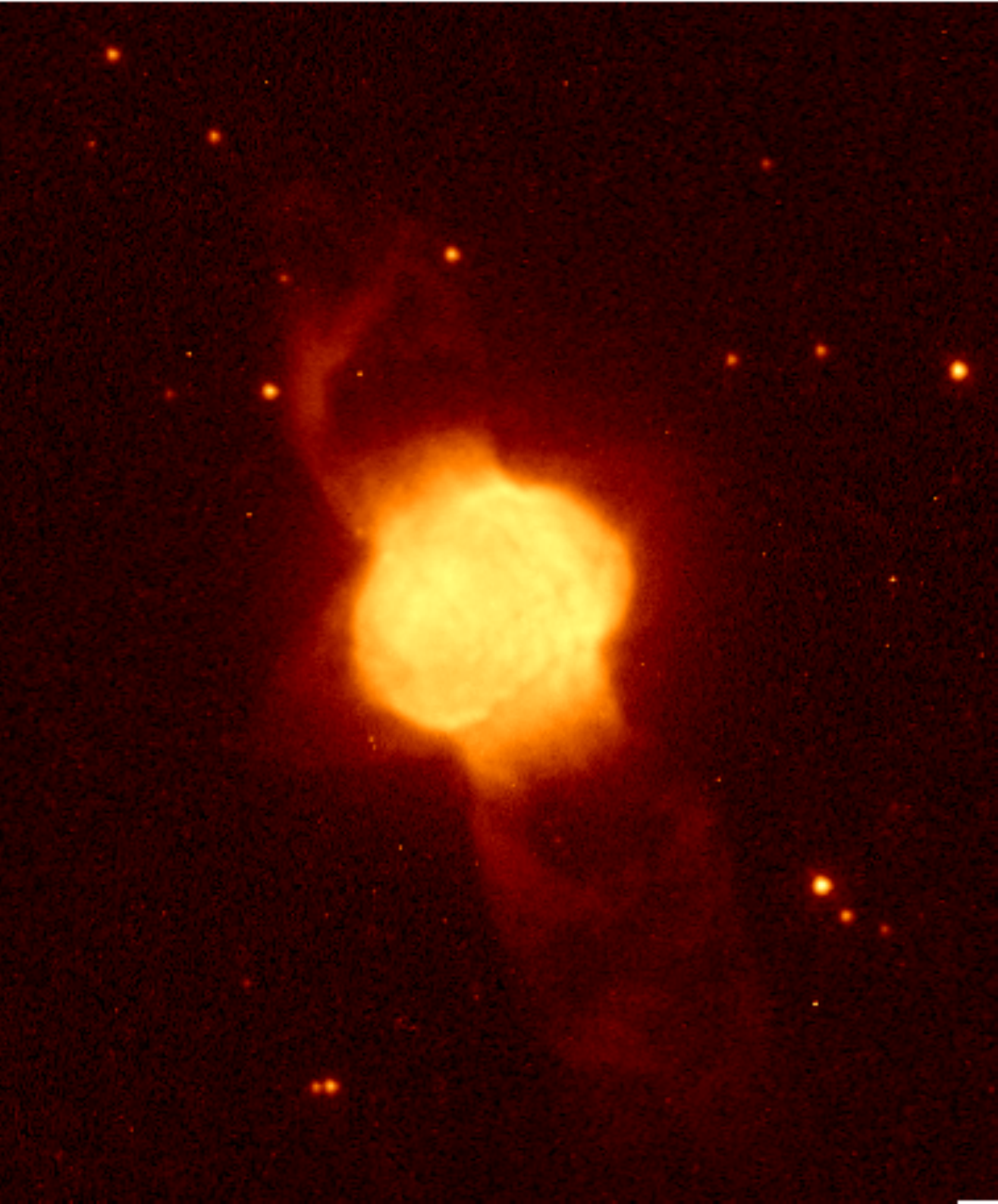}{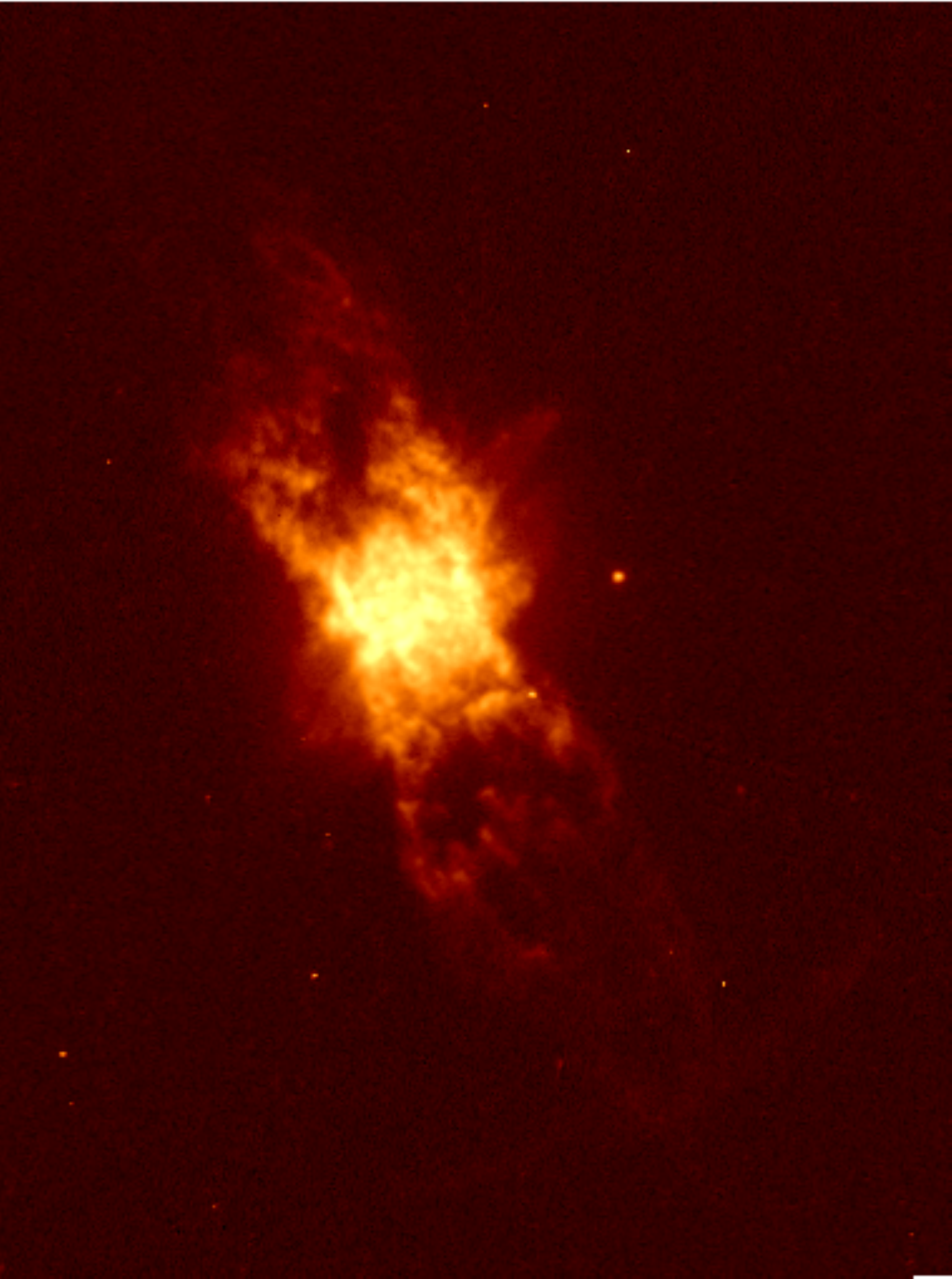}
\figcaption{
Extended bipolar PNe, PN~G021.1-05.9 and PN~G285.4+01.5 respectively.
}
\end{figure}

\begin{figure}
\epsscale{0.5}
\plotone{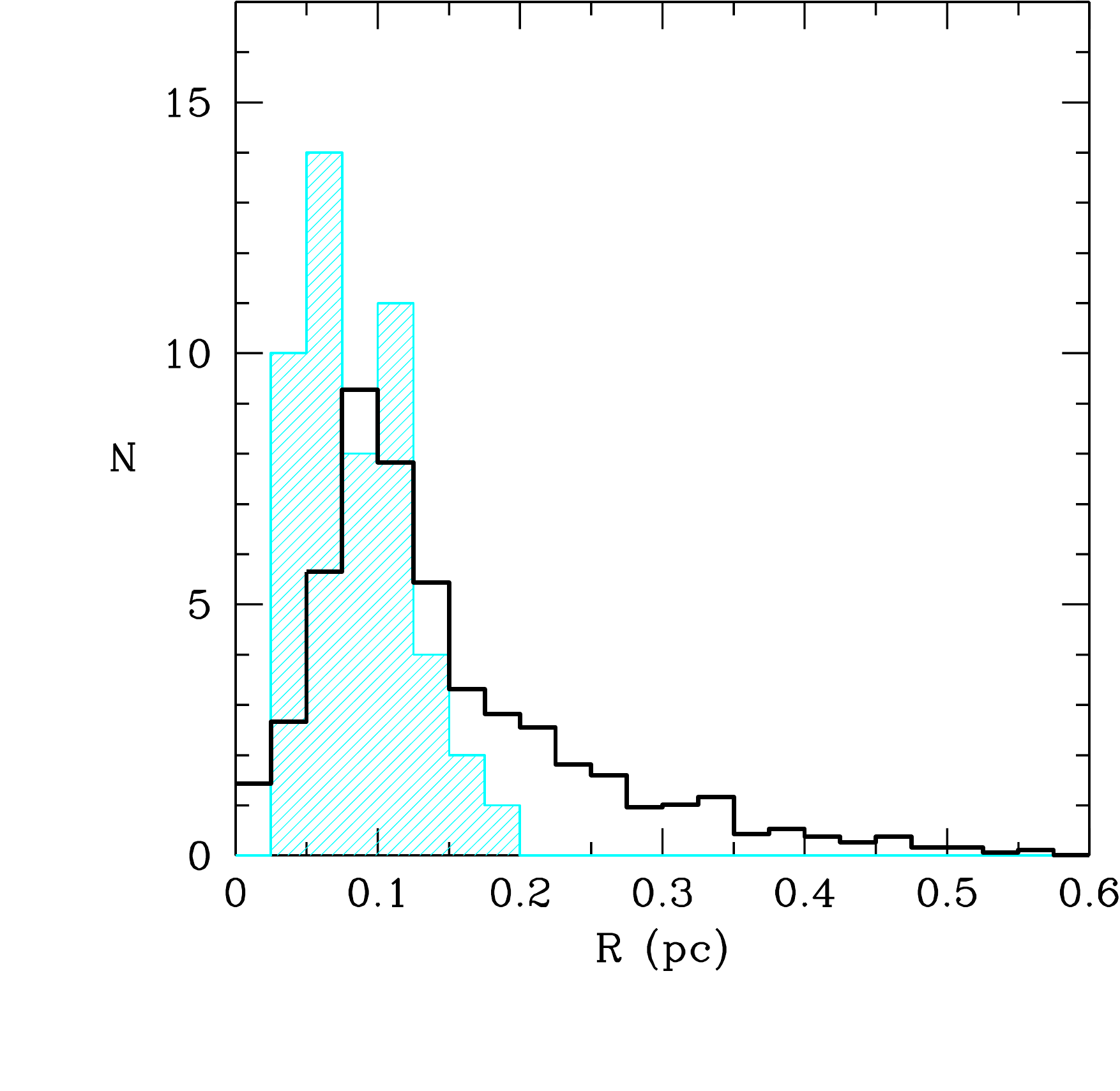}
\figcaption{
Histogram of the physical, radial size of compact Galactic PNe (shaded, cyan) compared to that of all Galactic PNe (black). 
The total number of PNe have been scaled for easier comparison. 
Physical radii have been determined from photometric radii and statistical distances.
}
\end{figure}

\begin{figure}
\epsscale{0.5}
\plotone{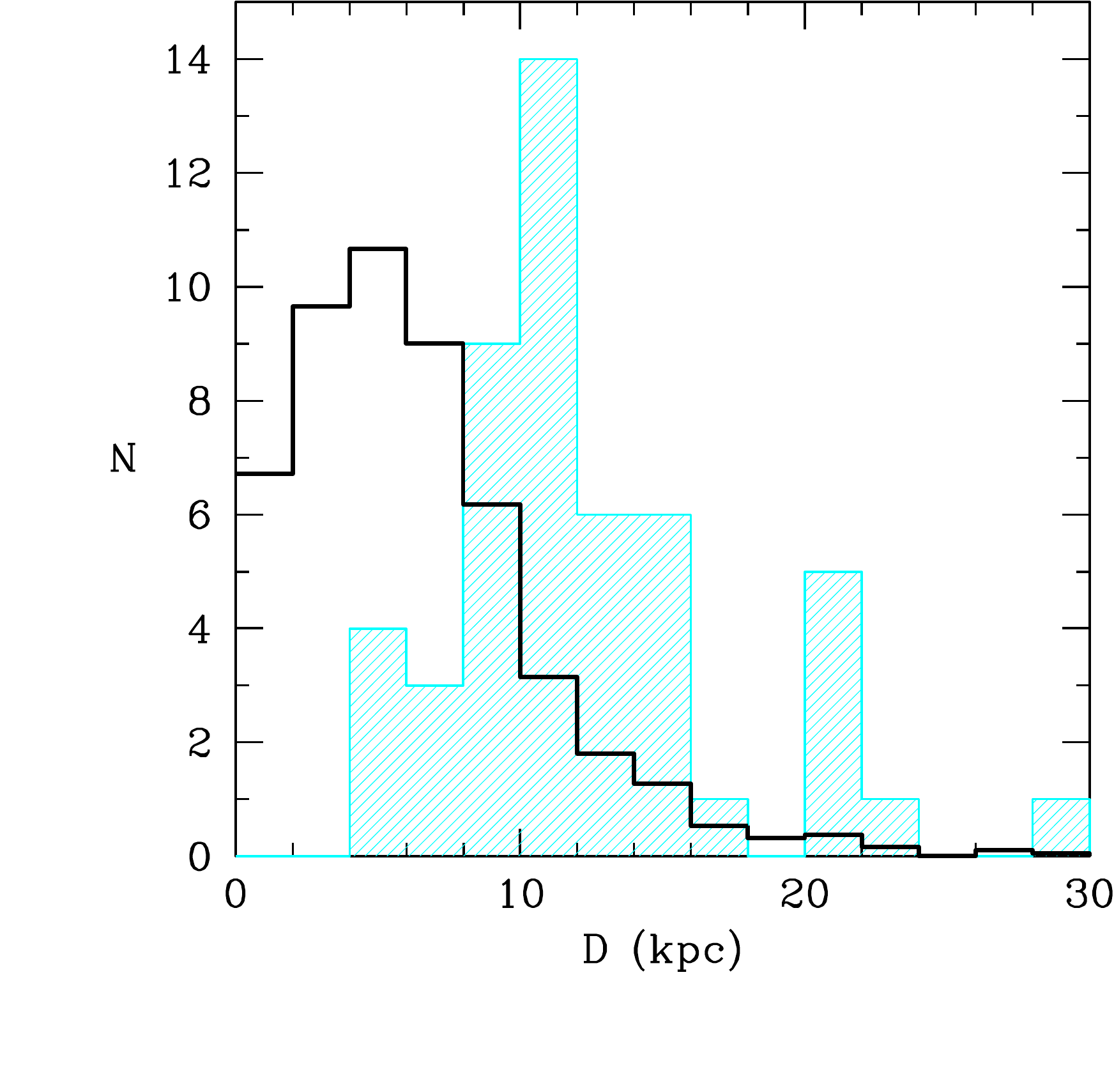}
\figcaption{
Distance distribution of compact Galactic PNe (shaded, cyan) compared to that of all Galactic PNe (black). 
The total number of PNe have been scaled for easier comparison.
}
\end{figure}

\begin{figure}
\epsscale{0.5}
\plotone{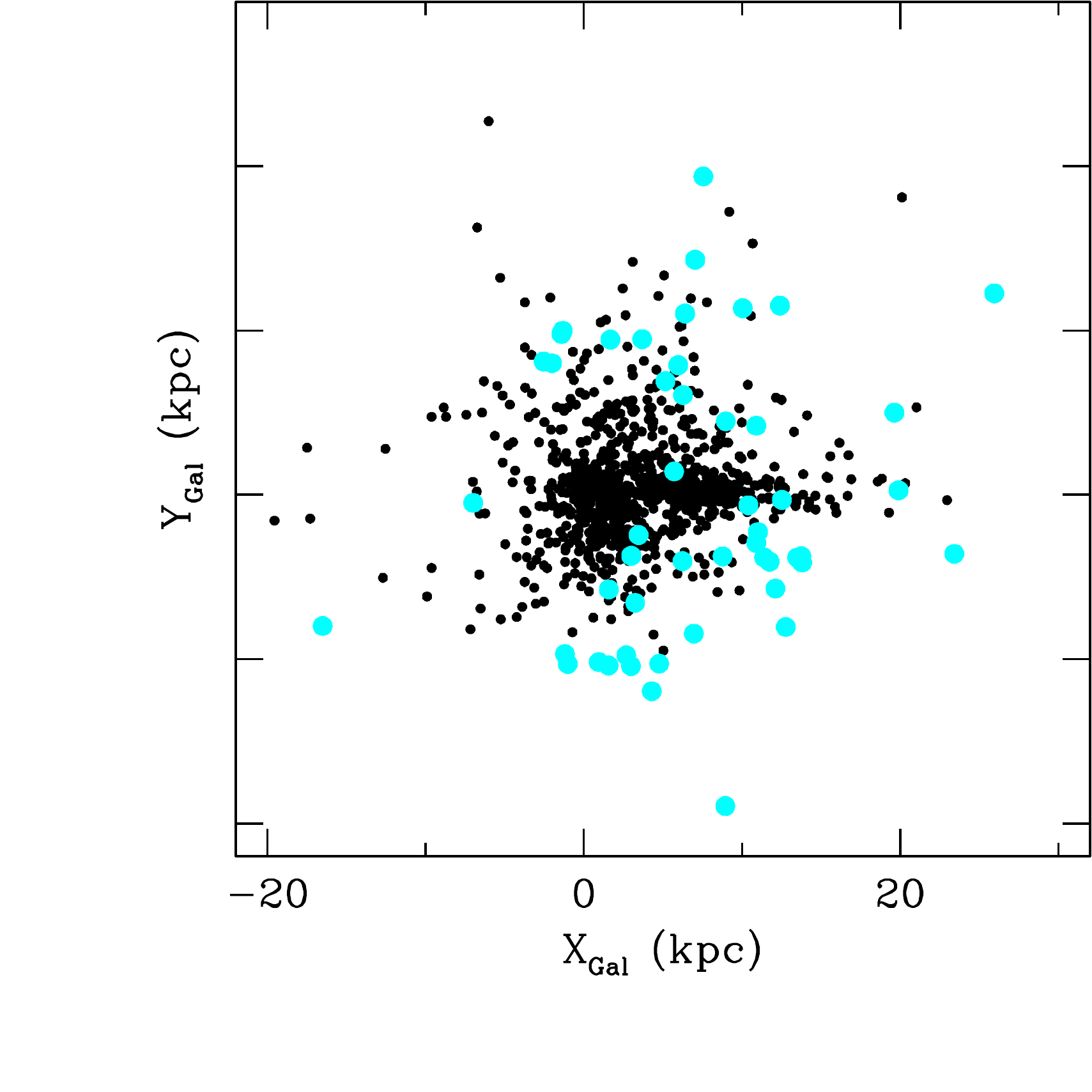}
\figcaption{
Spatial distribution of compact Galactic PNe (large, cyan dots) against the general distribution of Galactic PNe (small, black dots). 
}
\end{figure}

\begin{figure}
\epsscale{0.5}
\plotone{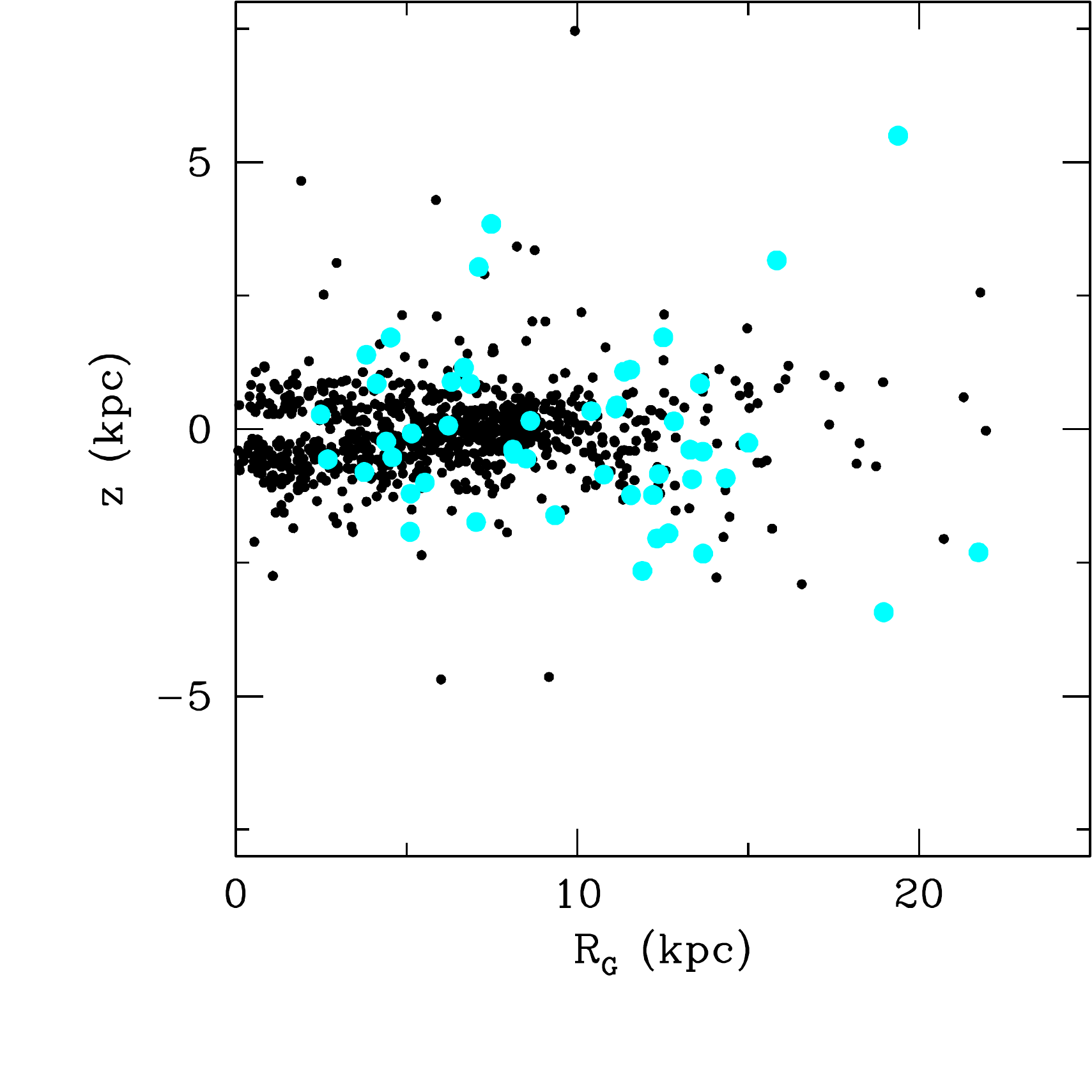}
\figcaption{
Spatial 
Radial distribution vs. distance from the galactic plane, z of compact Galactic PNe (large, cyan dots) against the general distribution of Galactic PNe (small, black dots). }
\end{figure}

\begin{figure}
\epsscale{0.5}
\plotone{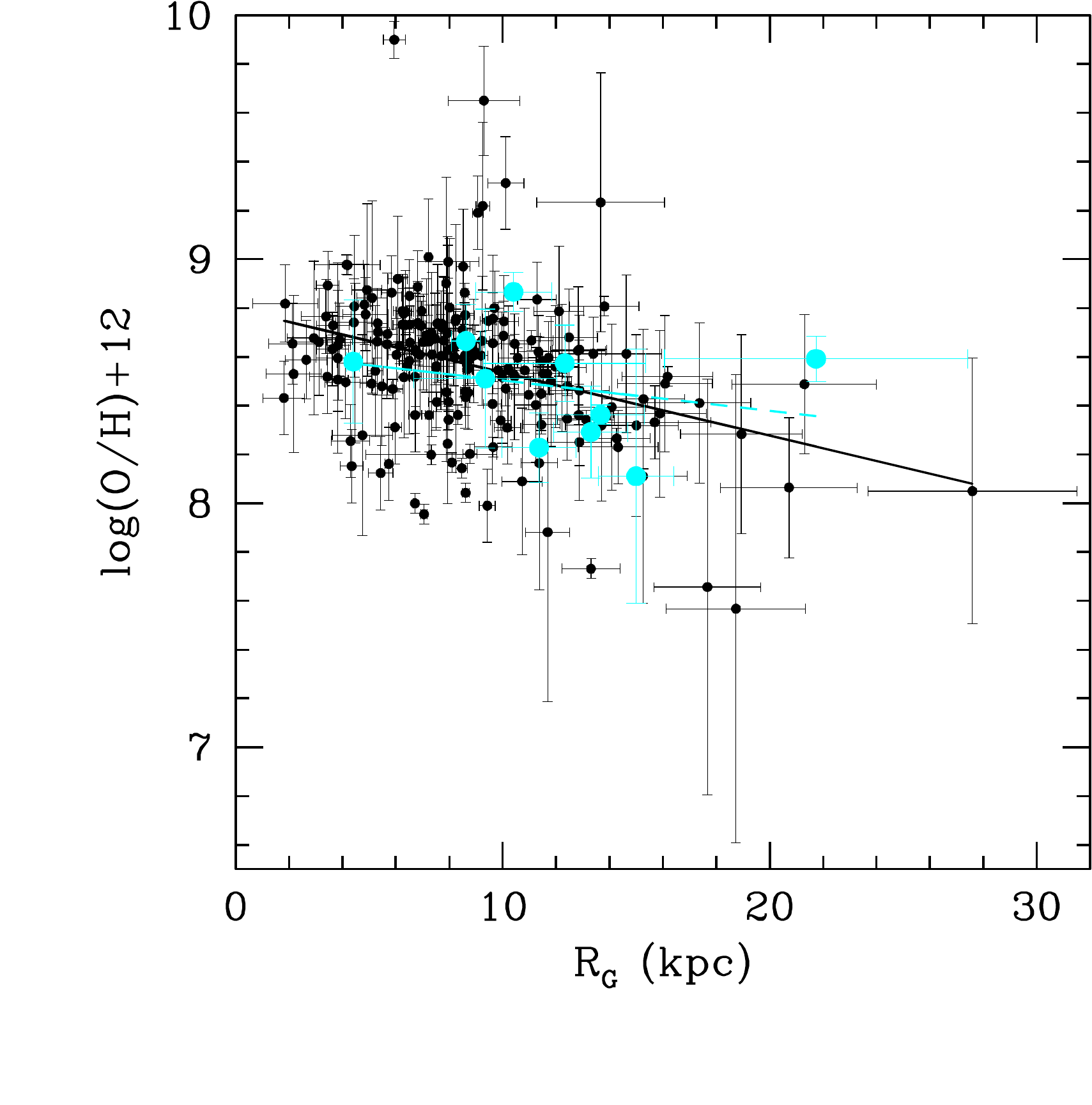}
\figcaption{
Radial oxygen gradient for Galactic PNe, excluding the bulge and halo population. 
The general PN population (filled, black dots) are contrasted with the compact PNe (large cyan dots). 
Gradients have been fit to the complete (solid line) and compact (broken line) populations. 
}
\end{figure}

\begin{figure}
\epsscale{0.5}
\plotone{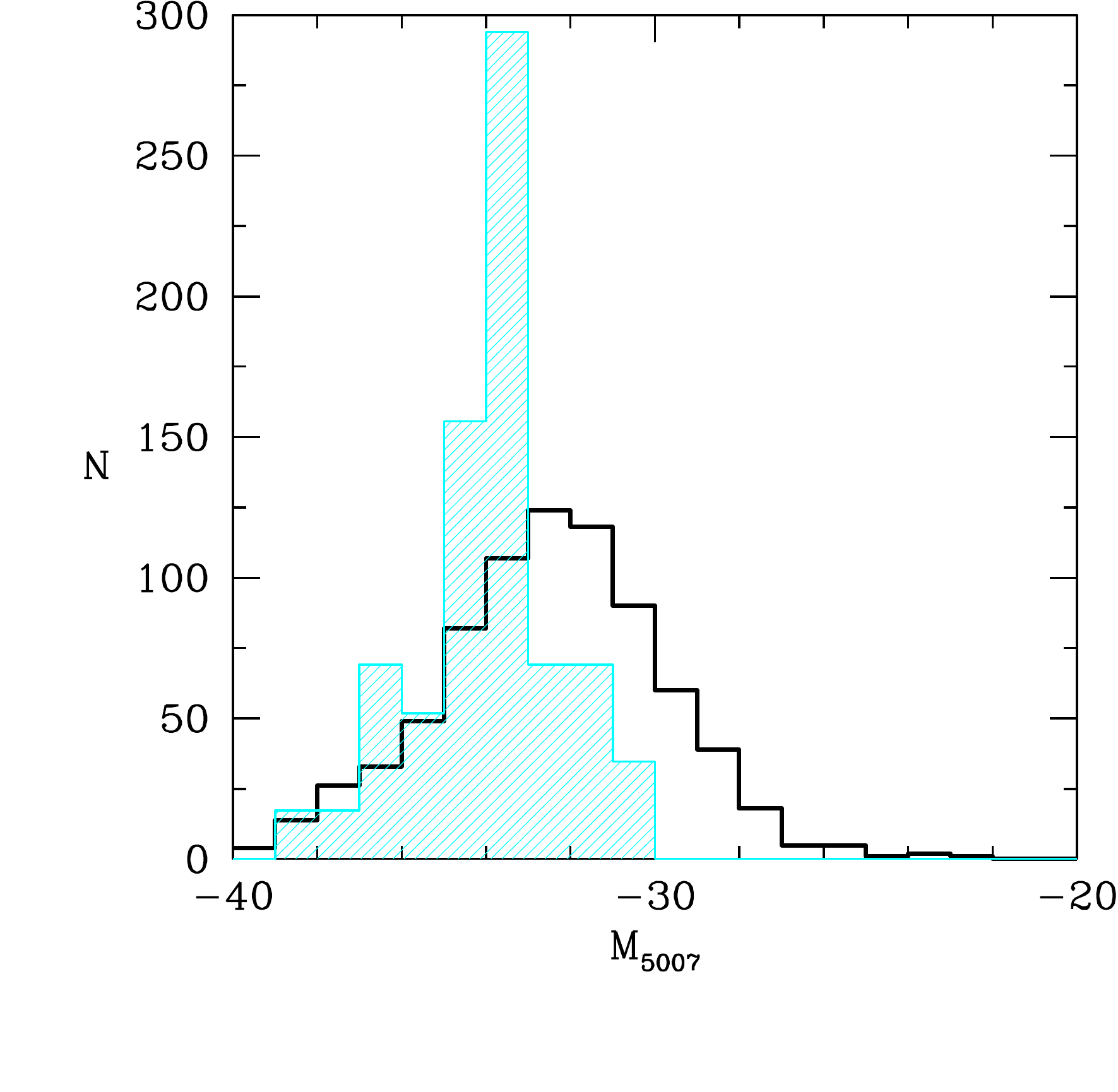}
\figcaption{
The distribution of absolute equivalent magnitudes for Galactic PNe with known [\ion{O}{3}] $\lambda5007$ flux and distance. 
Compact Galactic PNe are represented by the shaded histogram, while the general Galactic sample is in black. 
The two histograms have been scaled for direct comparison.
}
\end{figure}

\figsetstart
\figsetnum{12}
\figsettitle{All Compact Galactic PNe observed}

\figsetgrpstart
\figsetgrpnum{12.1}
\figsetgrptitle{PN~G000.8--07.6}
\figsetplot{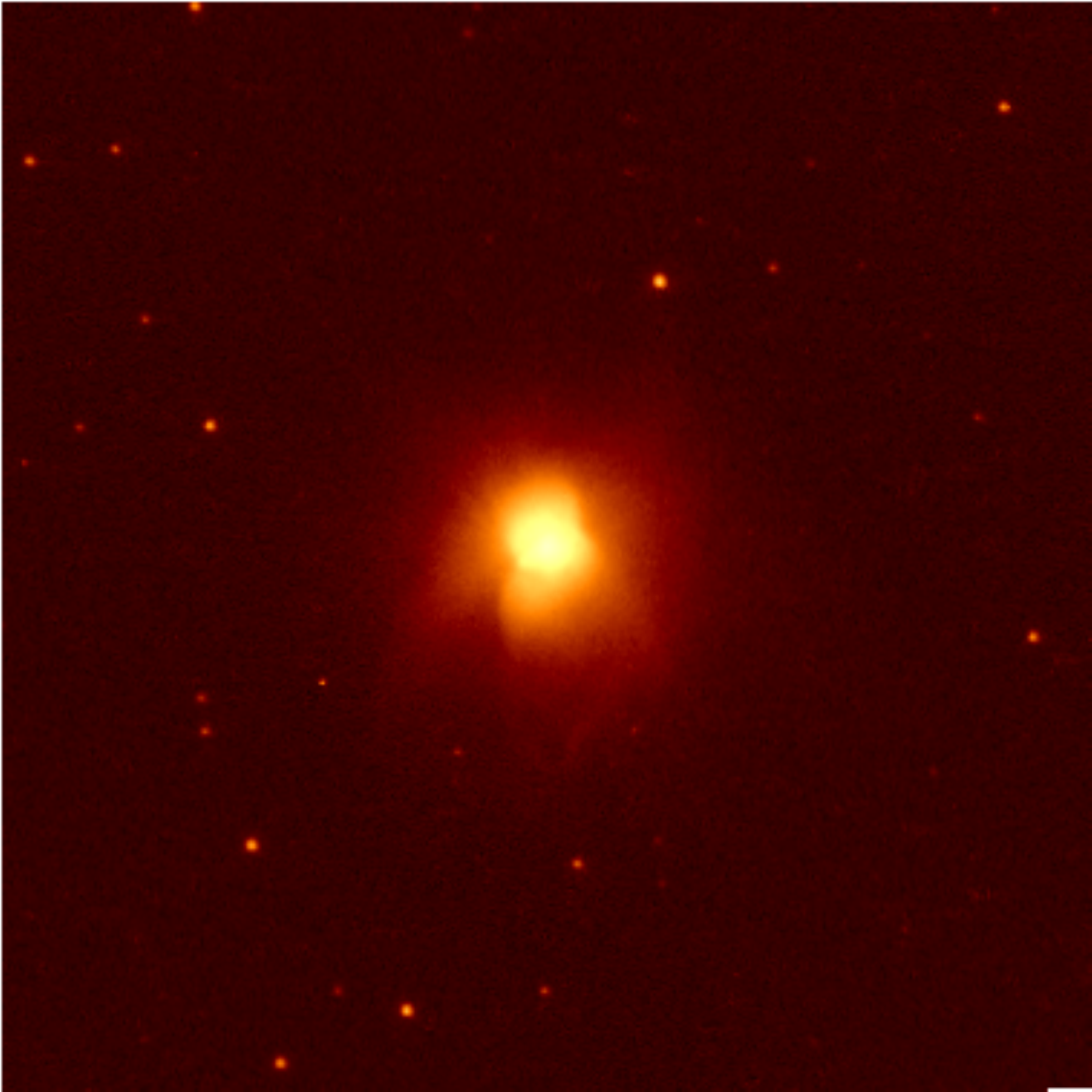}
\figsetgrpnote{False-color image in the F502N filter.}
\figsetgrpend

\figsetgrpstart
\figsetgrpnum{12.2}
\figsetgrptitle{PN~G014.0--05.5}
\figsetplot{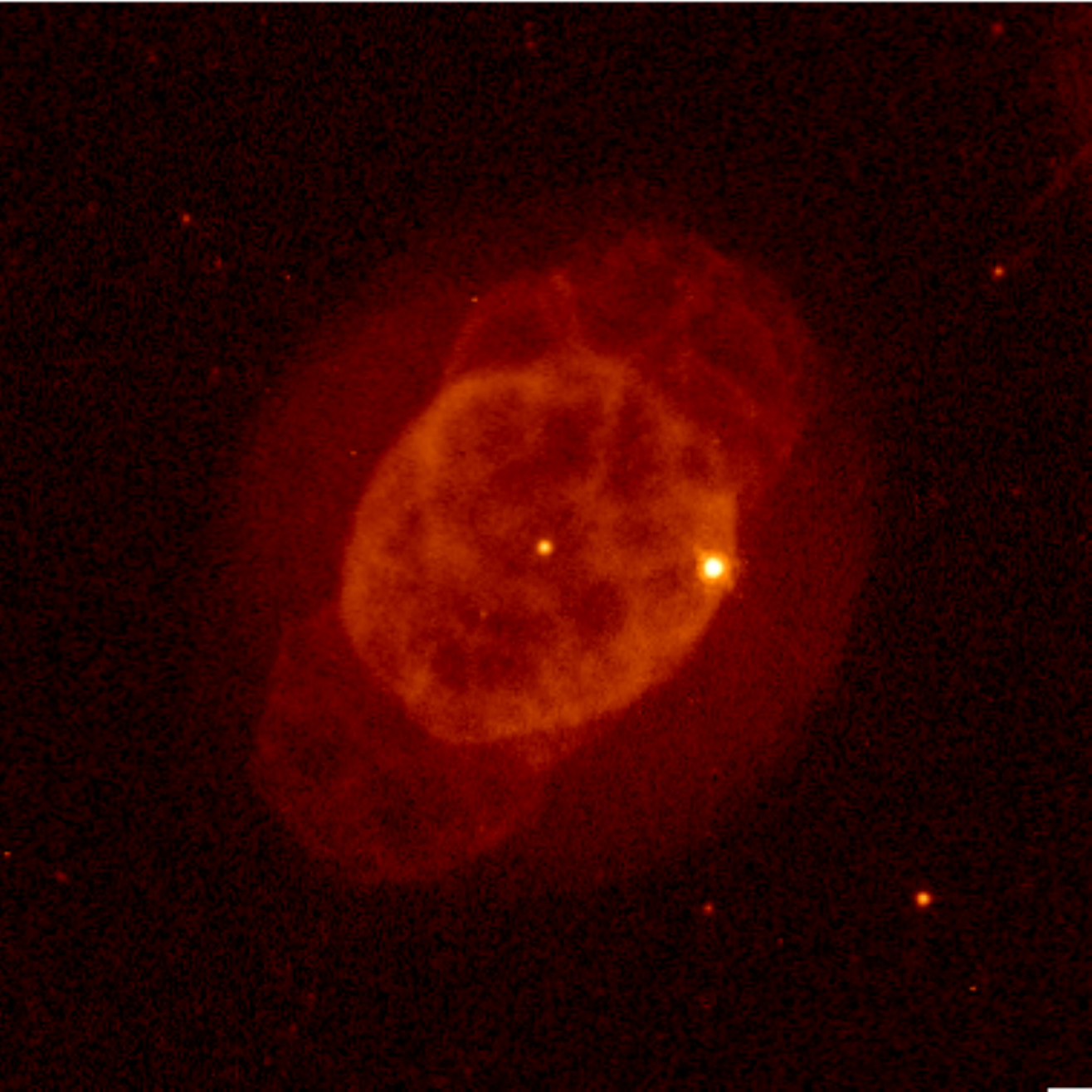}
\figsetgrpnote{False-color image in the F502N filter.}
\figsetgrpend

\figsetgrpstart
\figsetgrpnum{12.3}
\figsetgrptitle{PN~G014.3--05.5}
\figsetplot{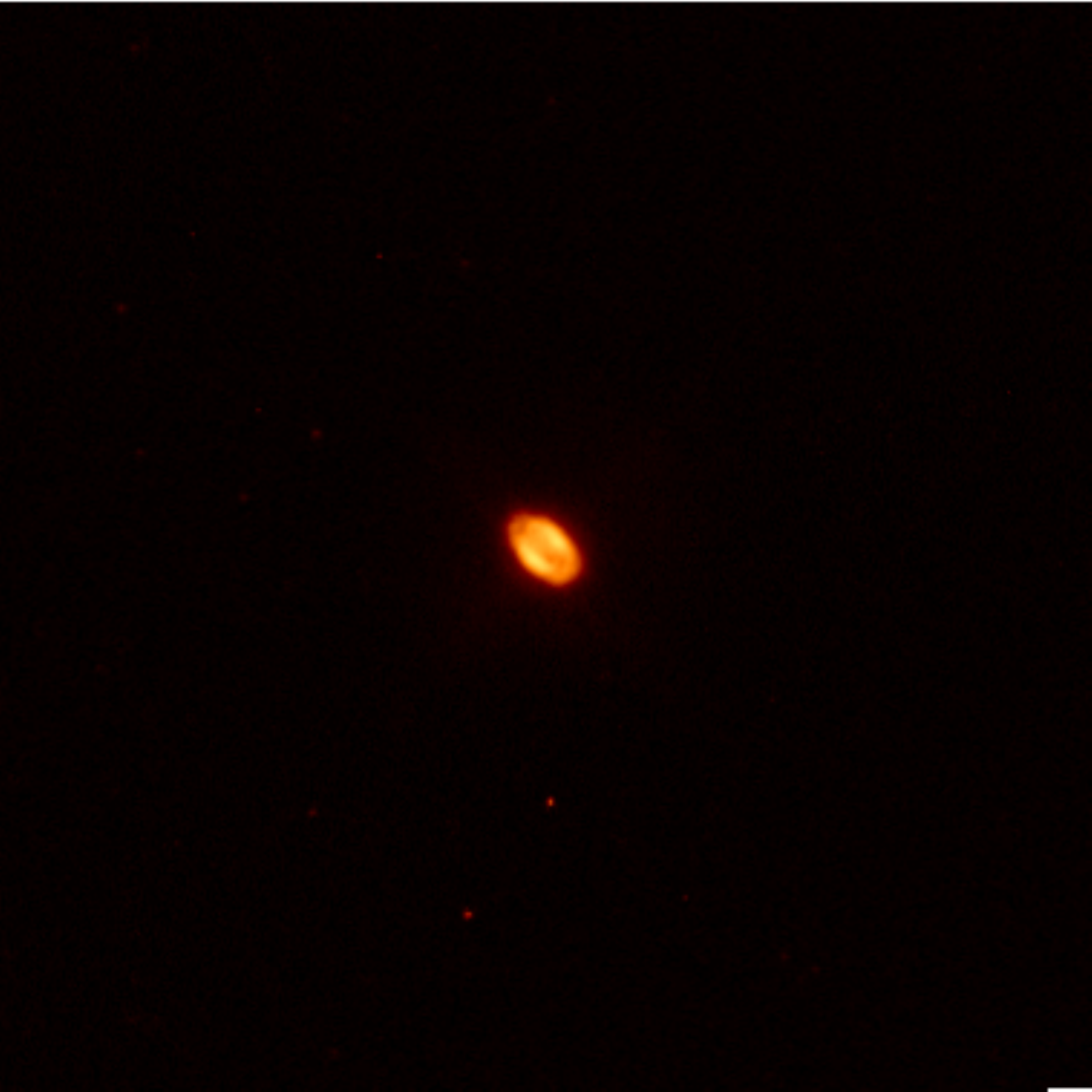}
\figsetgrpnote{False-color image in the F502N filter.}
\figsetgrpend

\figsetgrpstart
\figsetgrpnum{12.4}
\figsetgrptitle{PN~G021.1--05.9}
\figsetplot{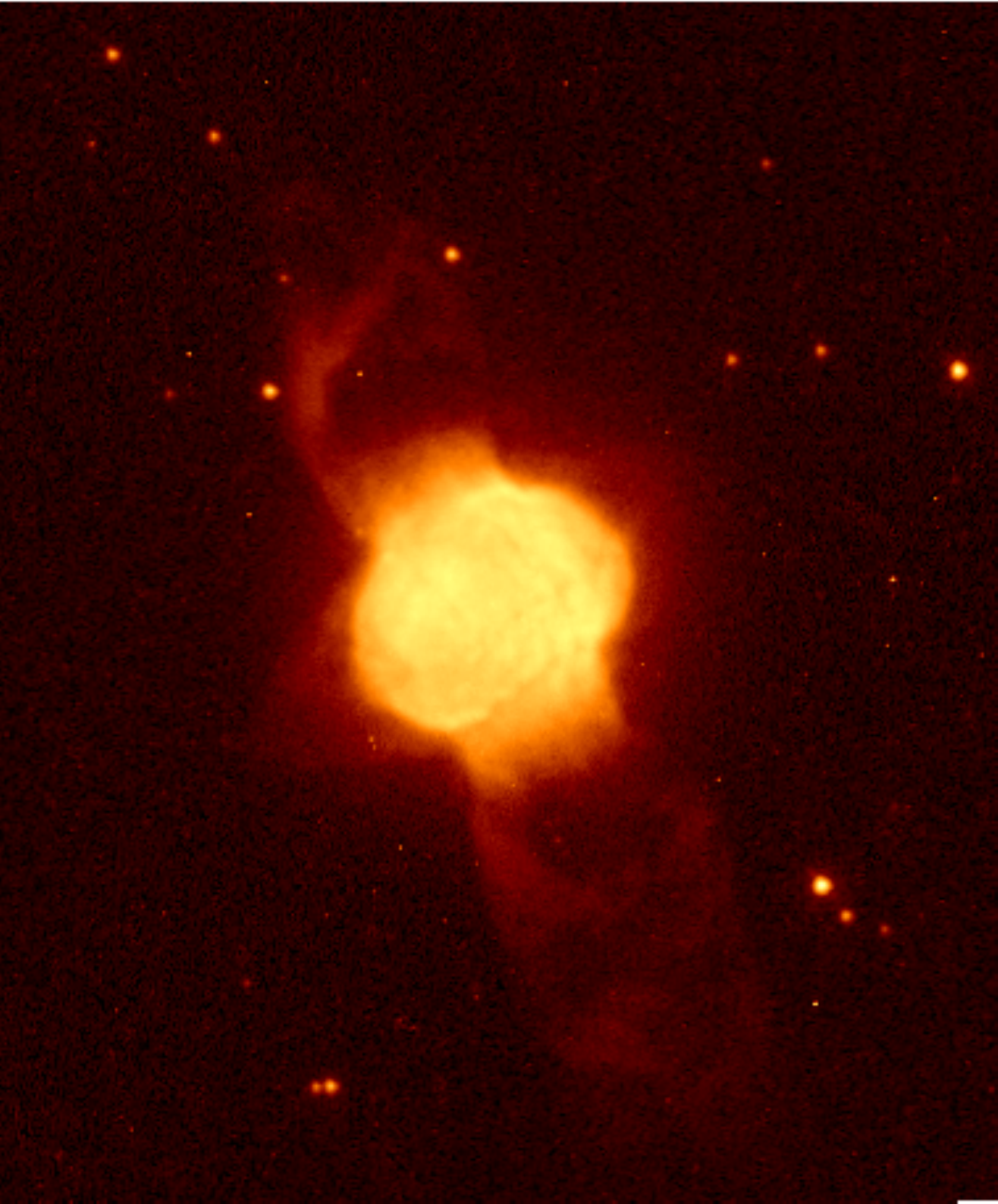}
\figsetgrpnote{False-color image in the F502N filter.}
\figsetgrpend

\figsetgrpstart
\figsetgrpnum{12.5}
\figsetgrptitle{PN~G025.3--04.6}
\figsetplot{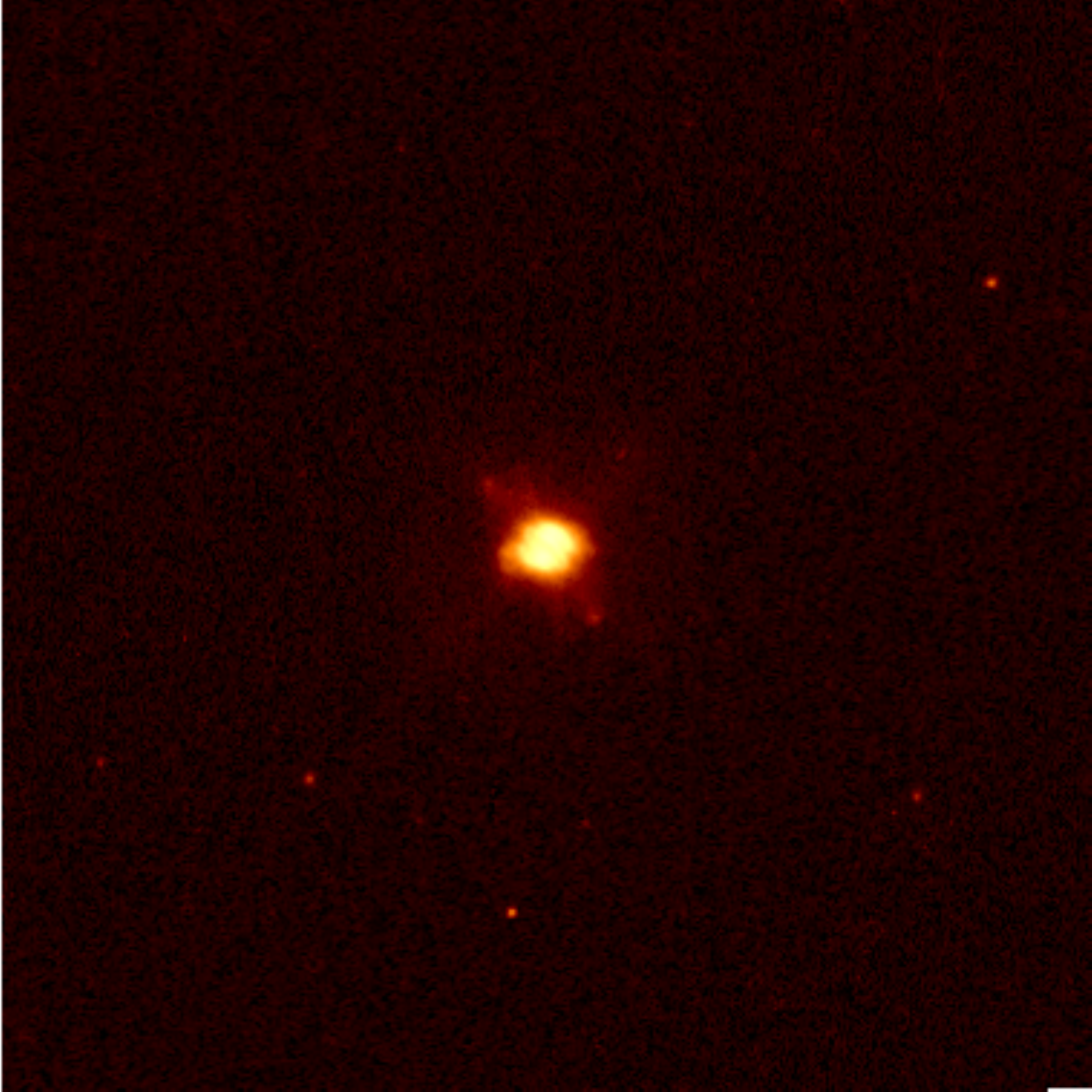}
\figsetgrpnote{False-color image in the F502N filter.}
\figsetgrpend

\figsetgrpstart
\figsetgrpnum{12.6}
\figsetgrptitle{PN~G026.5--03.0}
\figsetplot{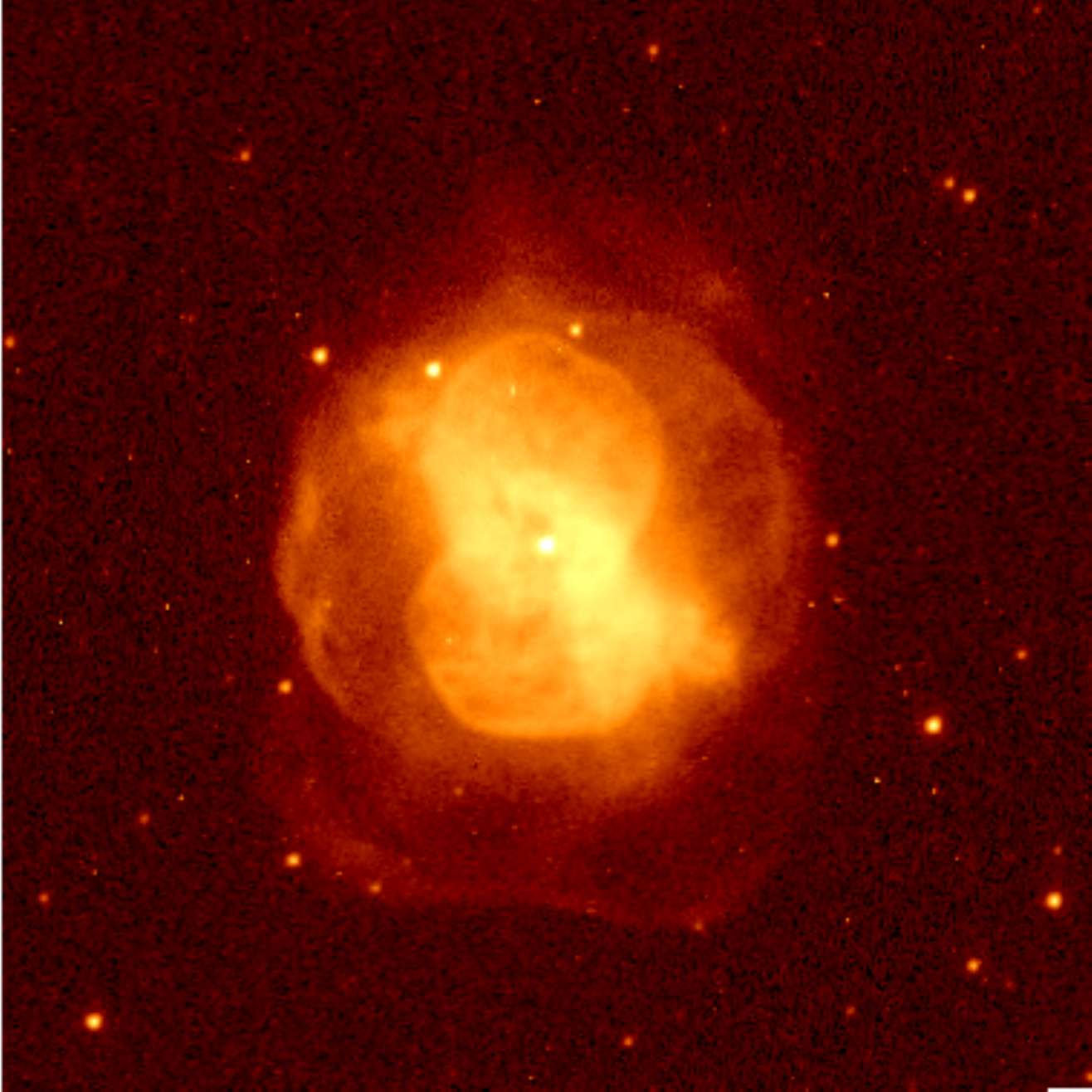}
\figsetgrpnote{False-color image in the F502N filter.}
\figsetgrpend

\figsetgrpstart
\figsetgrpnum{12.7}
\figsetgrptitle{PN~G042.9--06.9}
\figsetplot{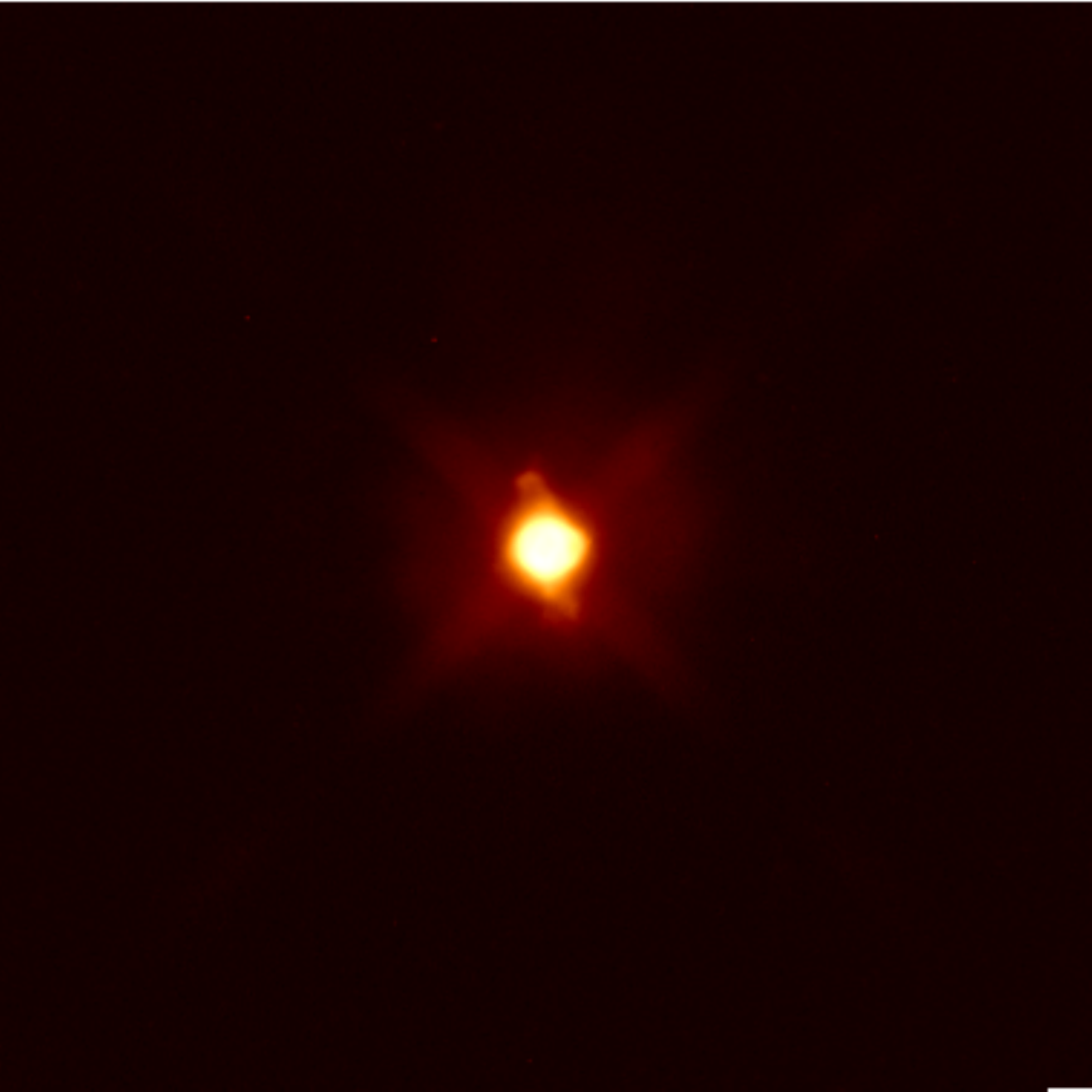}
\figsetgrpnote{False-color image in the F502N filter.}
\figsetgrpend

\figsetgrpstart
\figsetgrpnum{12.8}
\figsetgrptitle{PN~G044.1+05.8}
\figsetplot{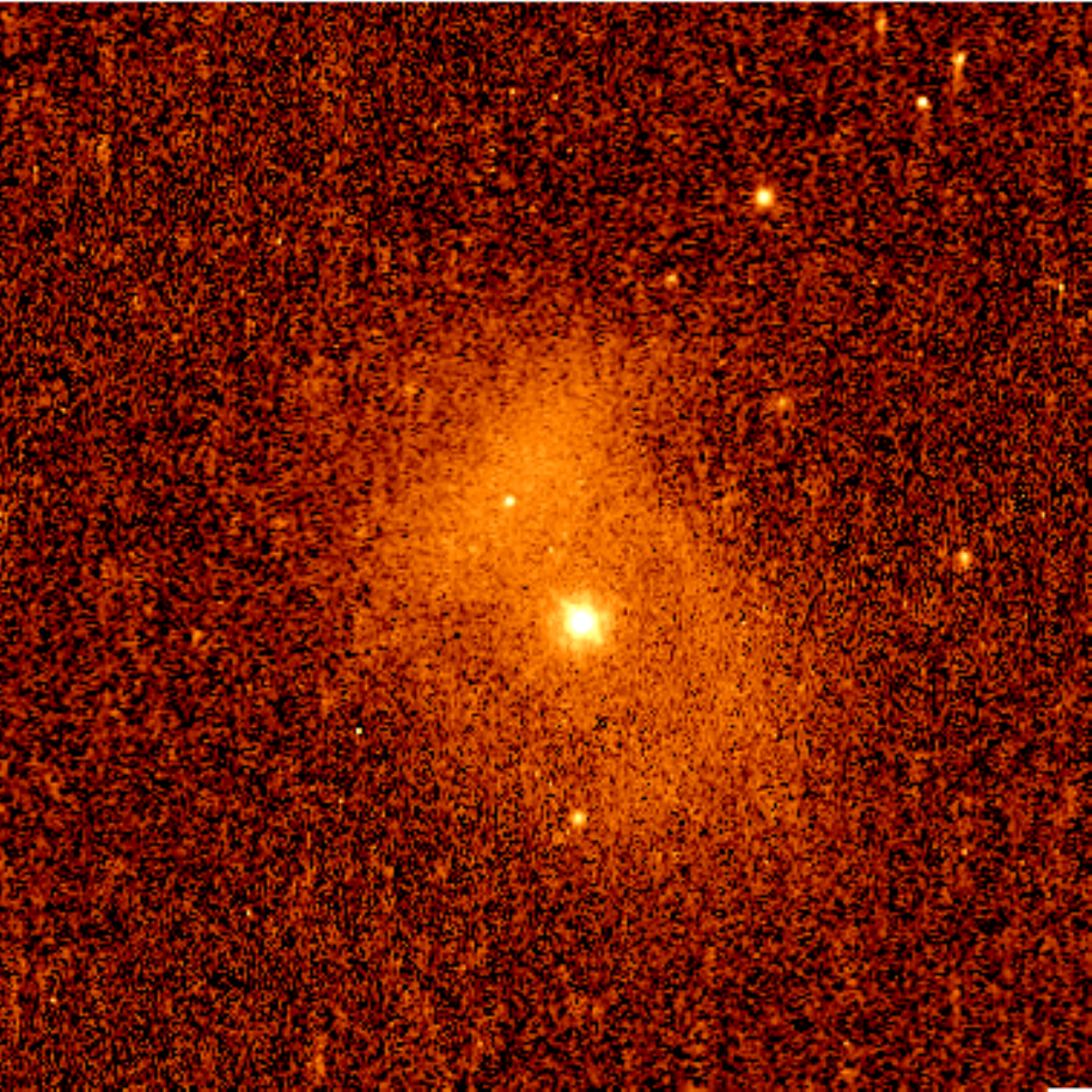}
\figsetgrpnote{False-color image in the F502N filter.}
\figsetgrpend

\figsetgrpstart
\figsetgrpnum{12.9}
\figsetgrptitle{PN~G048.5+04.2}
\figsetplot{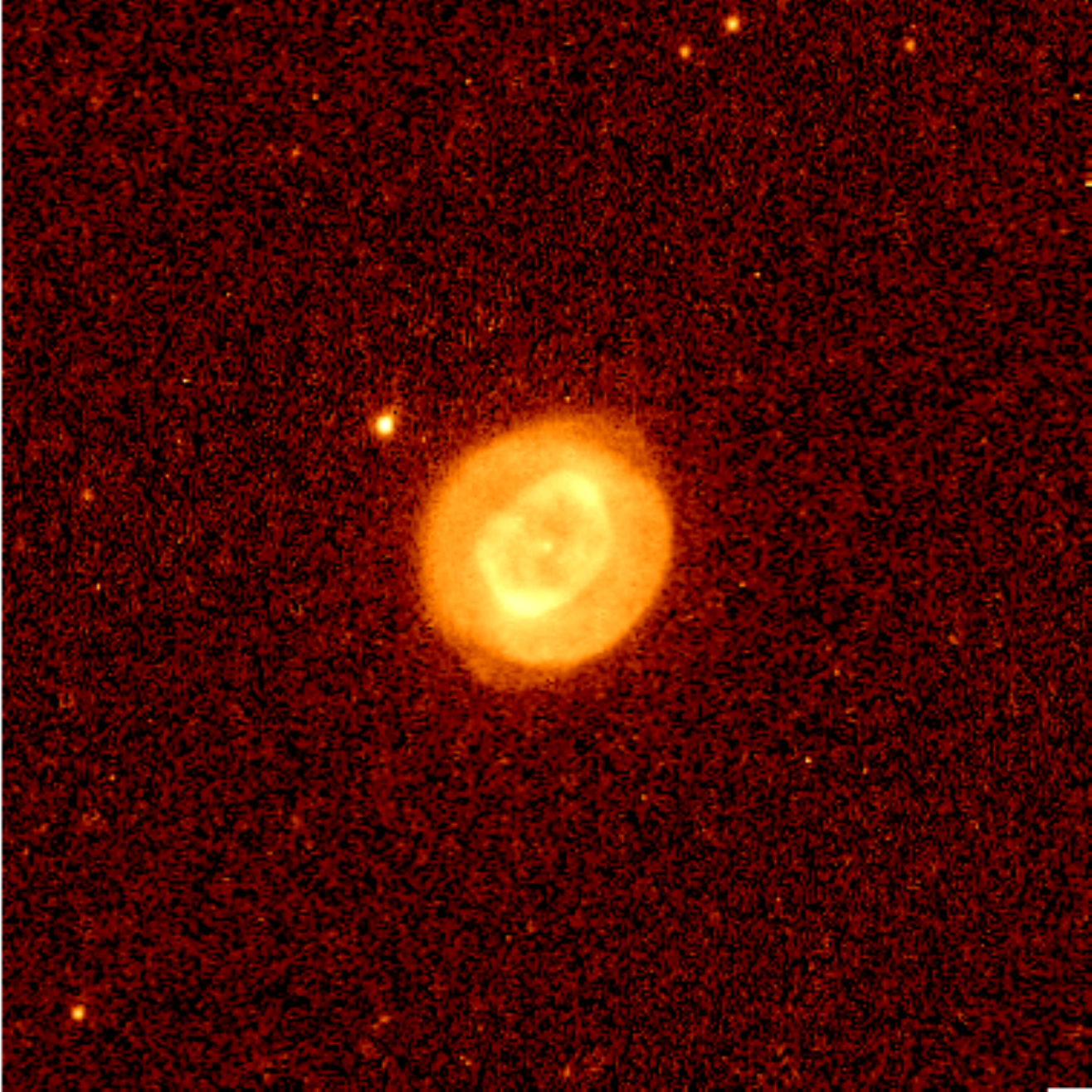}
\figsetgrpnote{False-color image in the F502N filter.}
\figsetgrpend

\figsetgrpstart
\figsetgrpnum{12.10}
\figsetgrptitle{PN~G052.9--02.7}
\figsetplot{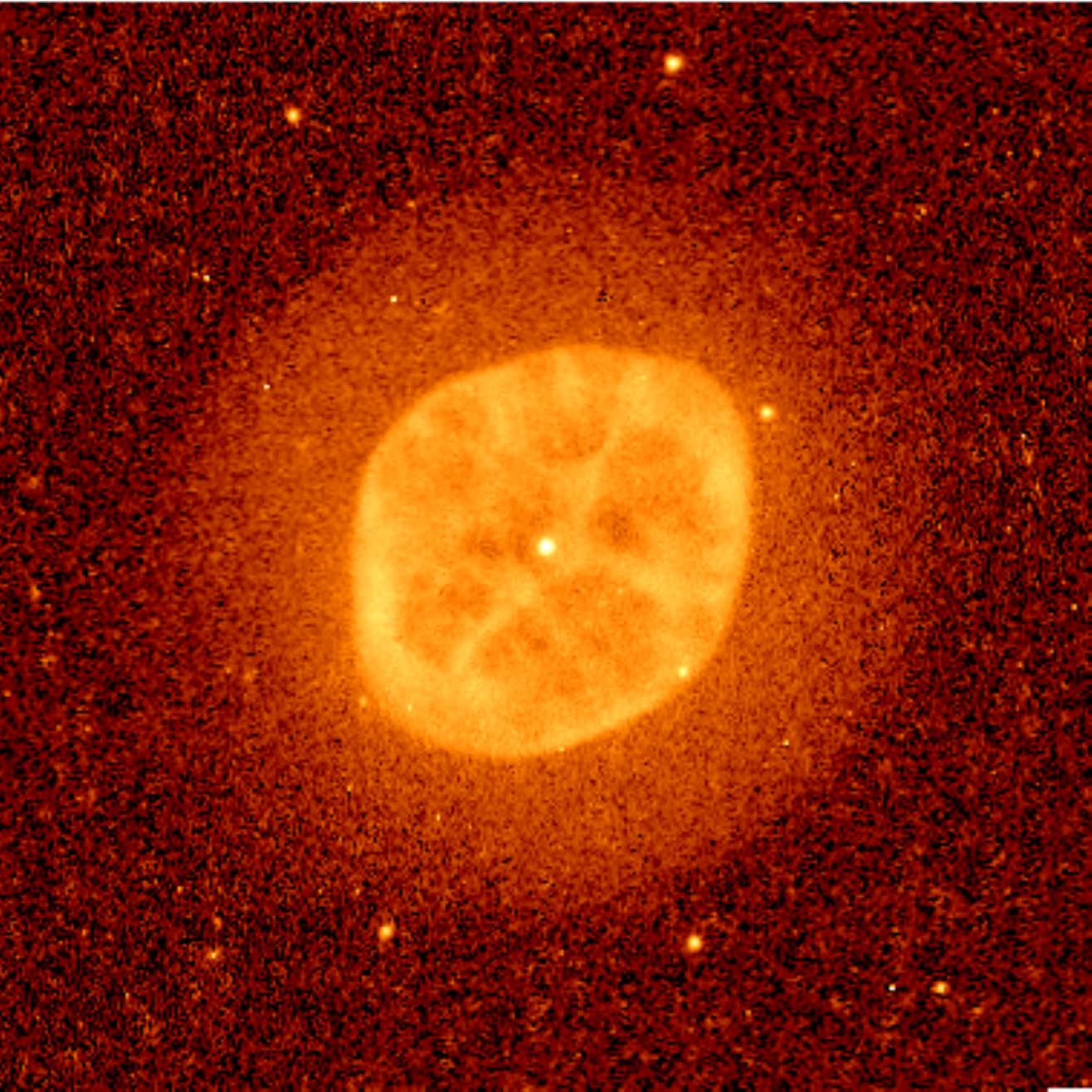}
\figsetgrpnote{False-color image in the F502N filter.}
\figsetgrpend

\figsetgrpstart
\figsetgrpnum{12.11}
\figsetgrptitle{PN~G053.3+24.0}
\figsetplot{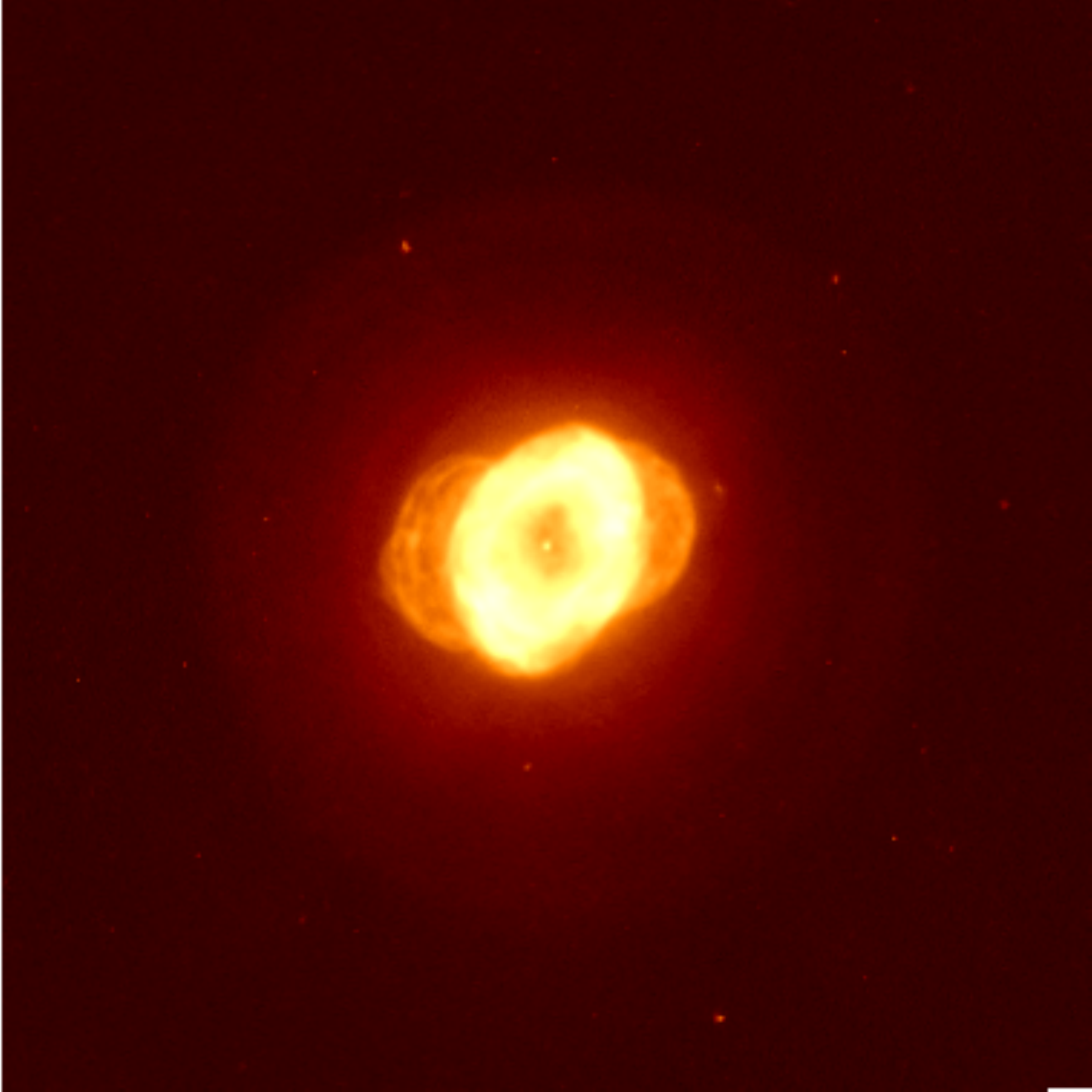}
\figsetgrpnote{False-color image in the F502N filter.}
\figsetgrpend

\figsetgrpstart
\figsetgrpnum{12.12}
\figsetgrptitle{PN~G059.9+02.0}
\figsetplot{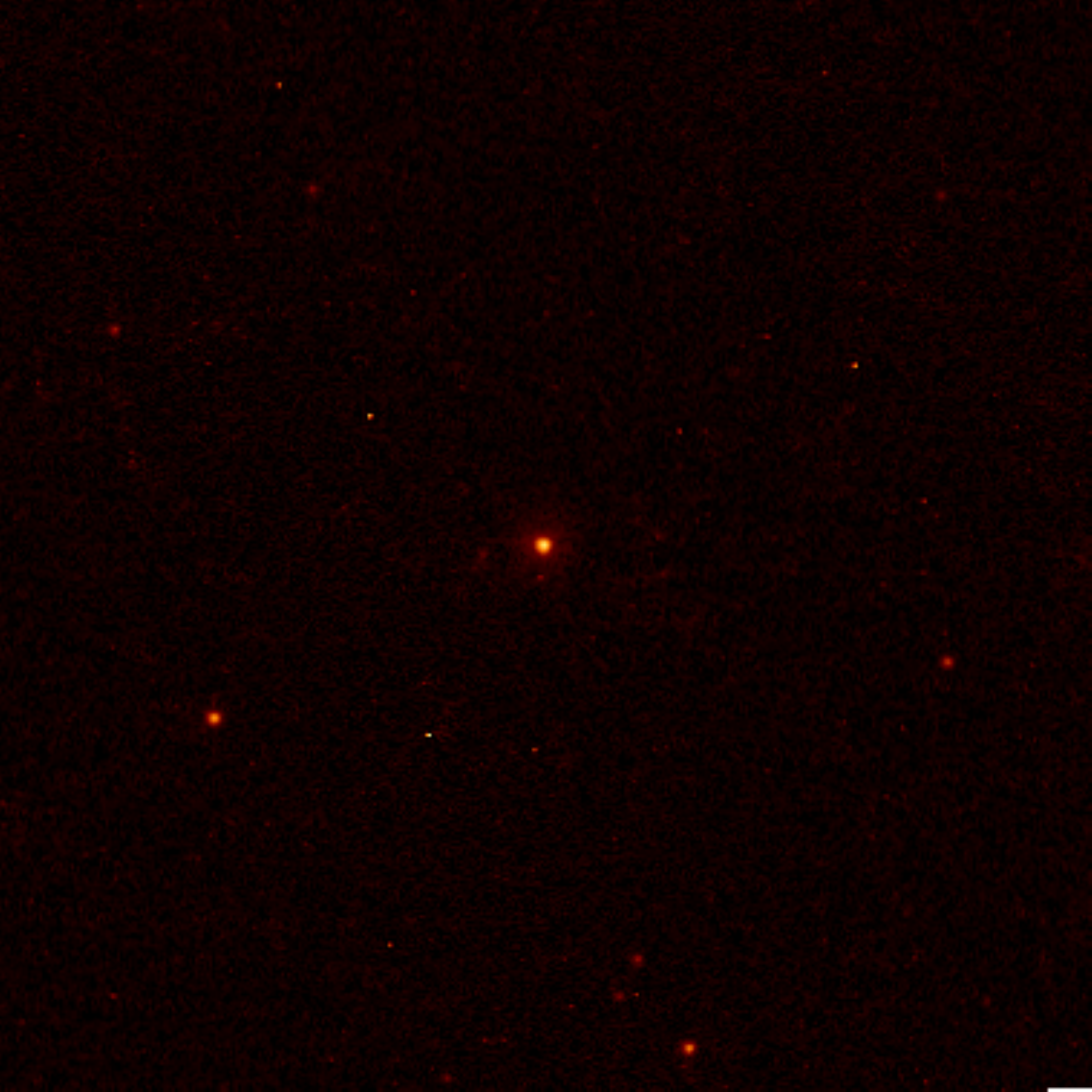}
\figsetgrpnote{False-color image in the F502N filter.}
\figsetgrpend

\figsetgrpstart
\figsetgrpnum{12.13}
\figsetgrptitle{PN~G063.8--03.3}
\figsetplot{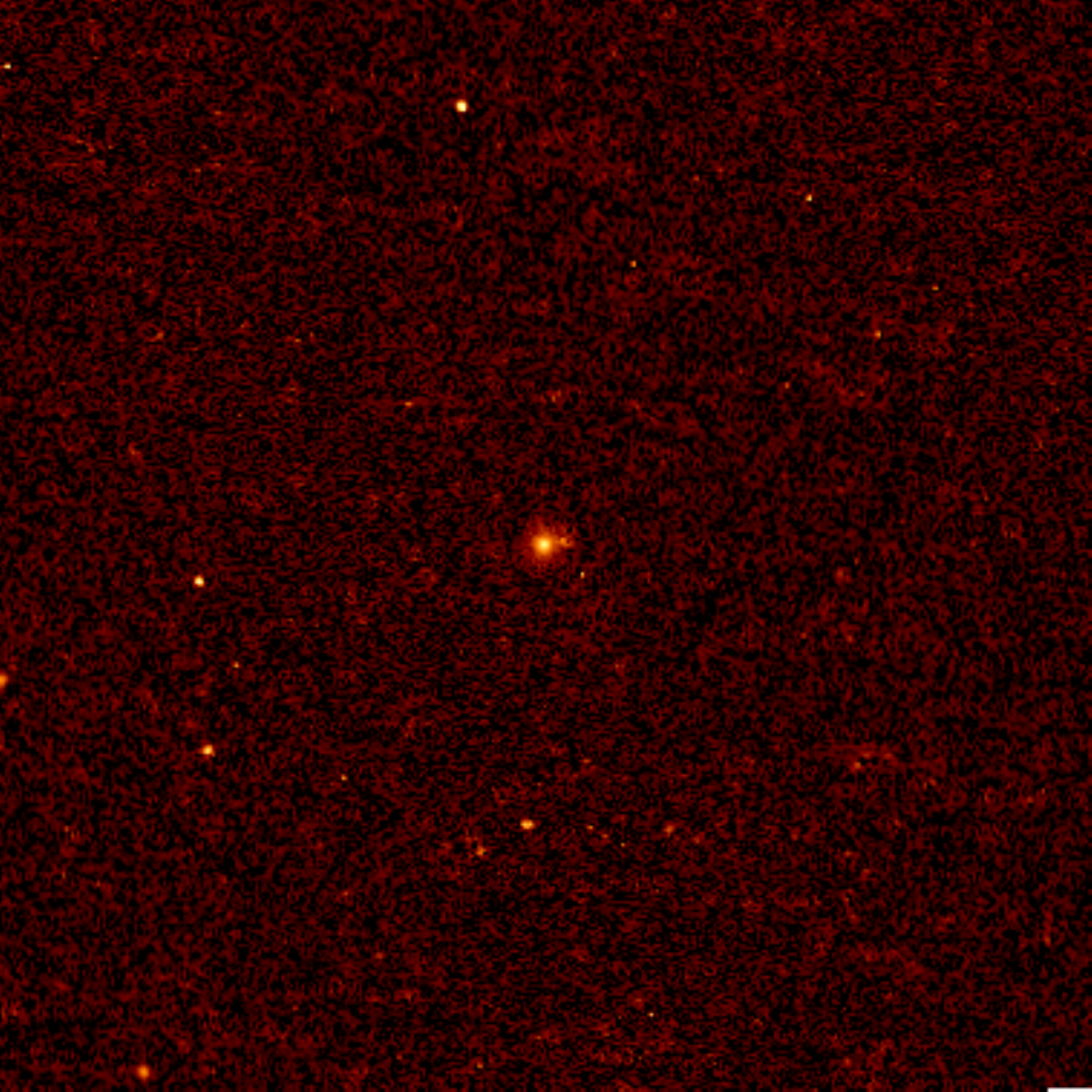}
\figsetgrpnote{False-color image in the F502N filter.}
\figsetgrpend

\figsetgrpstart
\figsetgrpnum{12.14}
\figsetgrptitle{PN~G068.7+01.9}
\figsetplot{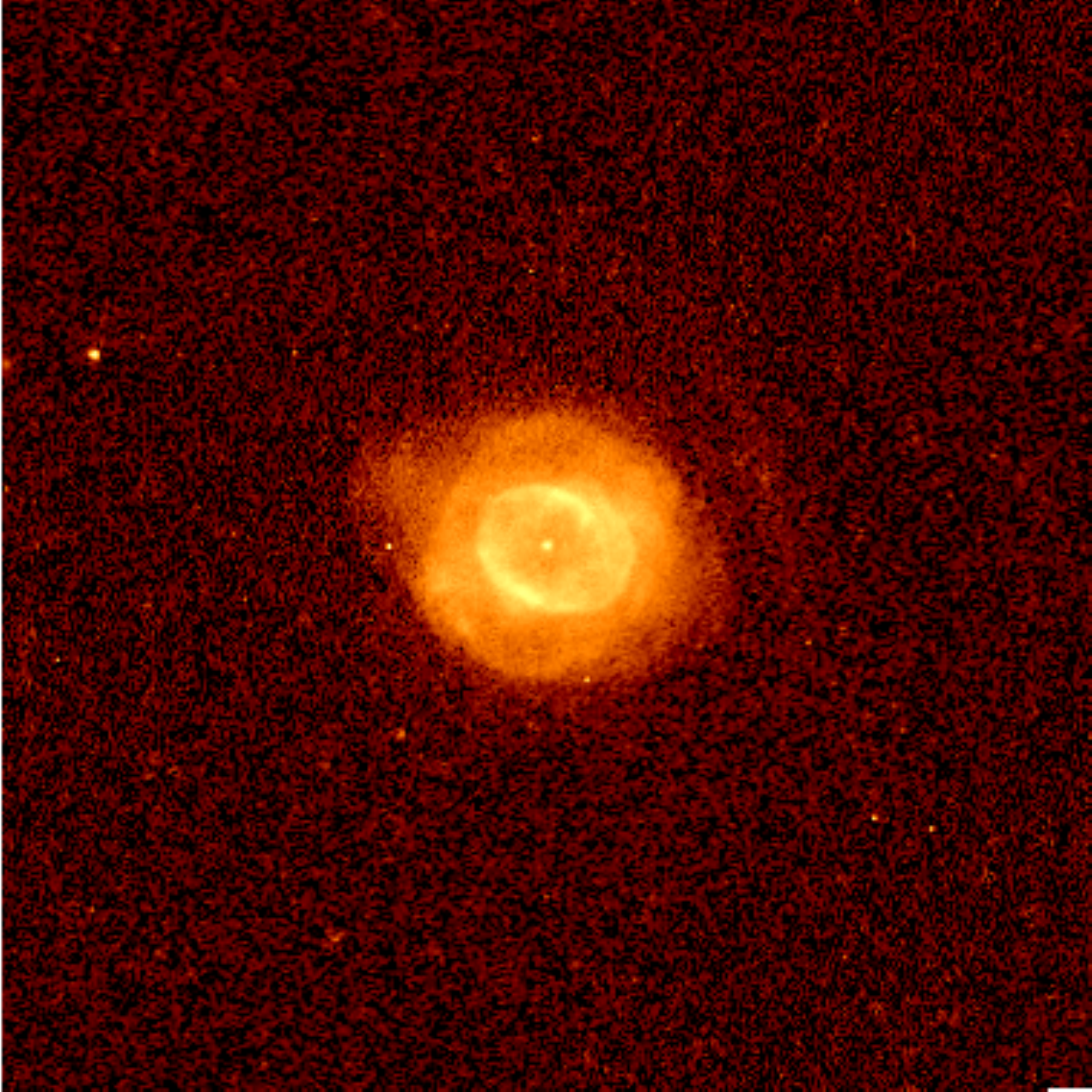}
\figsetgrpnote{False-color image in the F502N filter.}
\figsetgrpend

\figsetgrpstart
\figsetgrpnum{12.15}
\figsetgrptitle{PN~G068.7+14.8}
\figsetplot{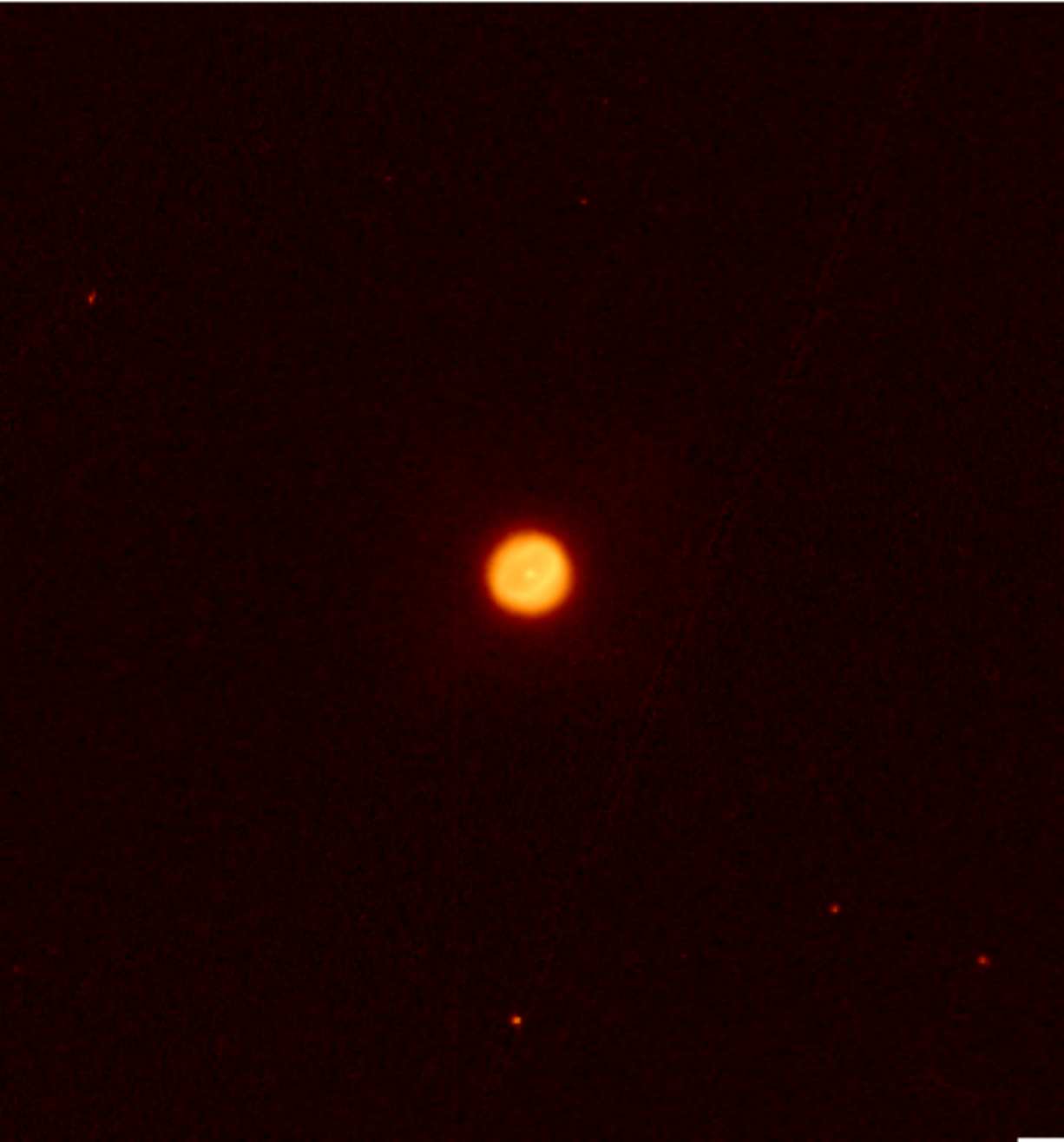}
\figsetgrpnote{False-color image in the F502N filter.}
\figsetgrpend

\figsetgrpstart
\figsetgrpnum{12.16}
\figsetgrptitle{PN~G079.9+06.4}
\figsetplot{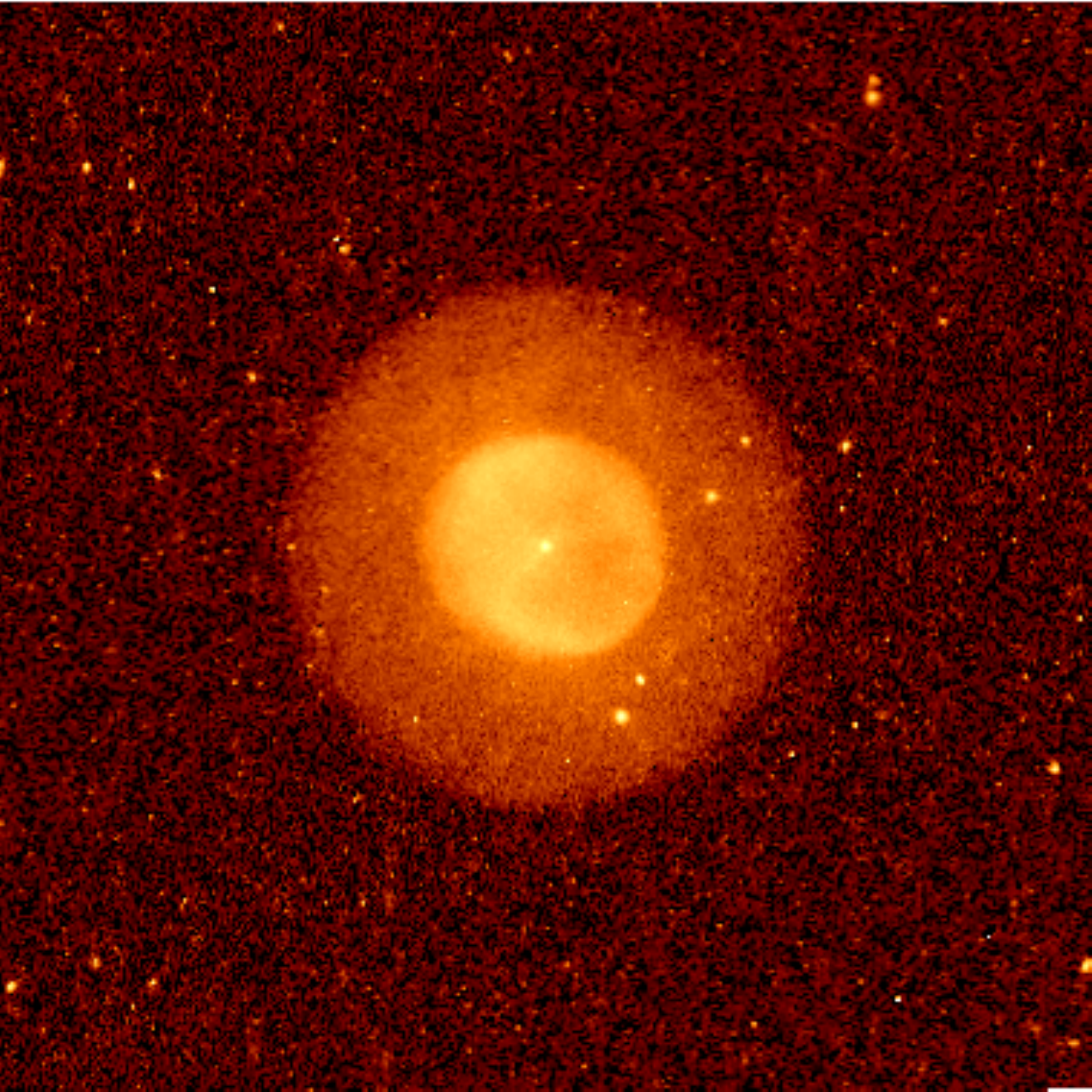}
\figsetgrpnote{False-color image in the F502N filter.}
\figsetgrpend

\figsetgrpstart
\figsetgrpnum{12.17}
\figsetgrptitle{PN~G097.6--02.4}
\figsetplot{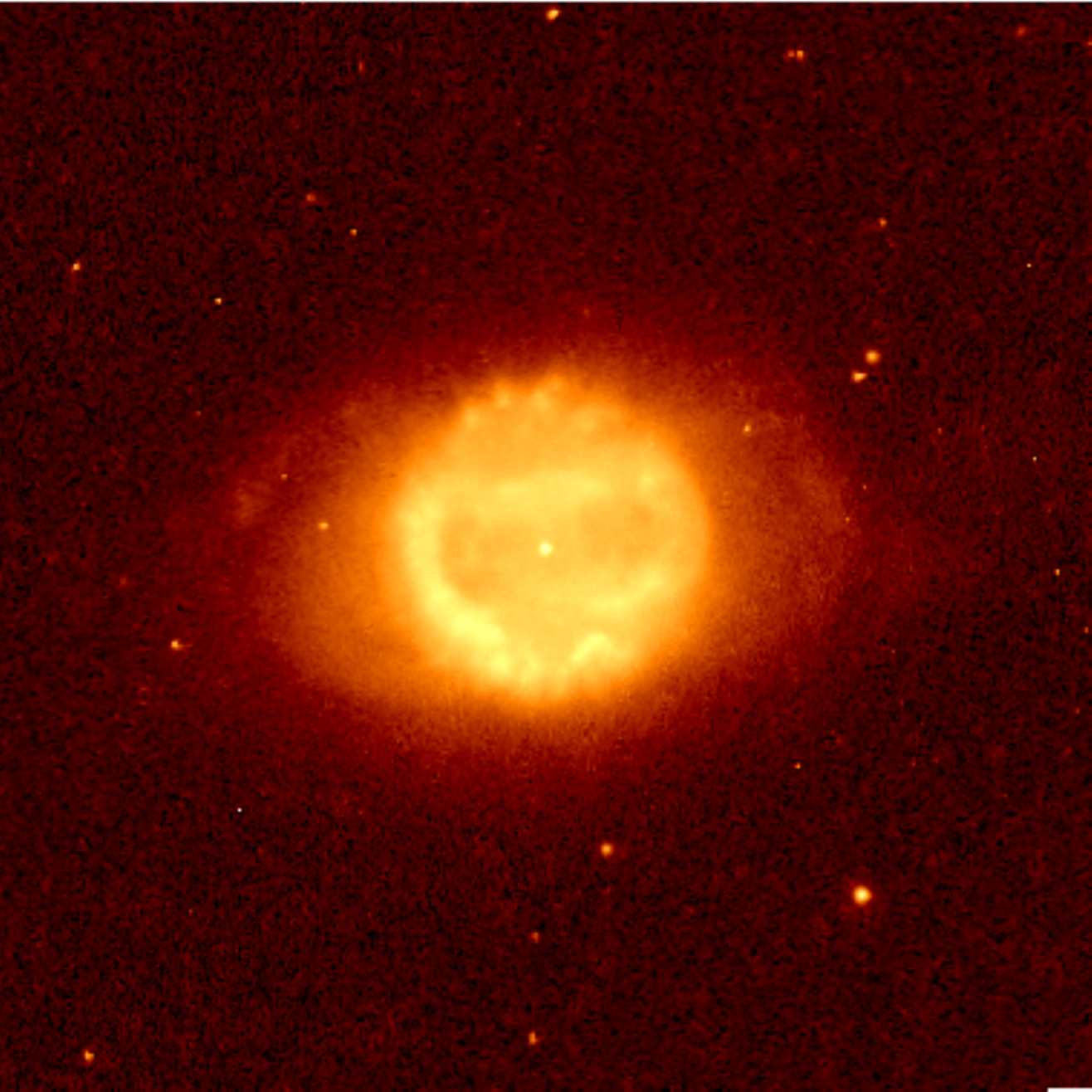}
\figsetgrpnote{False-color image in the F502N filter.}
\figsetgrpend

\figsetgrpstart
\figsetgrpnum{12.18}
\figsetgrptitle{PN~G098.2+04.9}
\figsetplot{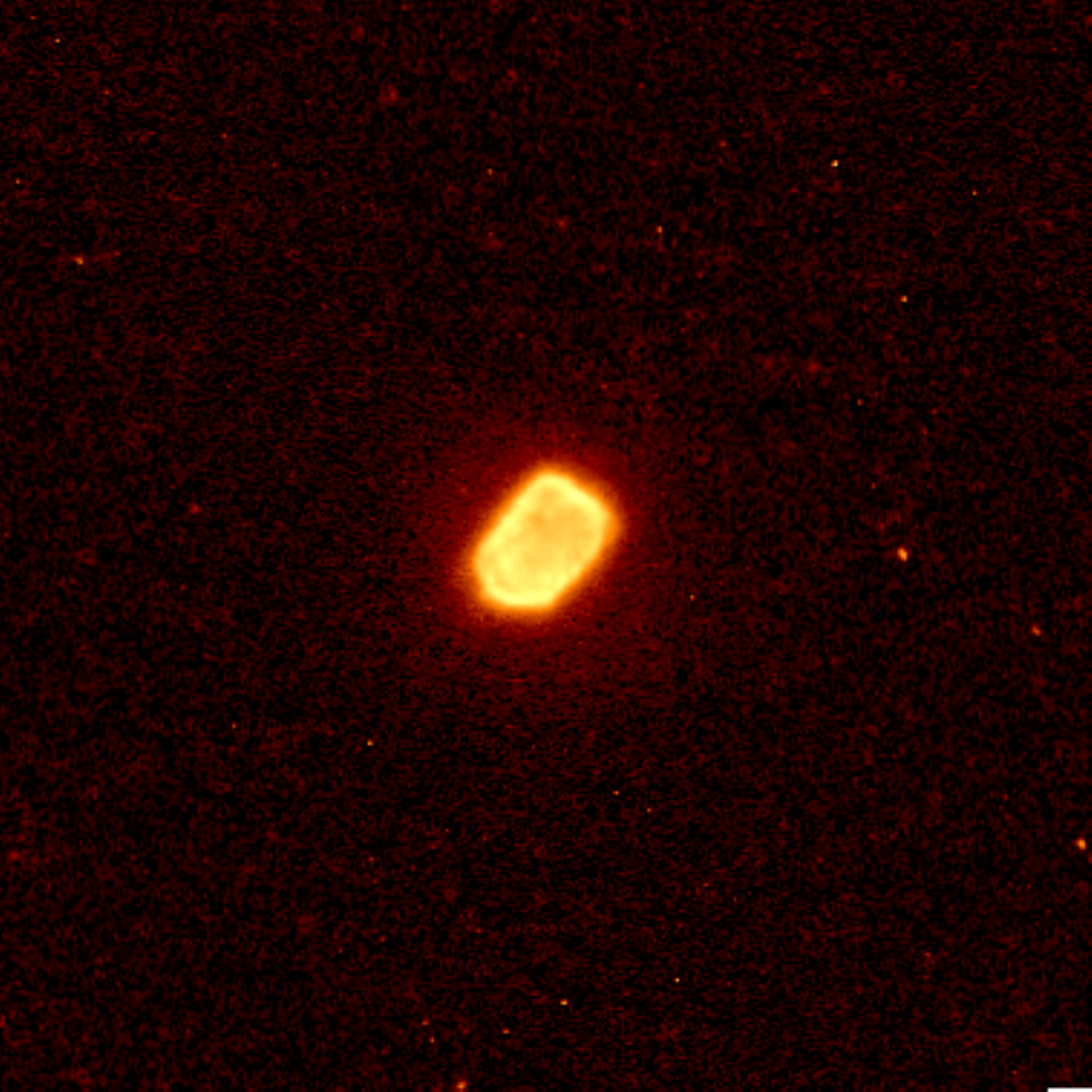}
\figsetgrpnote{False-color image in the F502N filter.}
\figsetgrpend

\figsetgrpstart
\figsetgrpnum{12.19}
\figsetgrptitle{PN~G104.1+01.0}
\figsetplot{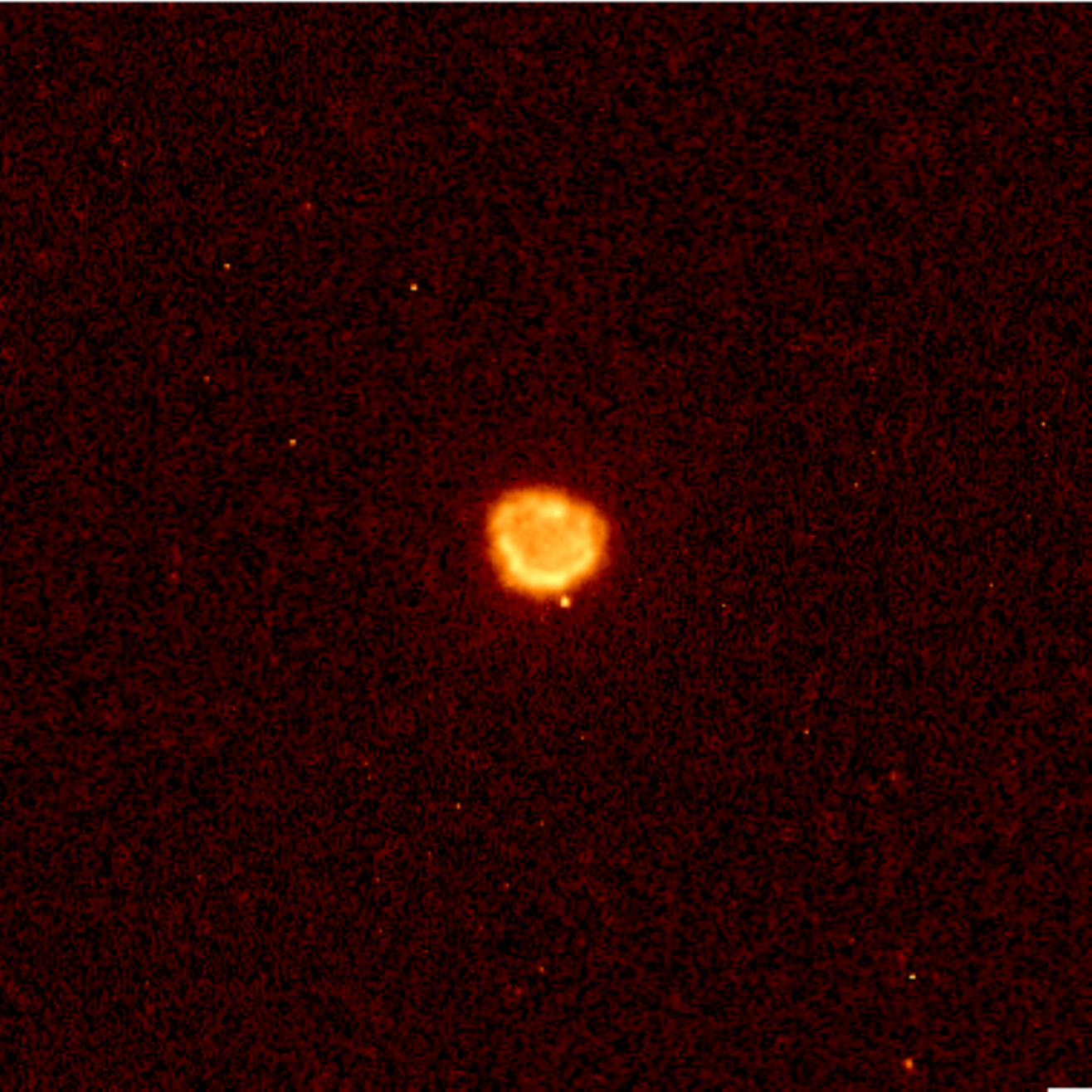}
\figsetgrpnote{False-color image in the F502N filter.}
\figsetgrpend

\figsetgrpstart
\figsetgrpnum{12.20}
\figsetgrptitle{PN~G107.4--02.6}
\figsetplot{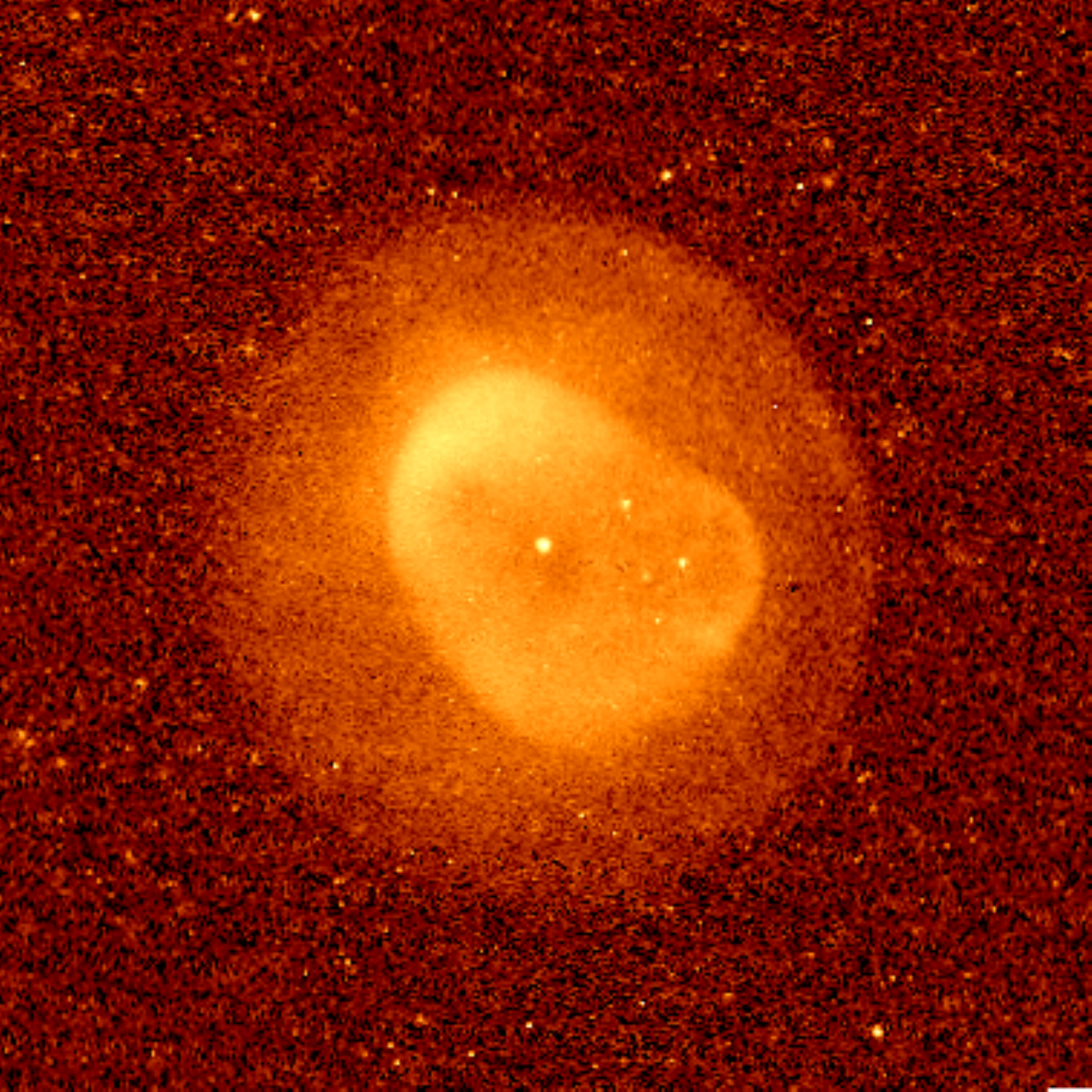}
\figsetgrpnote{False-color image in the F502N filter.}
\figsetgrpend

\figsetgrpstart
\figsetgrpnum{12.21}
\figsetgrptitle{PN~G184.0--02.1}
\figsetplot{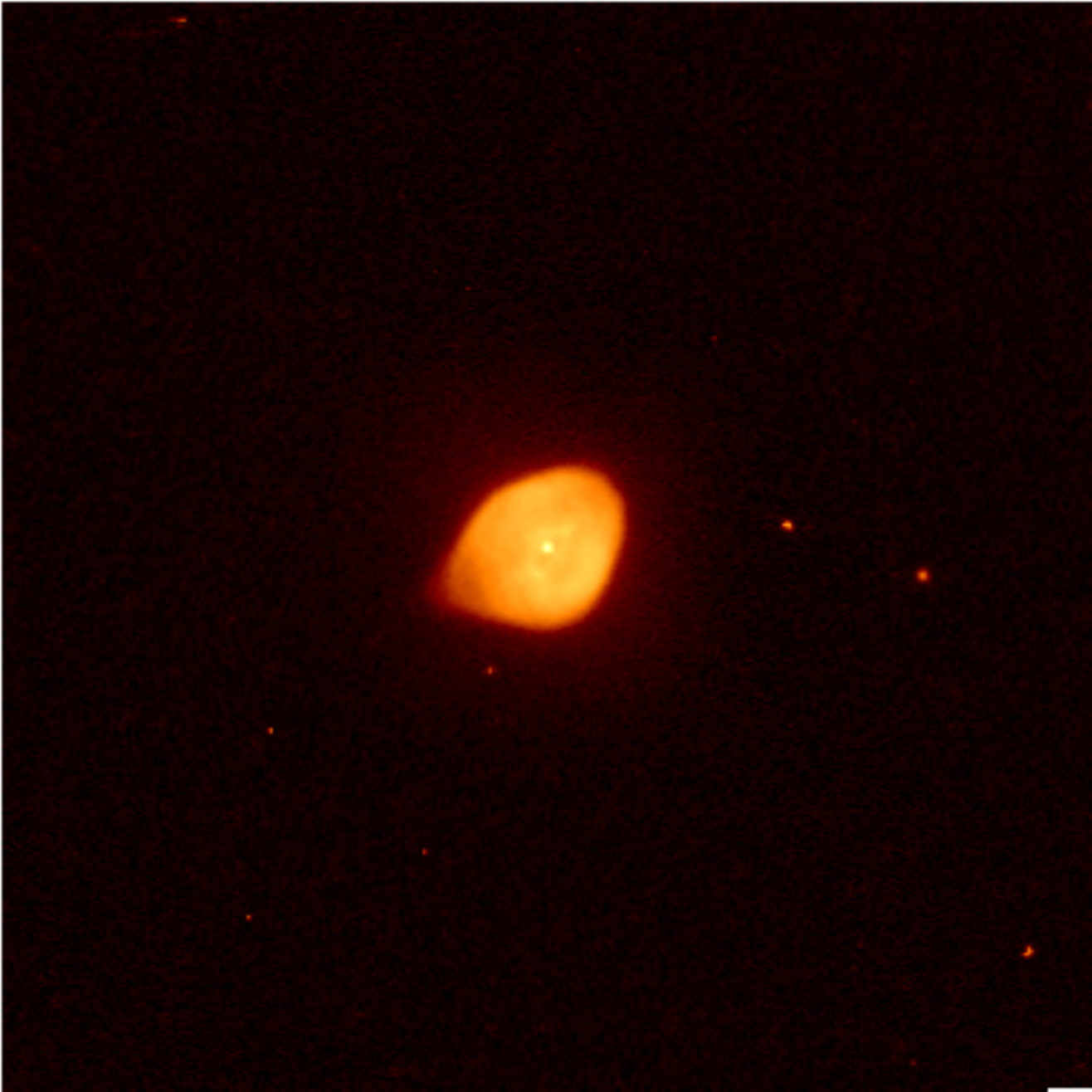}
\figsetgrpnote{False-color image in the F502N filter.}
\figsetgrpend

\figsetgrpstart
\figsetgrpnum{12.22}
\figsetgrptitle{PN~G205.8--26.7}
\figsetplot{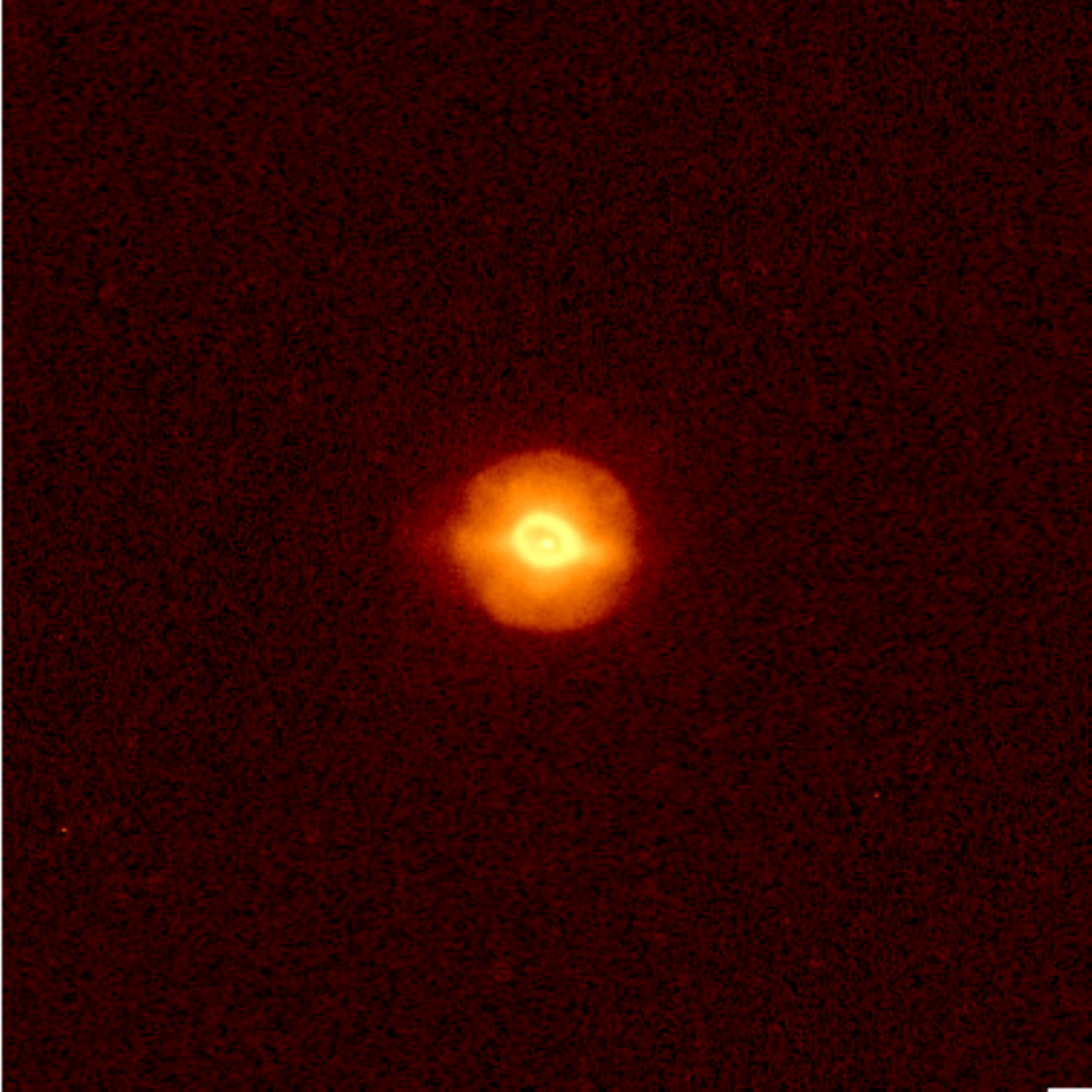}
\figsetgrpnote{False-color image in the F502N filter.}
\figsetgrpend

\figsetgrpstart
\figsetgrpnum{12.23}
\figsetgrptitle{PN~G263.0--05.5}
\figsetplot{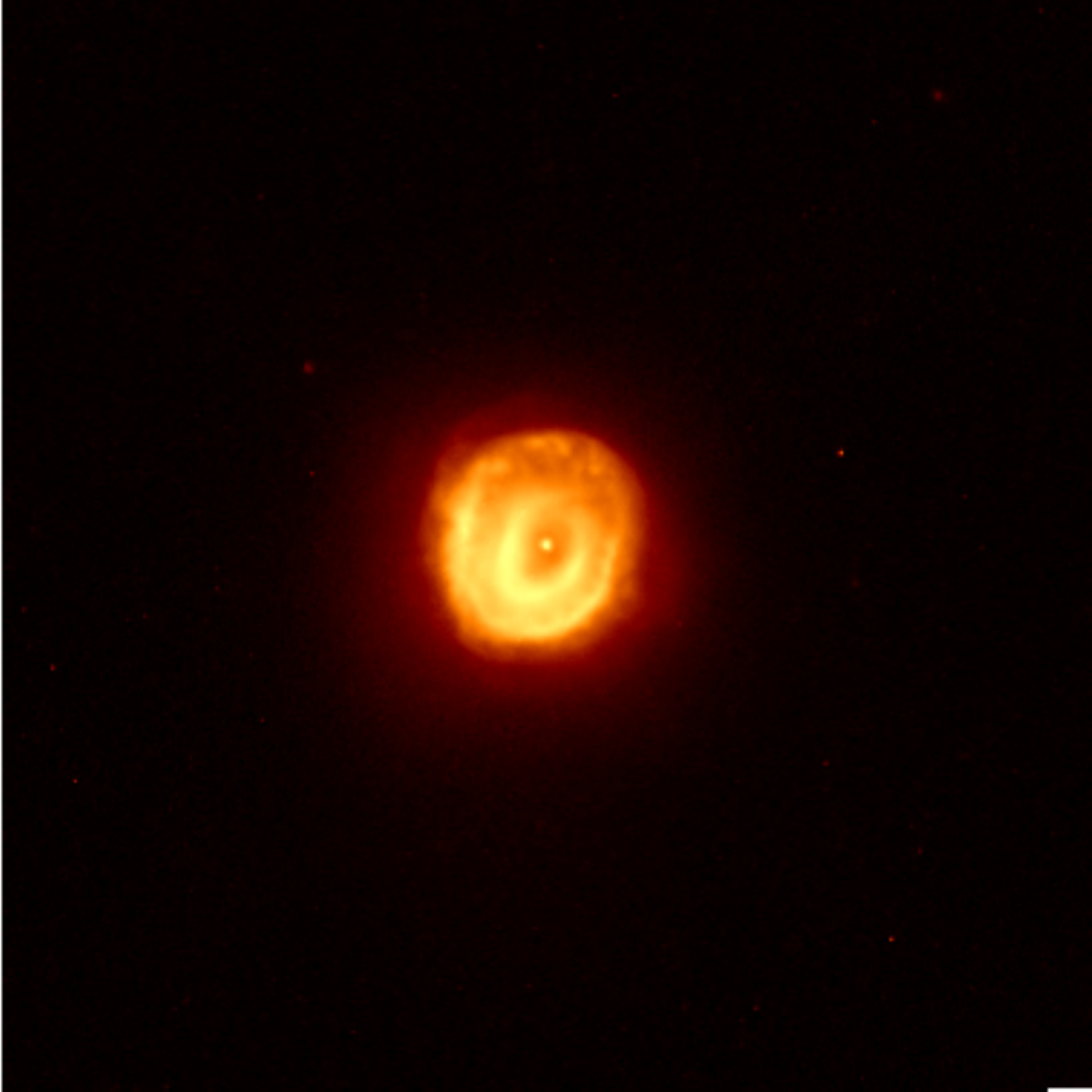}
\figsetgrpnote{False-color image in the F502N filter.}
\figsetgrpend

\figsetgrpstart
\figsetgrpnum{12.24}
\figsetgrptitle{PN~G264.4--12.7}
\figsetplot{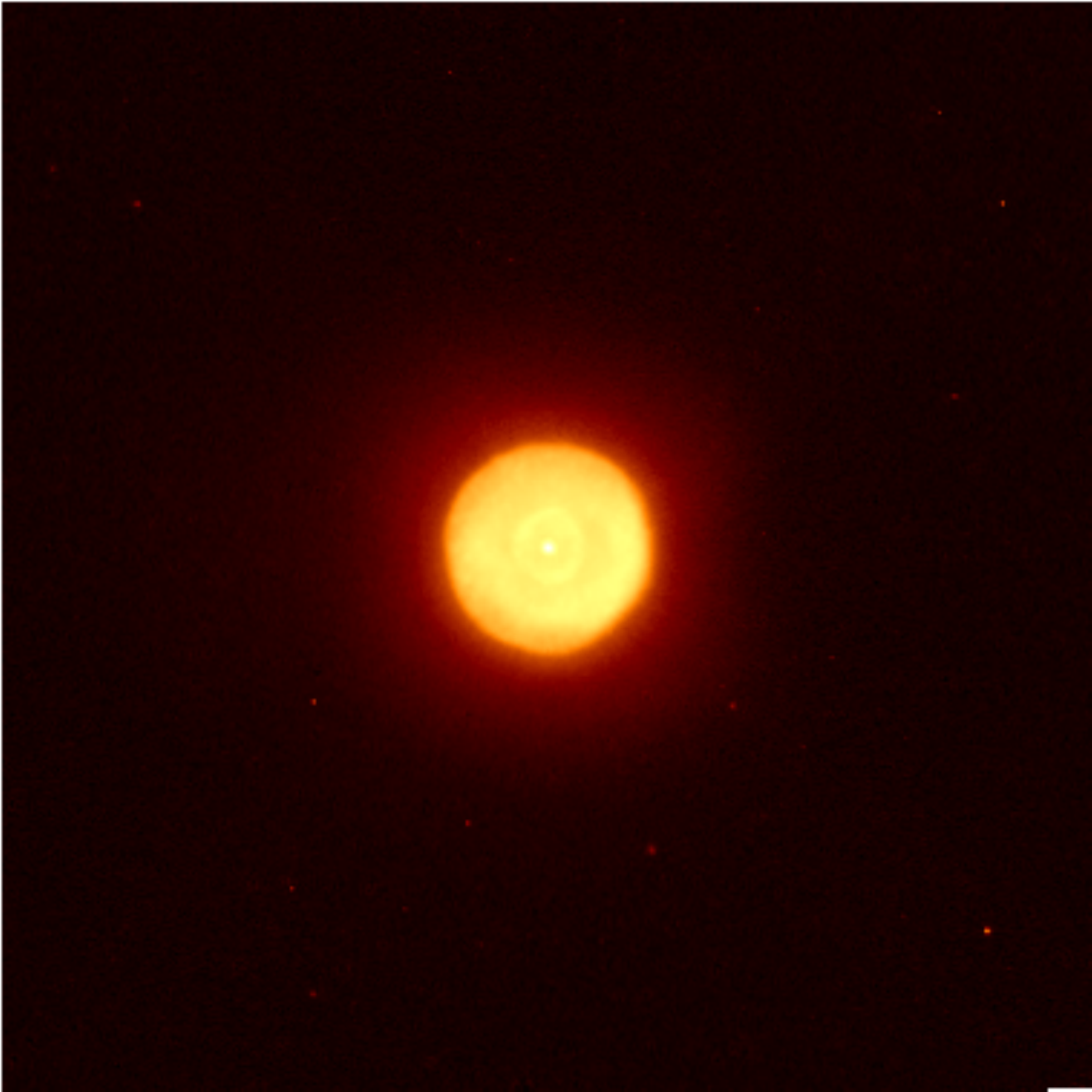}
\figsetgrpnote{False-color image in the F502N filter.}
\figsetgrpend

\figsetgrpstart
\figsetgrpnum{12.25}
\figsetgrptitle{PN~G275.3--04.7}
\figsetplot{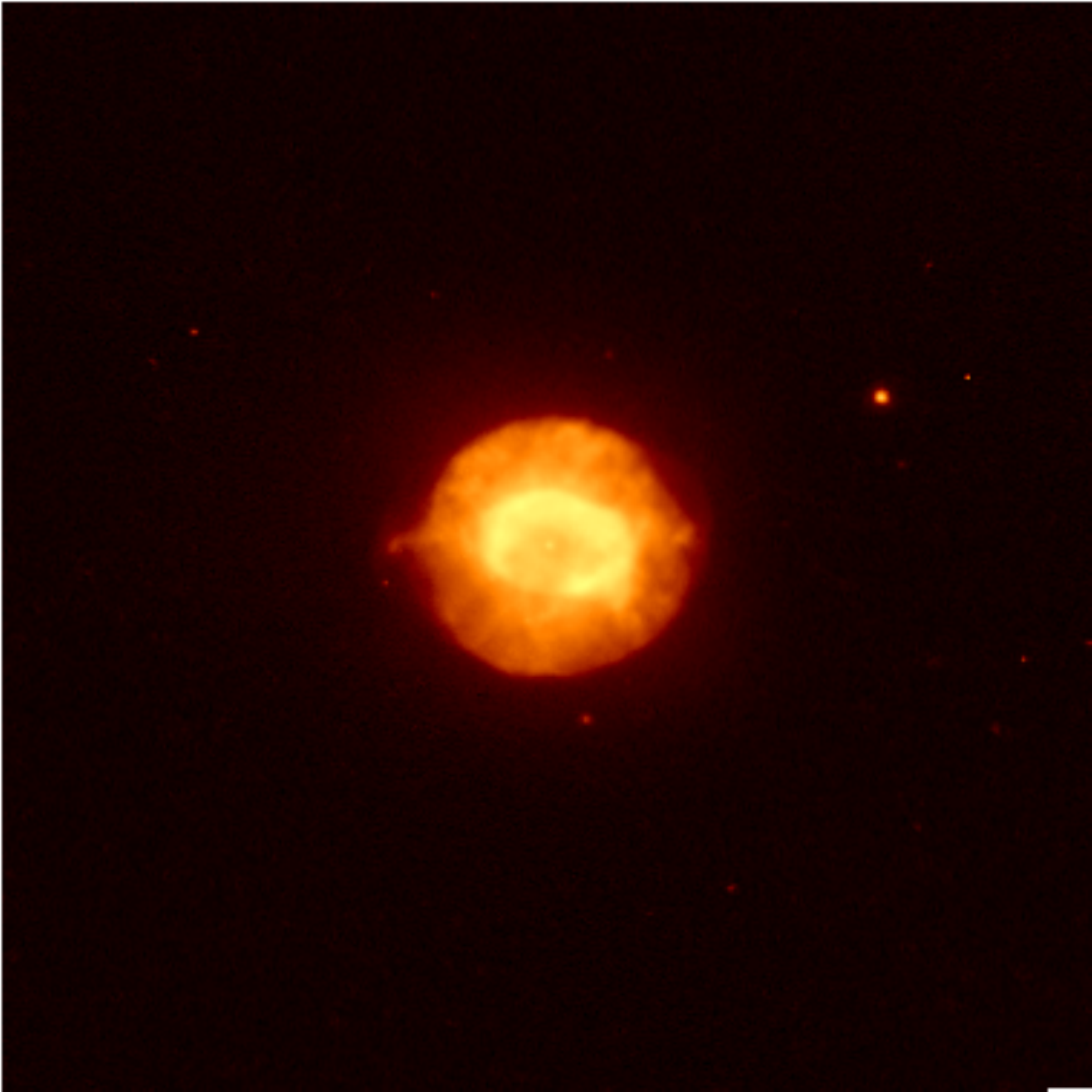}
\figsetgrpnote{False-color image in the F502N filter.}
\figsetgrpend

\figsetgrpstart
\figsetgrpnum{12.26}
\figsetgrptitle{PN~G278.6--06.7}
\figsetplot{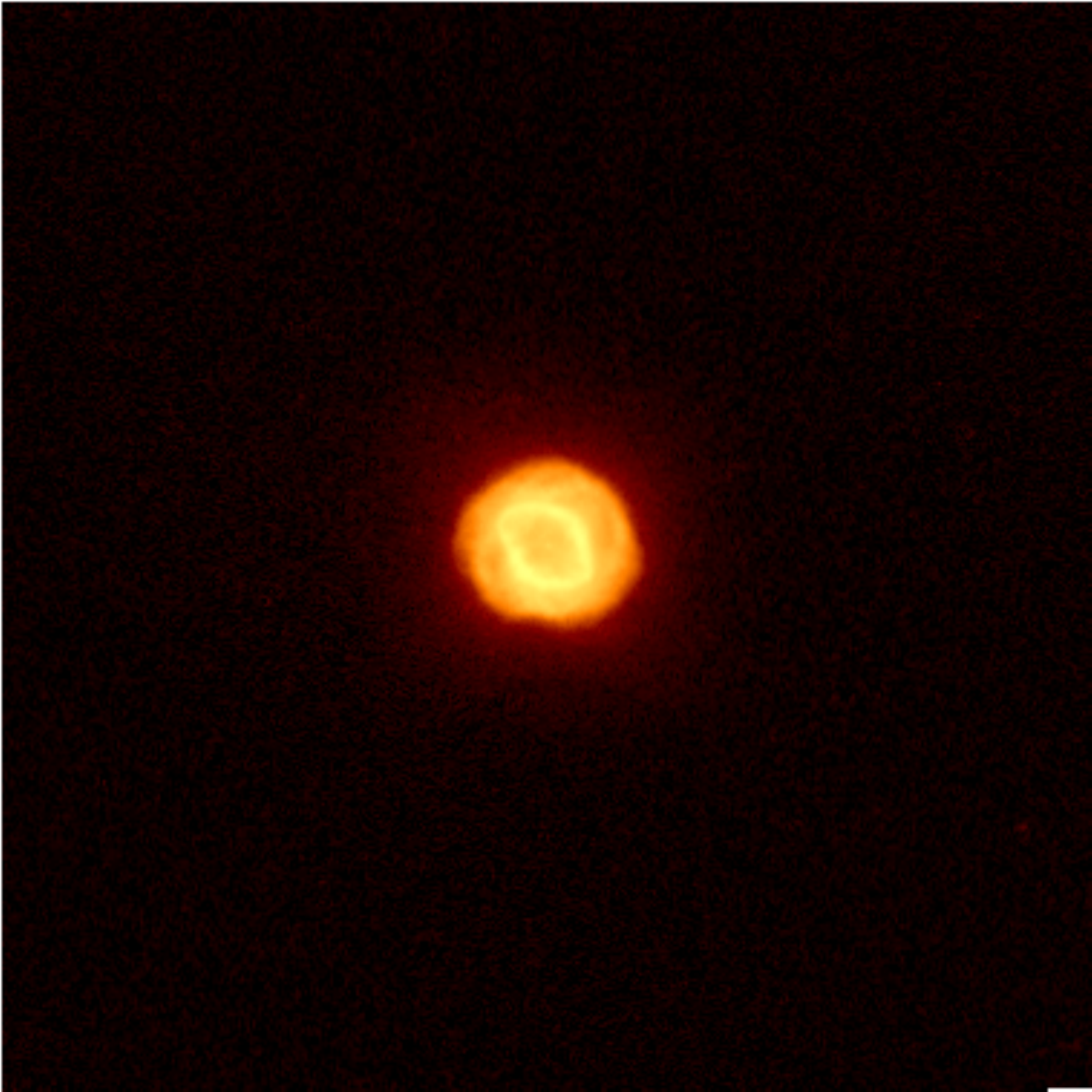}
\figsetgrpnote{False-color image in the F502N filter.}
\figsetgrpend

\figsetgrpstart
\figsetgrpnum{12.27}
\figsetgrptitle{PN~G285.4+01.5}
\figsetplot{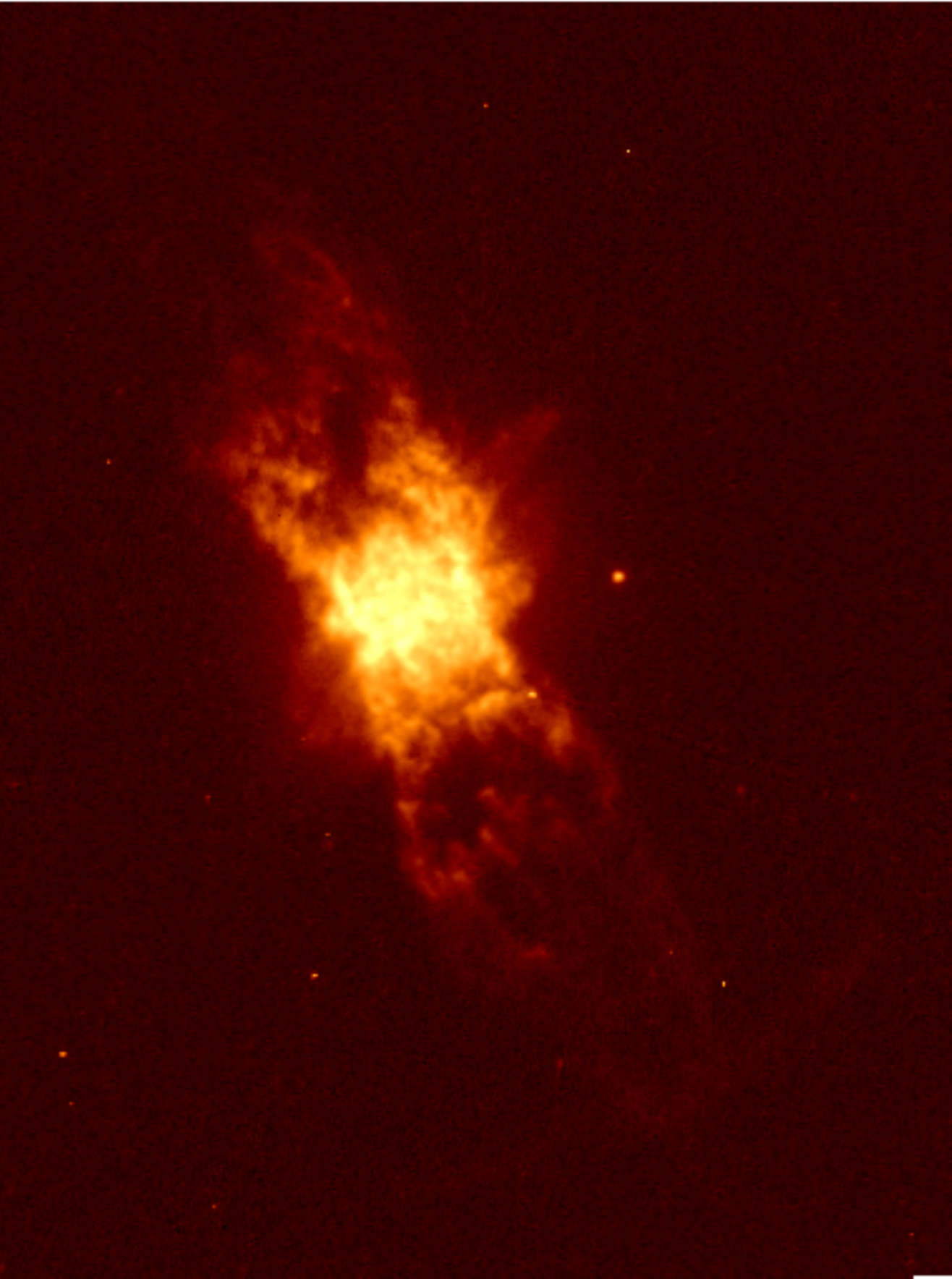}
\figsetgrpnote{False-color image in the F502N filter.}
\figsetgrpend

\figsetgrpstart
\figsetgrpnum{12.28}
\figsetgrptitle{PN~G285.4+02.2}
\figsetplot{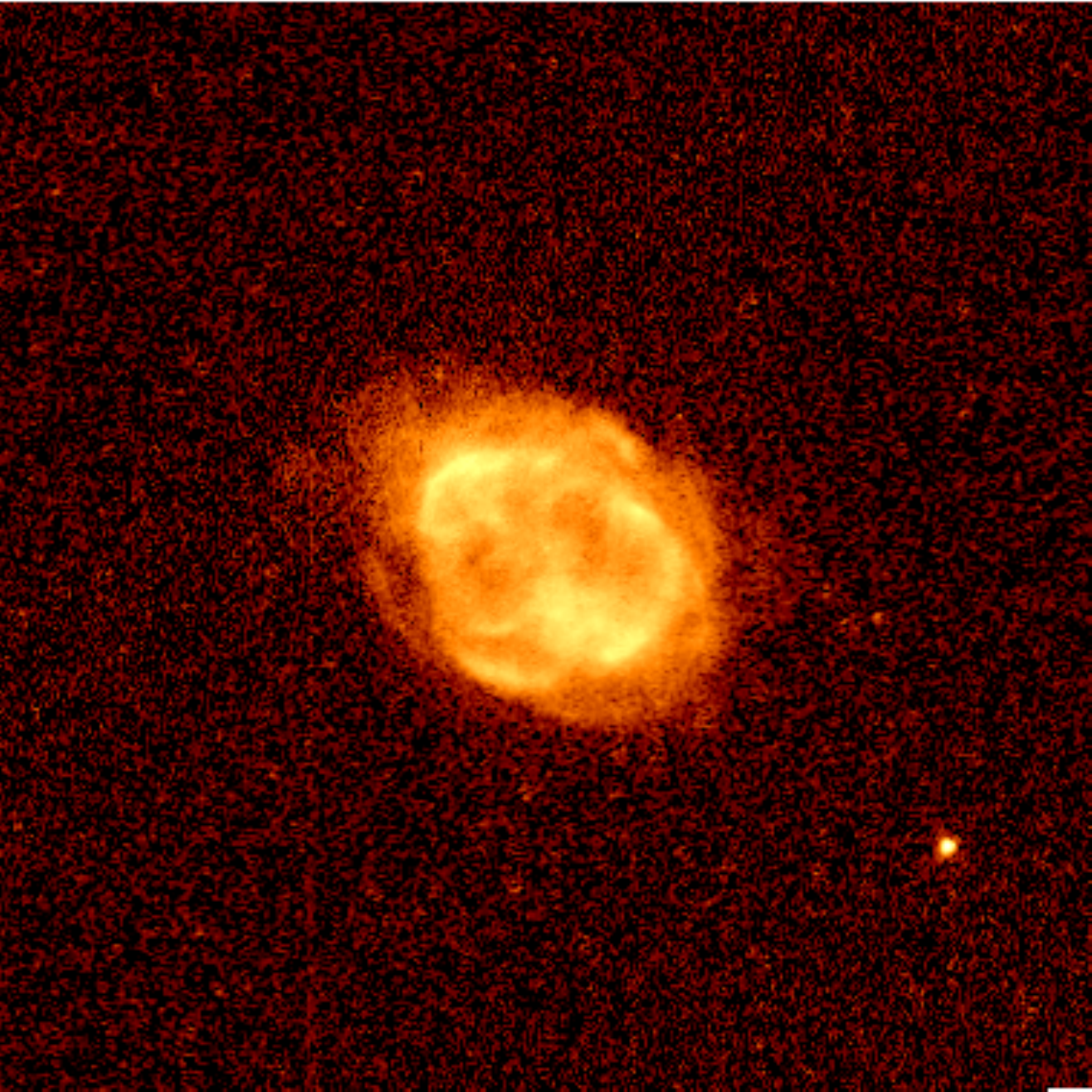}
\figsetgrpnote{False-color image in the F502N filter.}
\figsetgrpend

\figsetgrpstart
\figsetgrpnum{12.29}
\figsetgrptitle{PN~G286.0--06.5}
\figsetplot{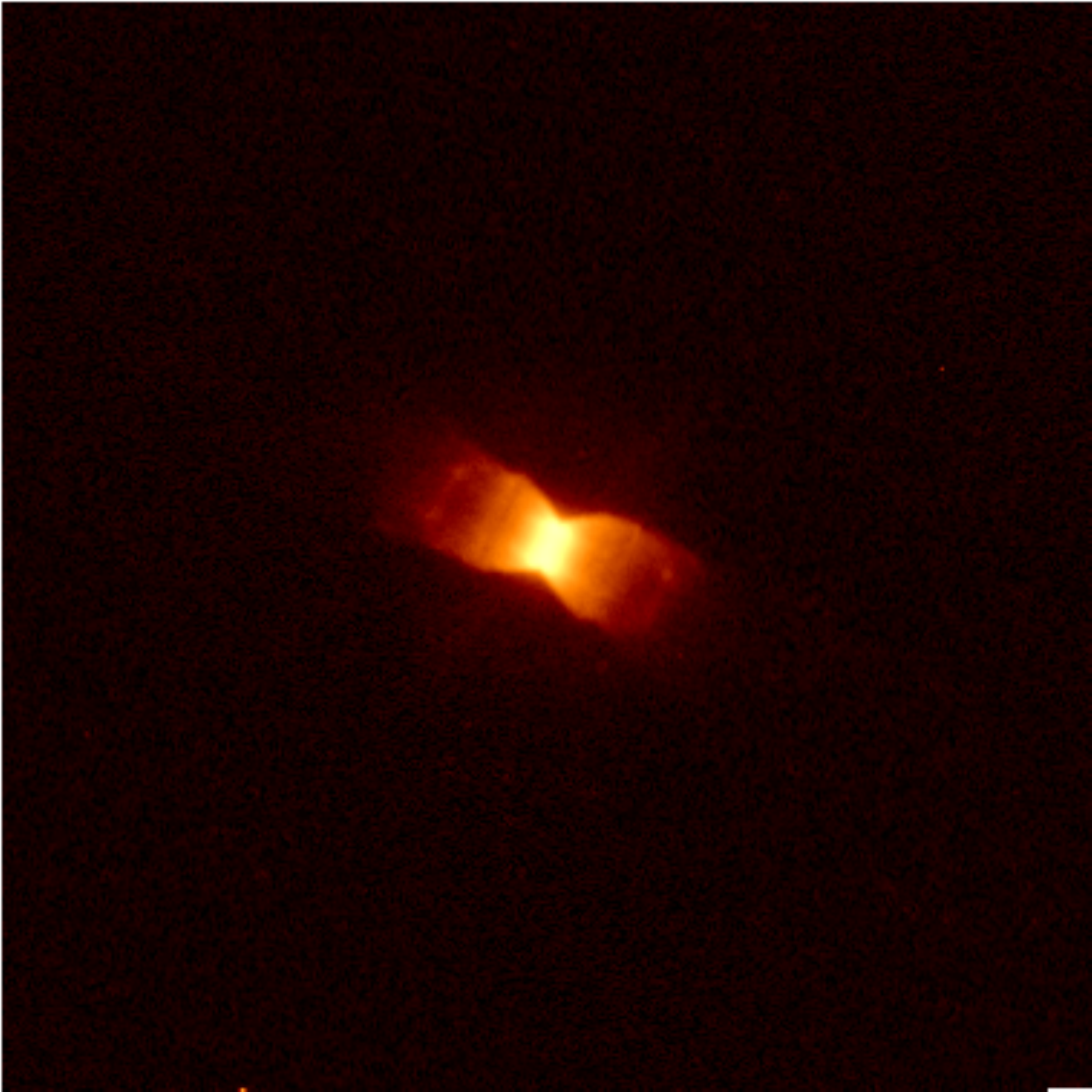}
\figsetgrpnote{False-color image in the F502N filter.}
\figsetgrpend

\figsetgrpstart
\figsetgrpnum{12.30}
\figsetgrptitle{PN~G289.8+07.7}
\figsetplot{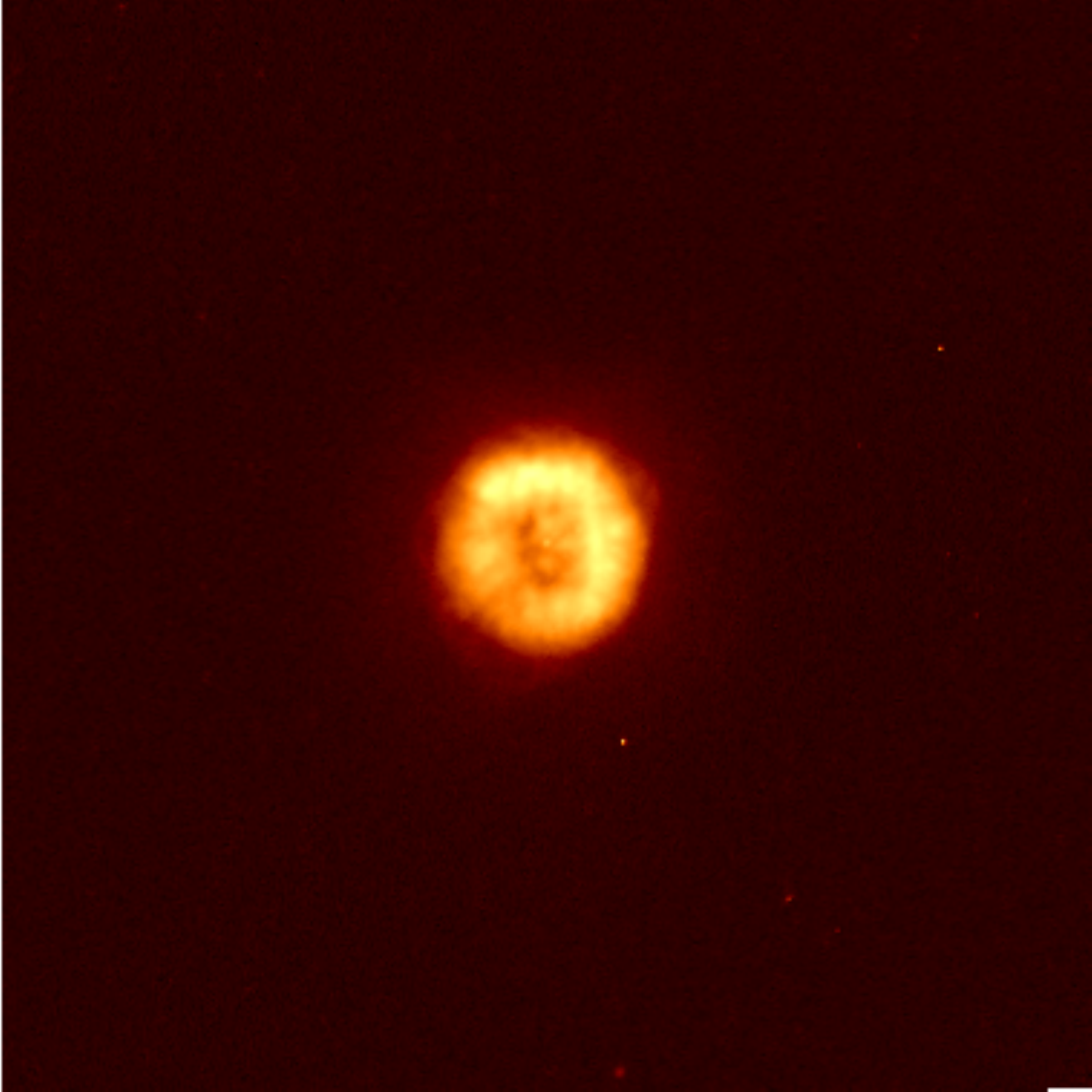}
\figsetgrpnote{False-color image in the F502N filter.}
\figsetgrpend

\figsetgrpstart
\figsetgrpnum{12.31}
\figsetgrptitle{PN~G294.9--04.3}
\figsetplot{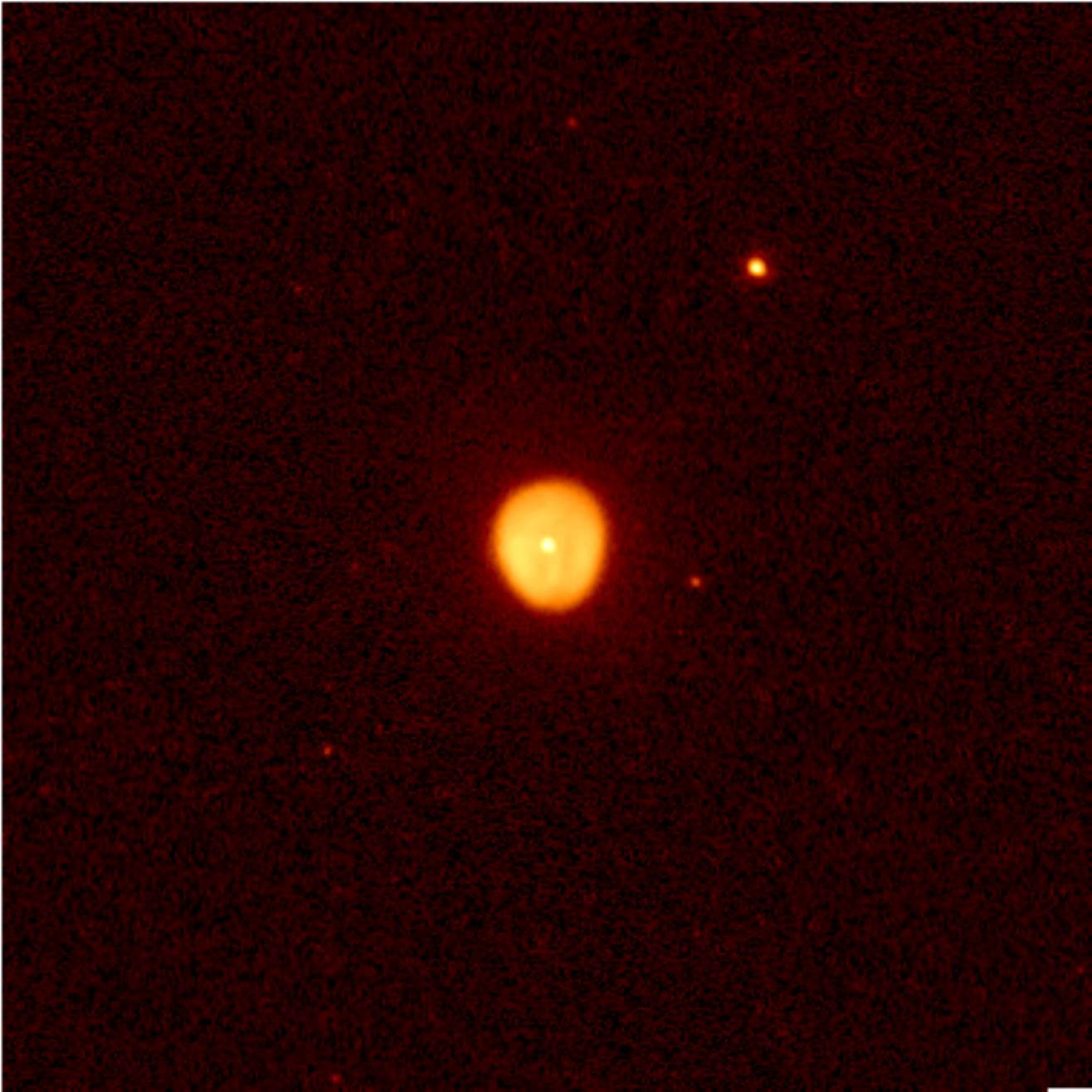}
\figsetgrpnote{False-color image in the F502N filter.}
\figsetgrpend

\figsetgrpstart
\figsetgrpnum{12.32}
\figsetgrptitle{PN~G295.3--09.3}
\figsetplot{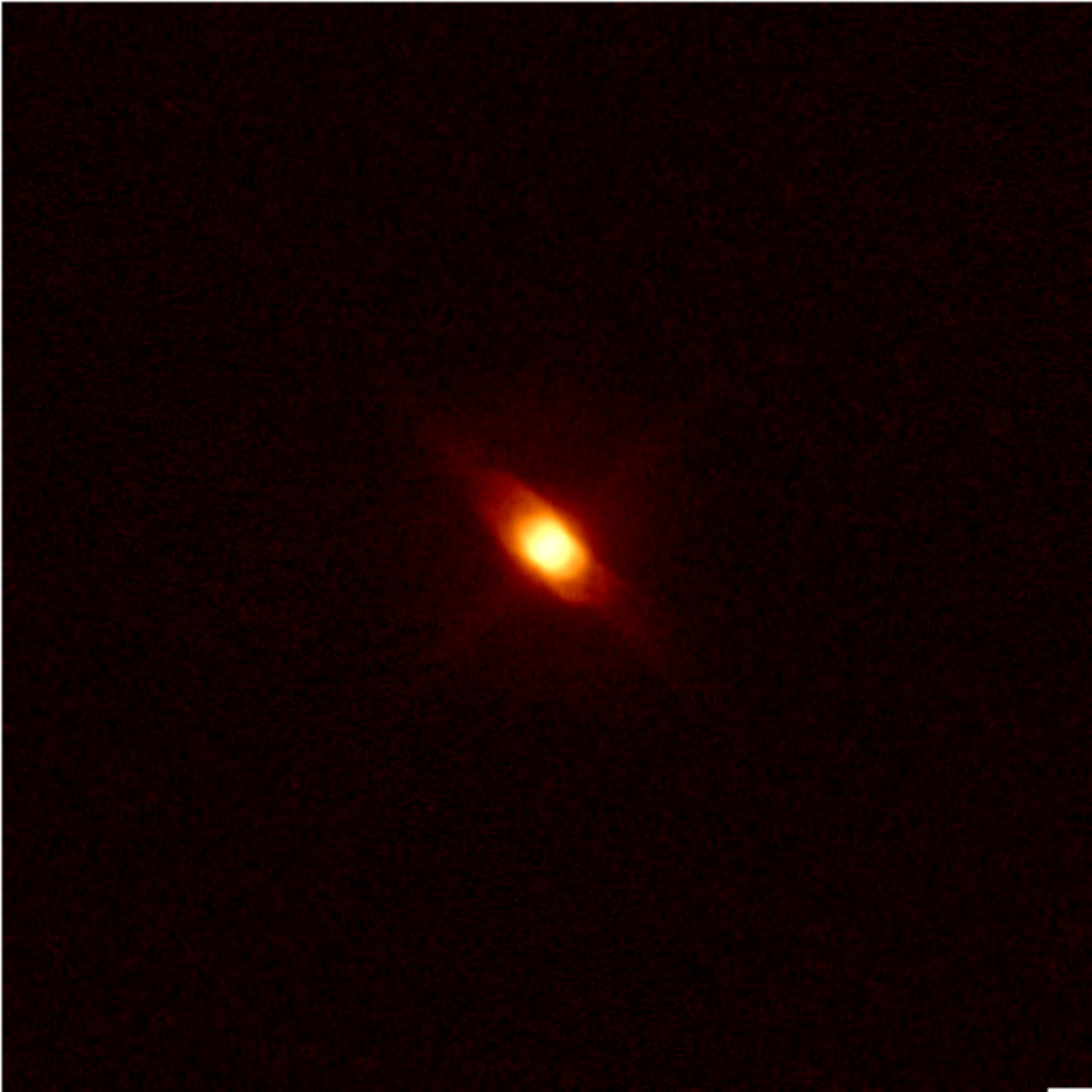}
\figsetgrpnote{False-color image in the F502N filter.}
\figsetgrpend

\figsetgrpstart
\figsetgrpnum{12.33}
\figsetgrptitle{PN~G296.3--03.0}
\figsetplot{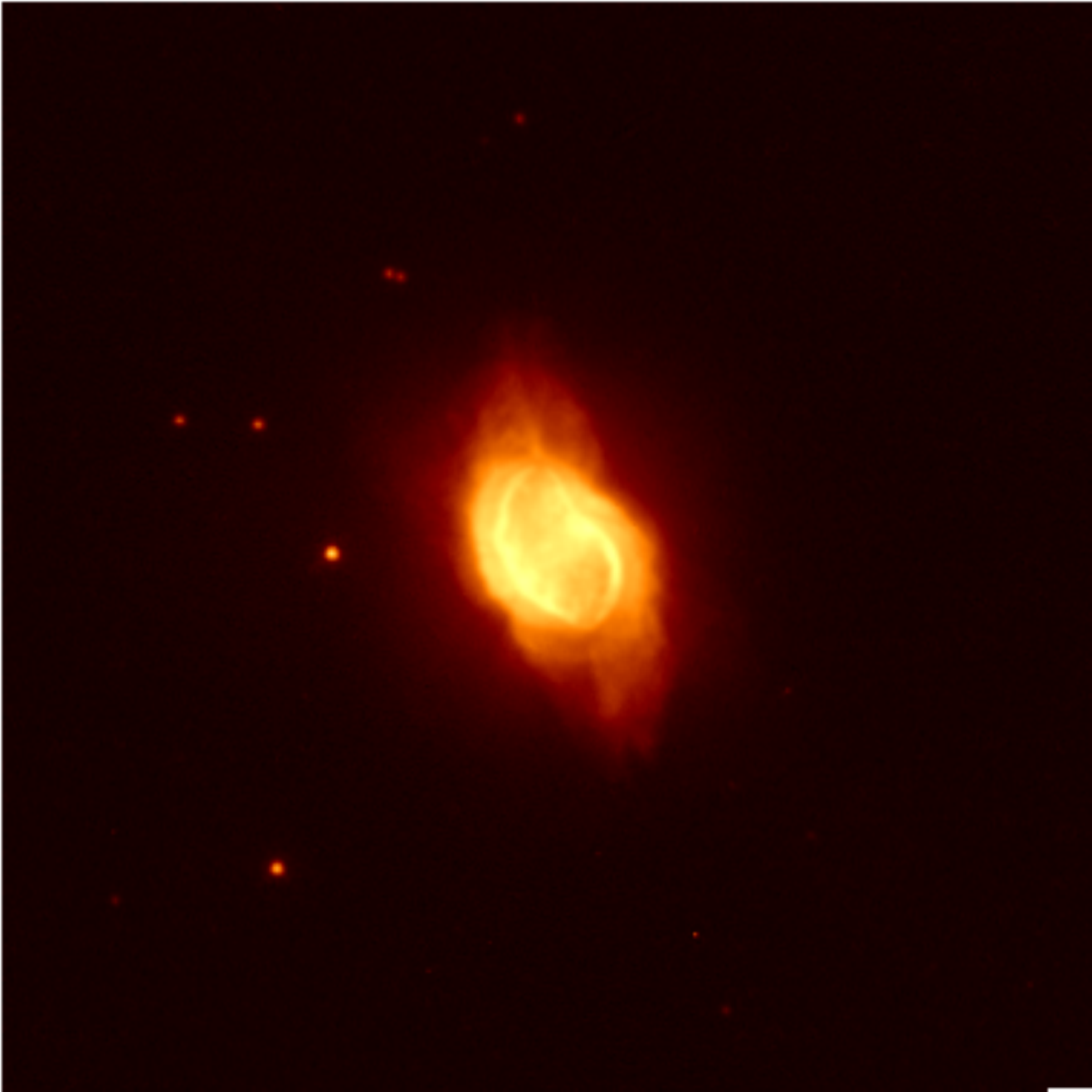}
\figsetgrpnote{False-color image in the F502N filter.}
\figsetgrpend

\figsetgrpstart
\figsetgrpnum{12.34}
\figsetgrptitle{PN~G309.0+00.8}
\figsetplot{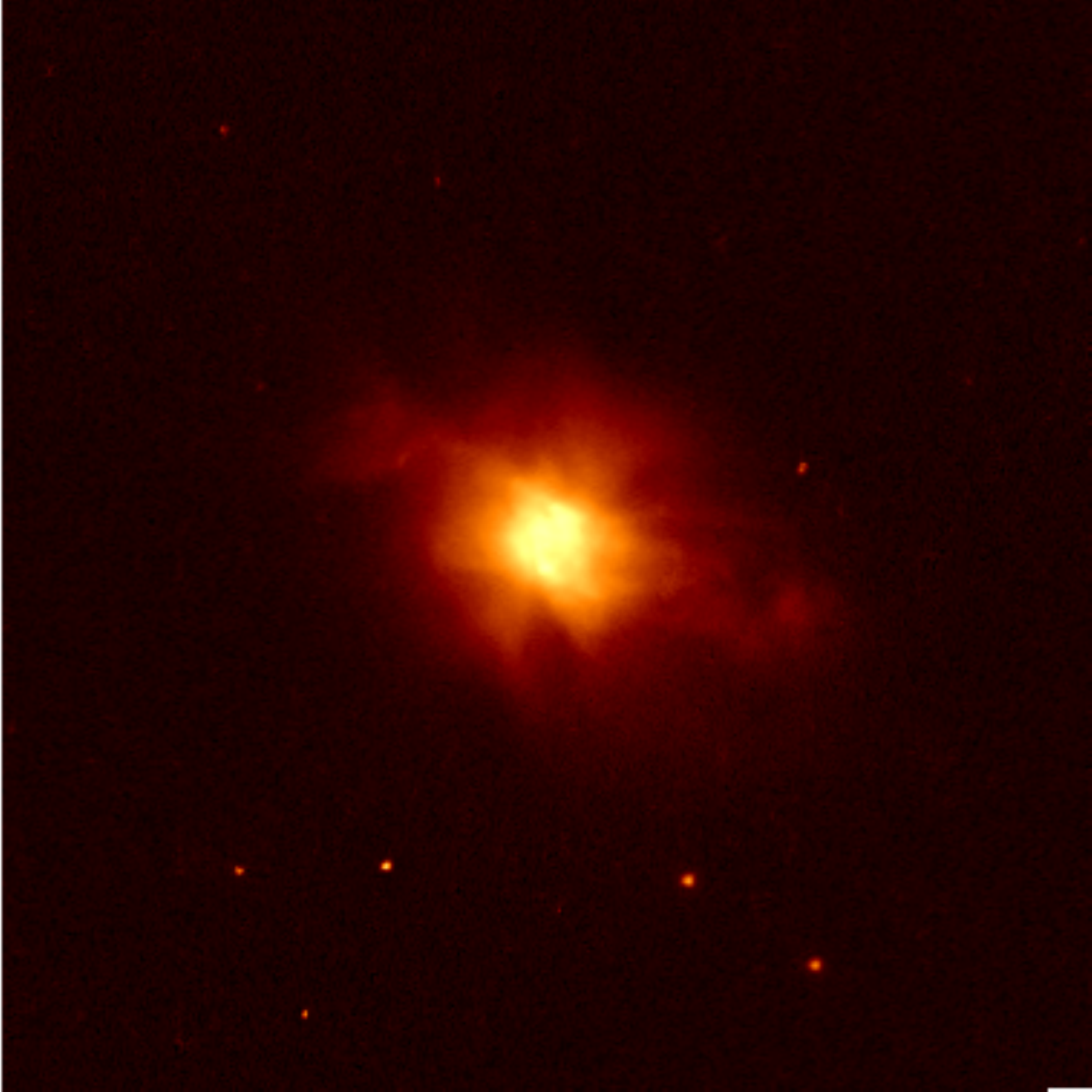}
\figsetgrpnote{False-color image in the F502N filter.}
\figsetgrpend

\figsetgrpstart
\figsetgrpnum{12.35}
\figsetgrptitle{PN~G309.5--02.9}
\figsetplot{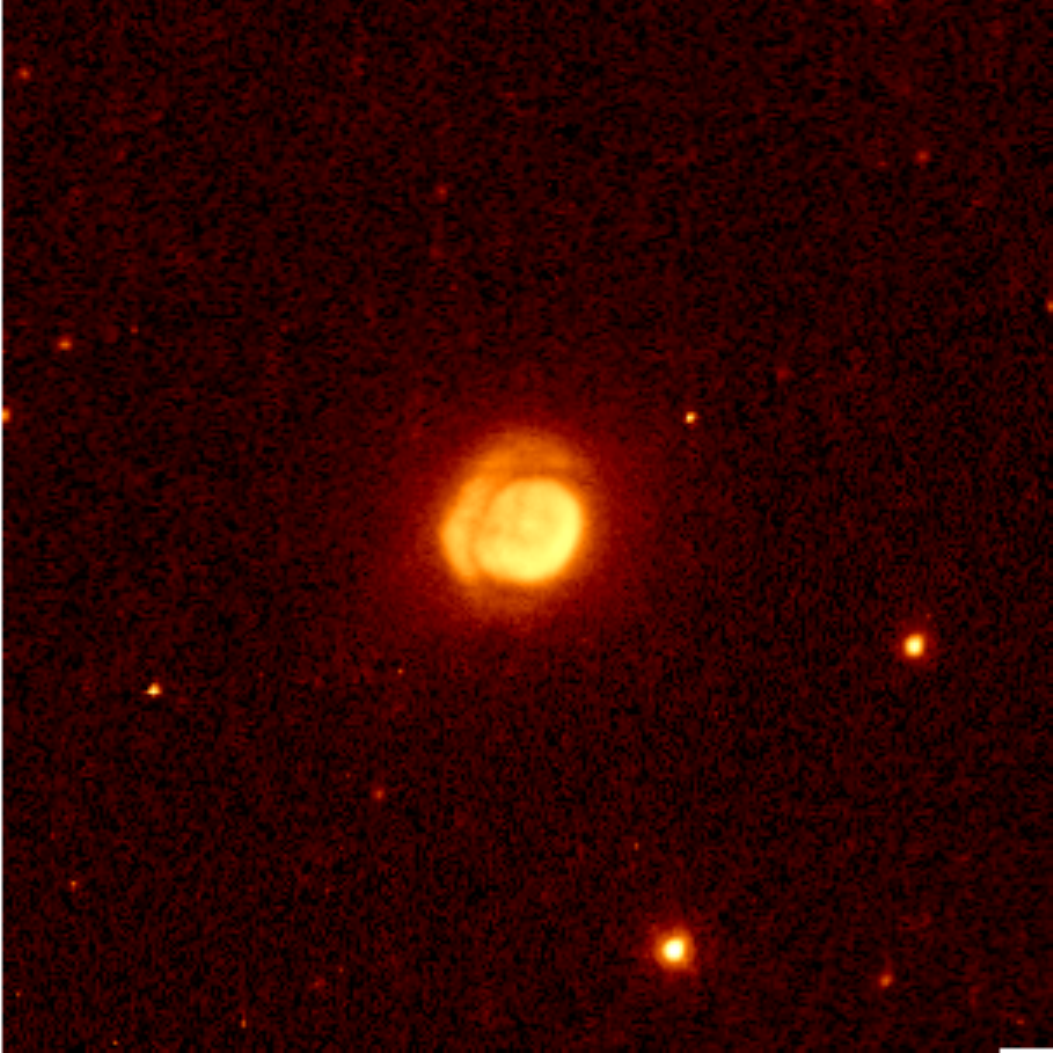}
\figsetgrpnote{False-color image in the F502N filter.}
\figsetgrpend

\figsetgrpstart
\figsetgrpnum{12.36}
\figsetgrptitle{PN~G324.8--01.1}
\figsetplot{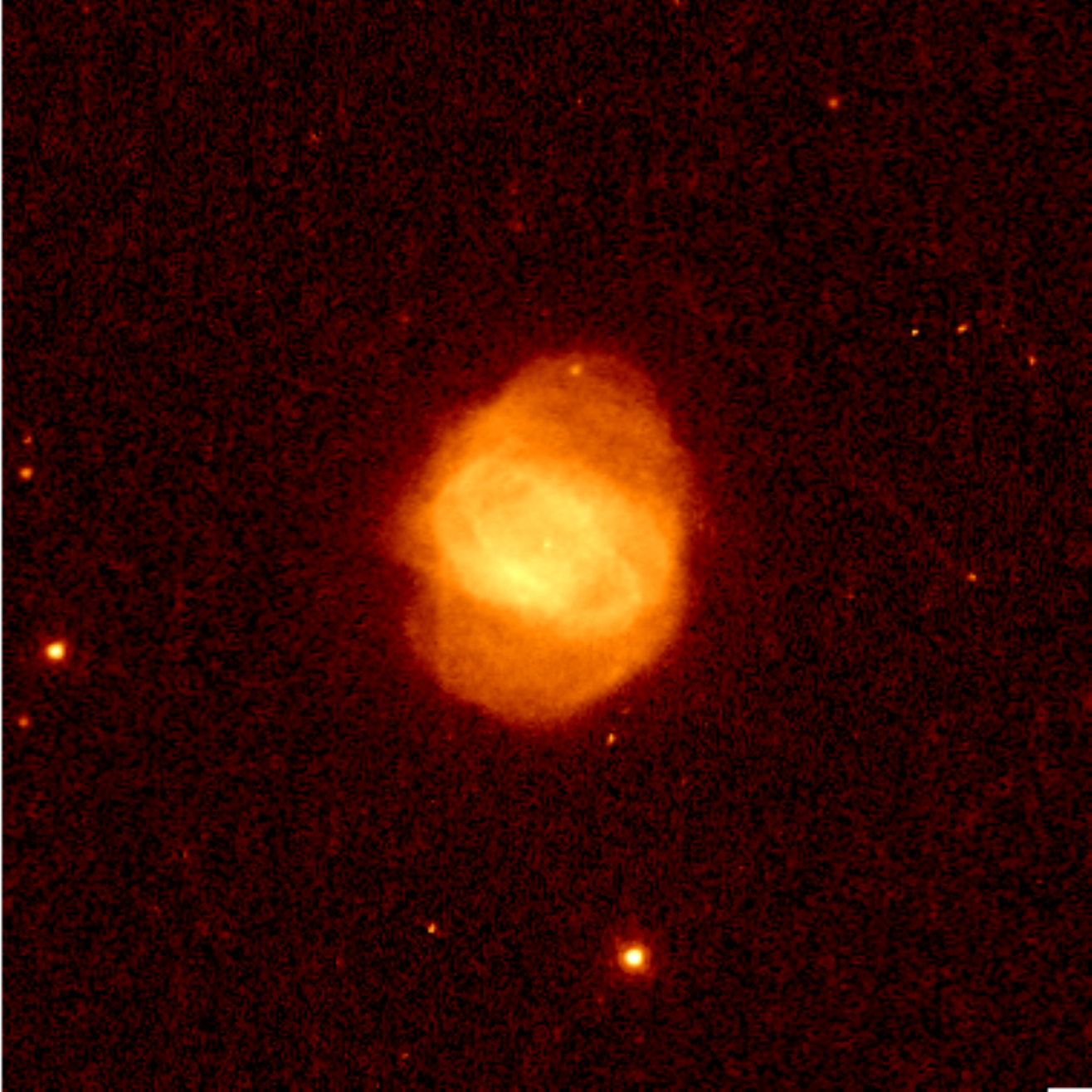}
\figsetgrpnote{False-color image in the F502N filter.}
\figsetgrpend

\figsetgrpstart
\figsetgrpnum{12.37}
\figsetgrptitle{PN~G327.1--01.8}
\figsetplot{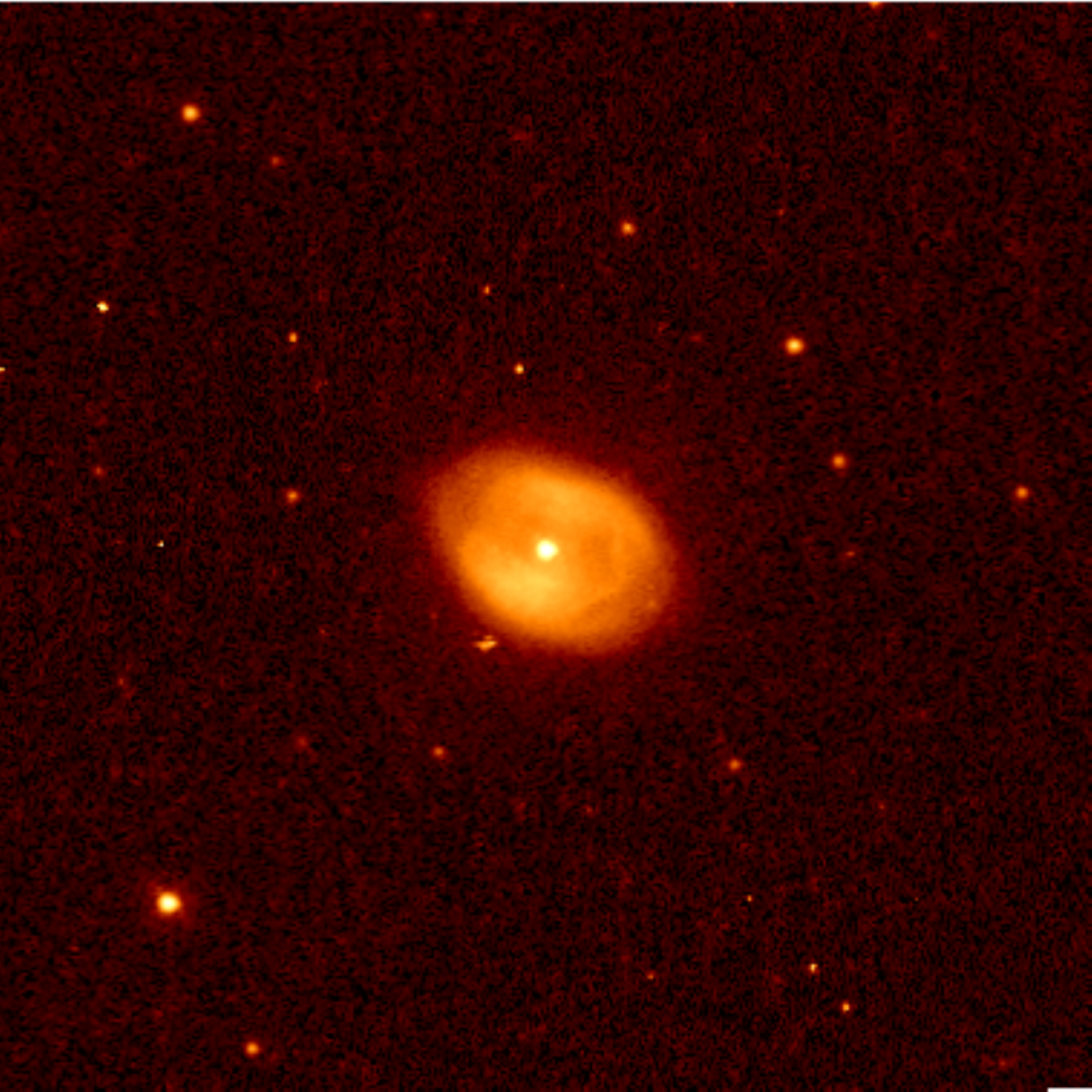}
\figsetgrpnote{False-color image in the F502N filter.}
\figsetgrpend

\figsetgrpstart
\figsetgrpnum{12.38}
\figsetgrptitle{PN~G327.8--06.1}
\figsetplot{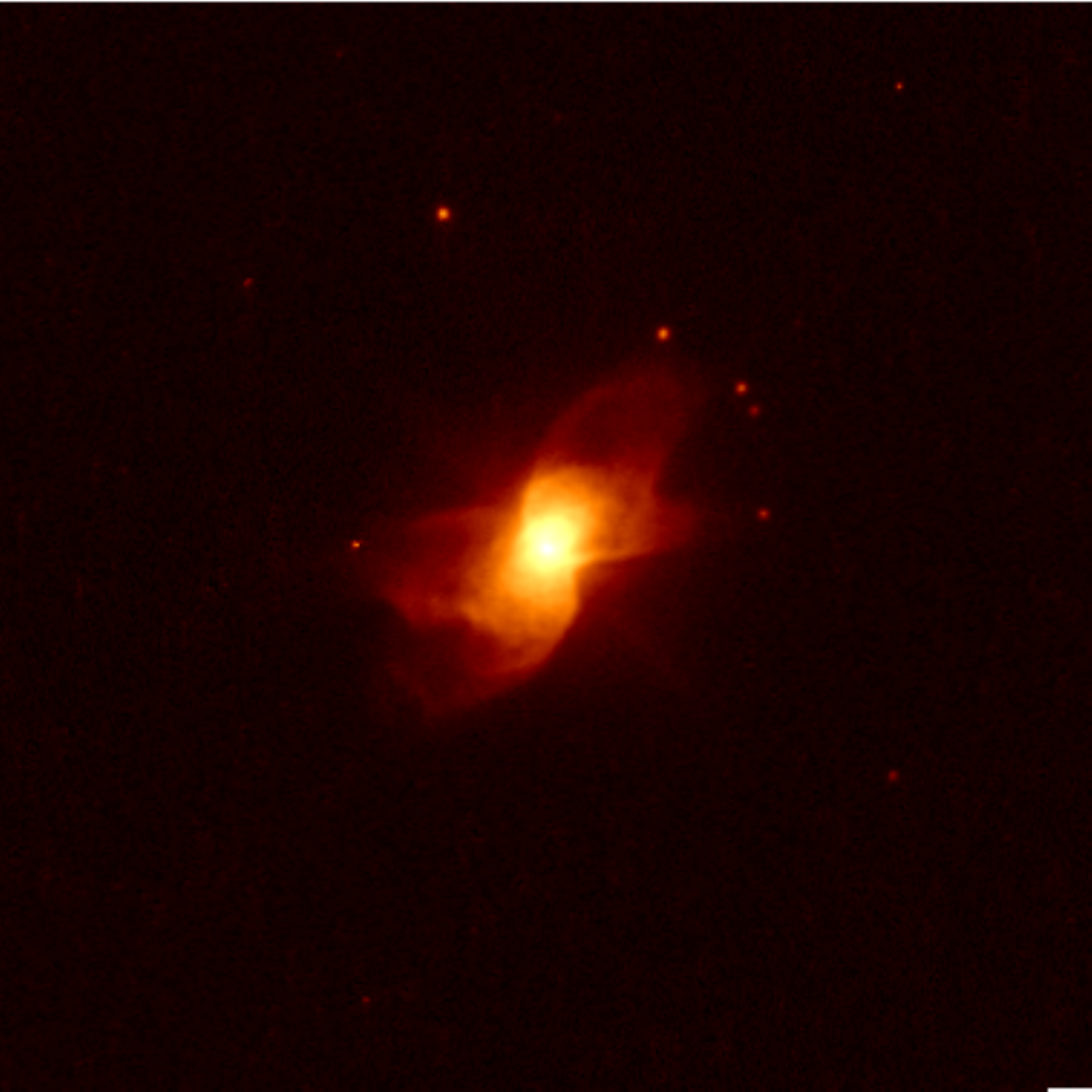}
\figsetgrpnote{False-color image in the F502N filter.}
\figsetgrpend

\figsetgrpstart
\figsetgrpnum{12.39}
\figsetgrptitle{PN~G334.8--07.4}
\figsetplot{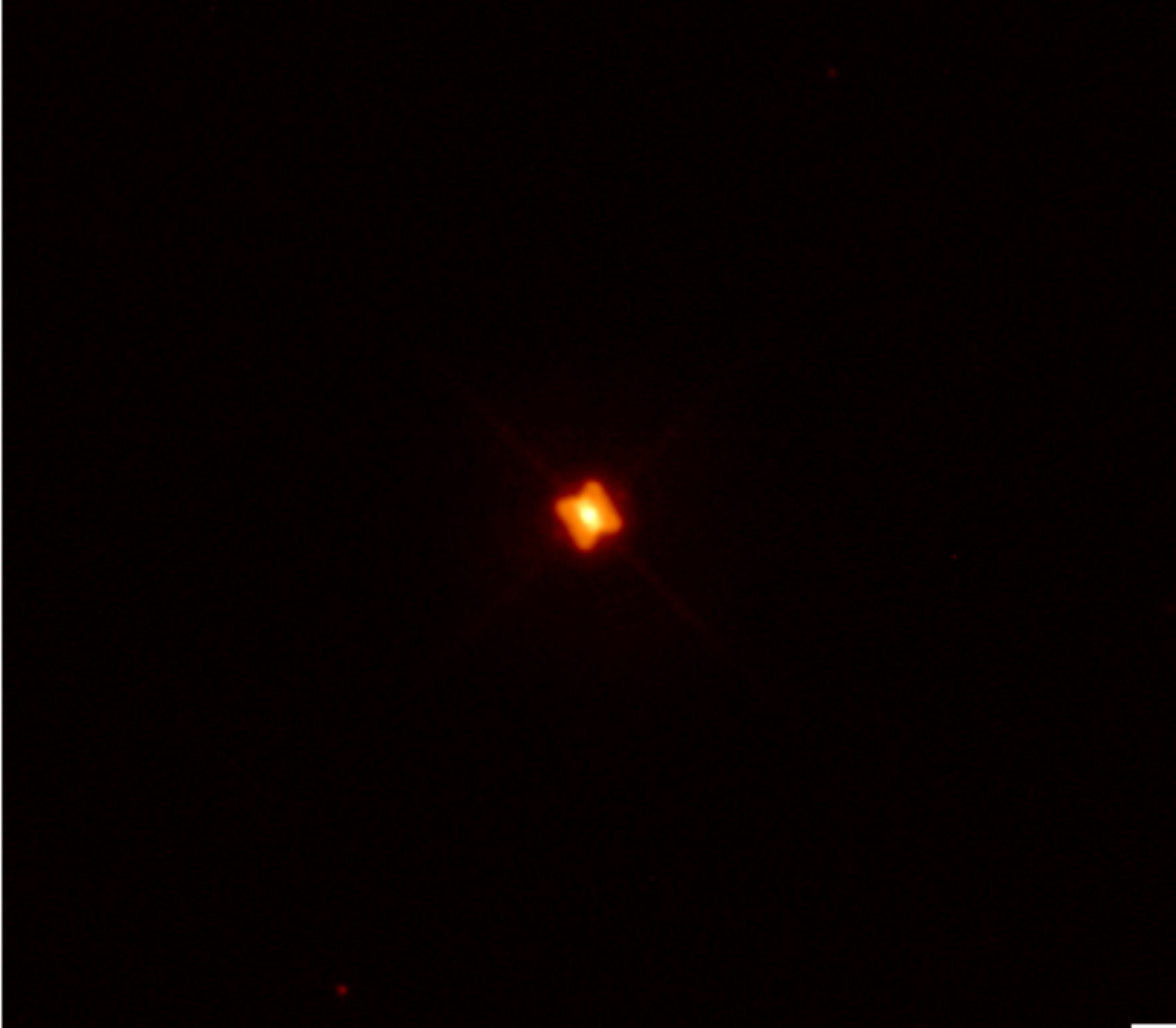}
\figsetgrpnote{False-color image in the F502N filter.}
\figsetgrpend

\figsetgrpstart
\figsetgrpnum{12.40}
\figsetgrptitle{PN~G336.9+08.3}
\figsetplot{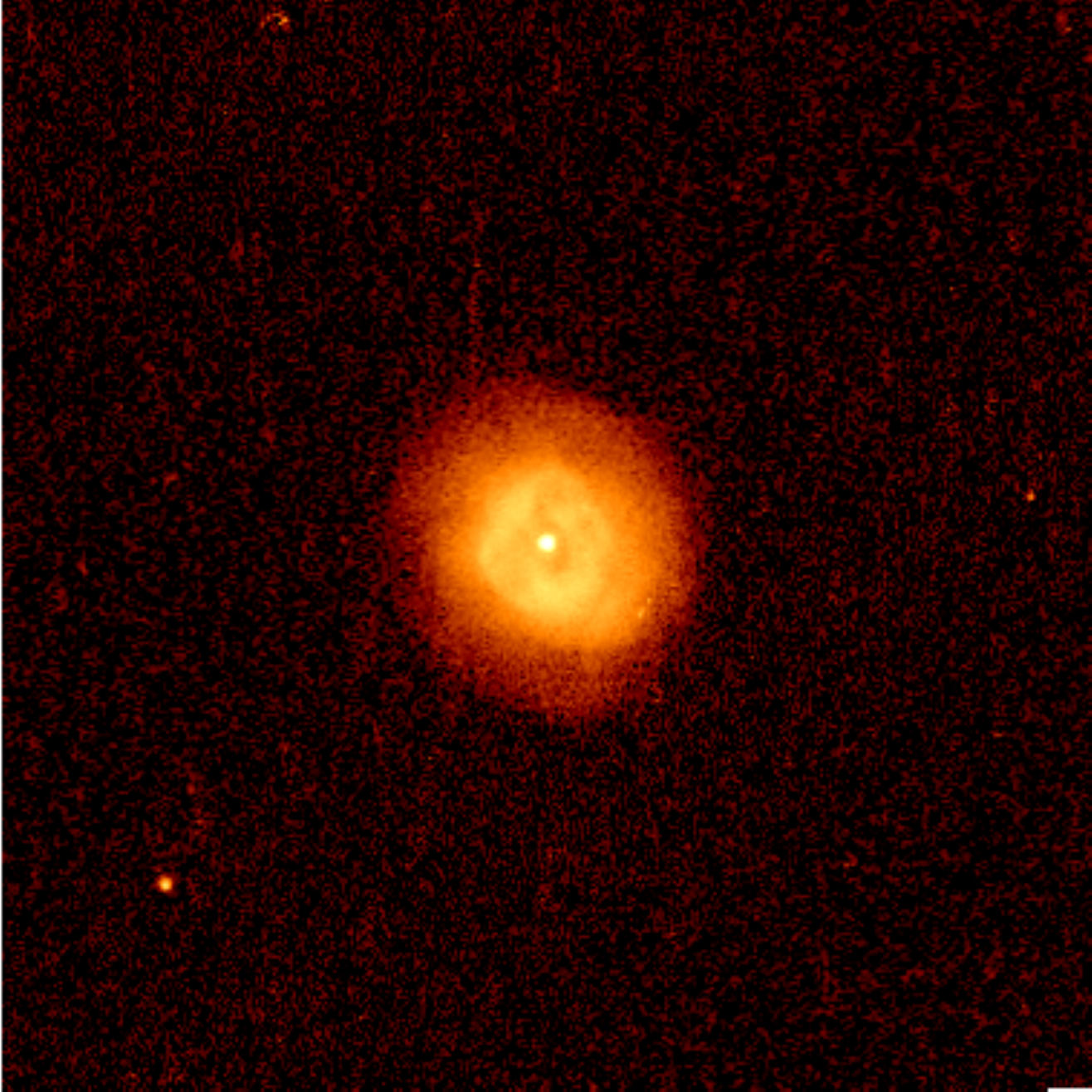}
\figsetgrpnote{False-color image in the F502N filter.}
\figsetgrpend

\figsetgrpstart
\figsetgrpnum{12.41}
\figsetgrptitle{PN~G340.9--04.6}
\figsetplot{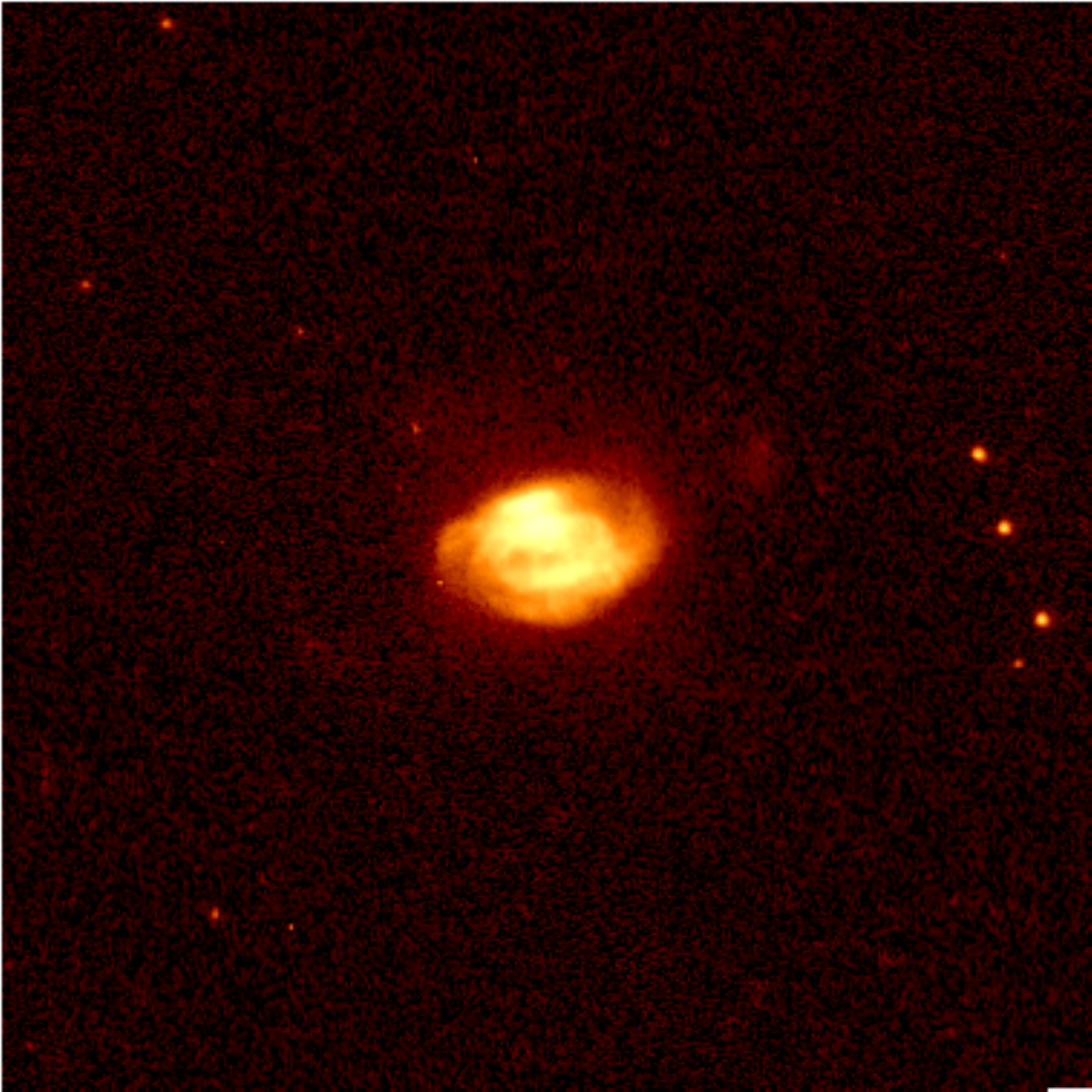}
\figsetgrpnote{False-color image in the F502N filter.}
\figsetgrpend

\figsetgrpstart
\figsetgrpnum{12.42}
\figsetgrptitle{PN~G341.5--09.1}
\figsetplot{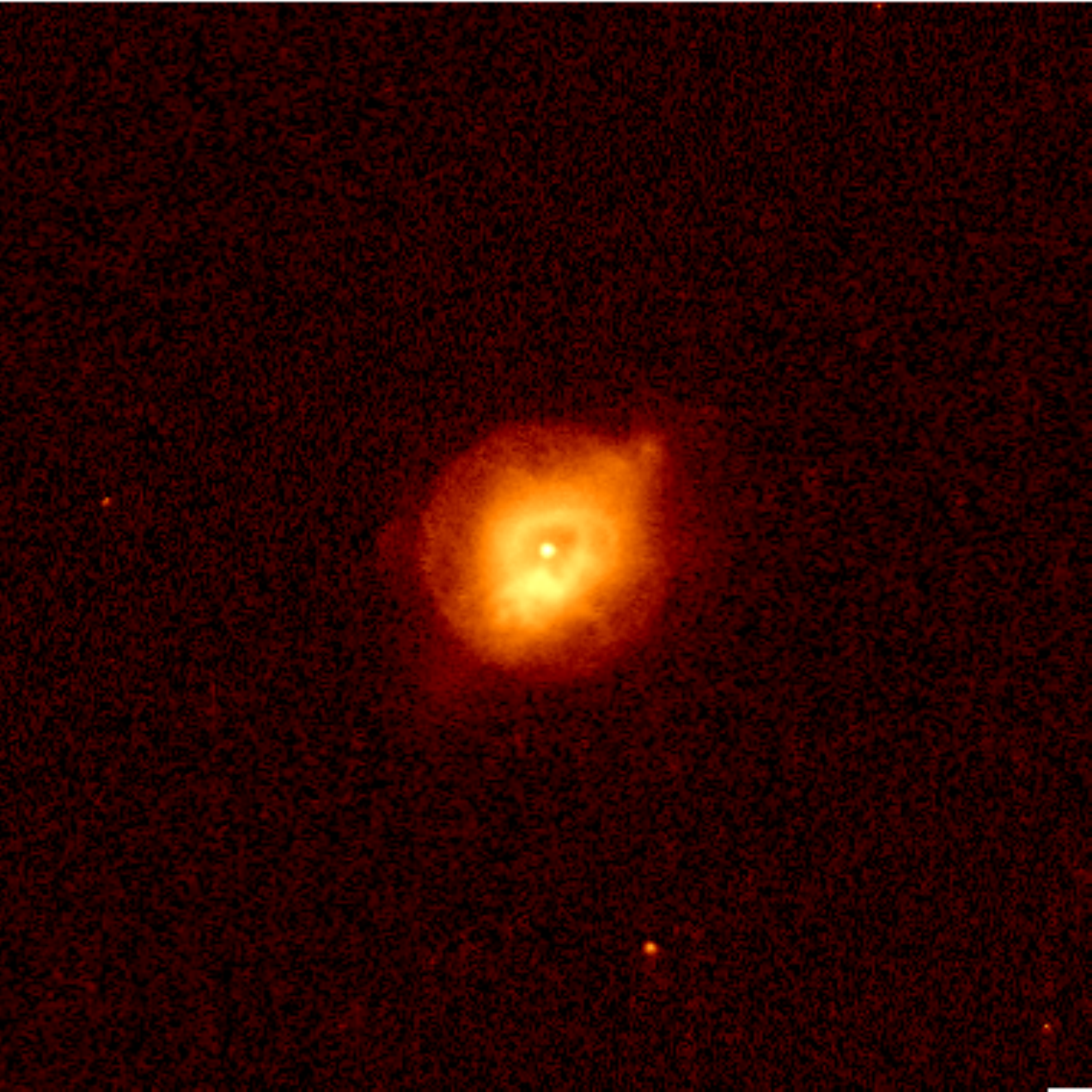}
\figsetgrpnote{False-color image in the F502N filter.}
\figsetgrpend

\figsetgrpstart
\figsetgrpnum{12.43}
\figsetgrptitle{PN~G343.4+11.9}
\figsetplot{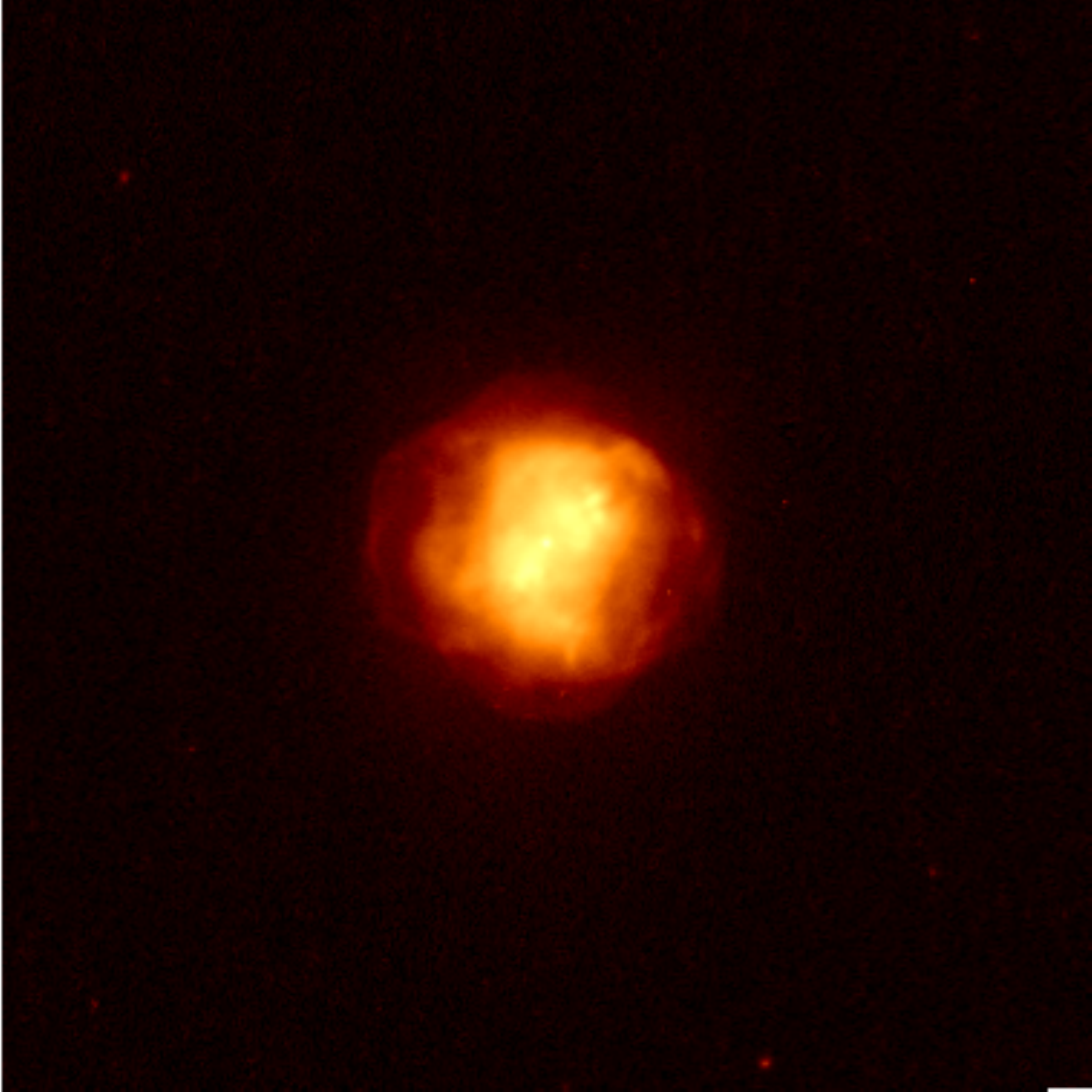}
\figsetgrpnote{False-color image in the F502N filter.}
\figsetgrpend

\figsetgrpstart
\figsetgrpnum{12.44}
\figsetgrptitle{PN~G344.2+04.7}
\figsetplot{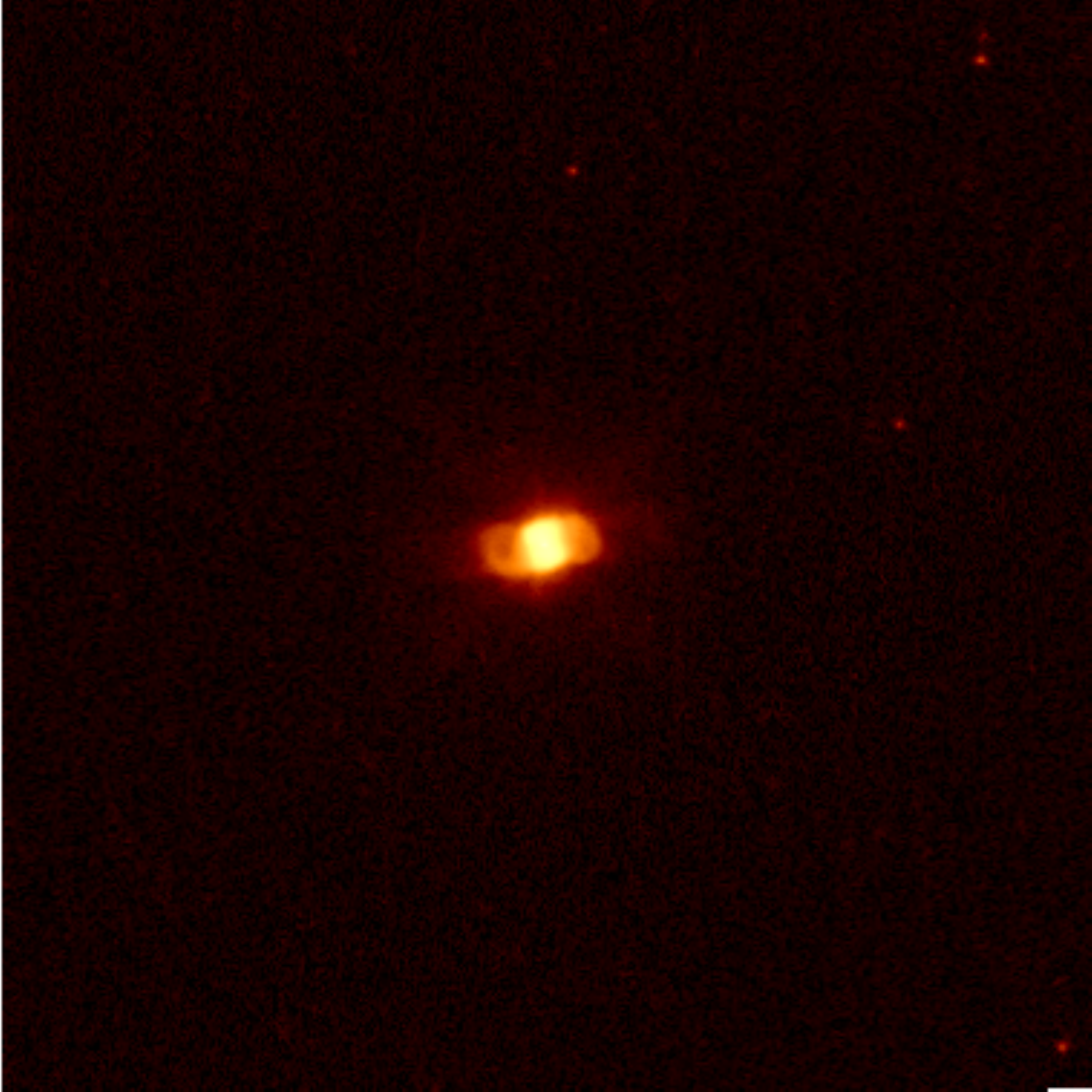}
\figsetgrpnote{False-color image in the F502N filter.}
\figsetgrpend

\figsetgrpstart
\figsetgrpnum{12.45}
\figsetgrptitle{PN~G344.8+03.4}
\figsetplot{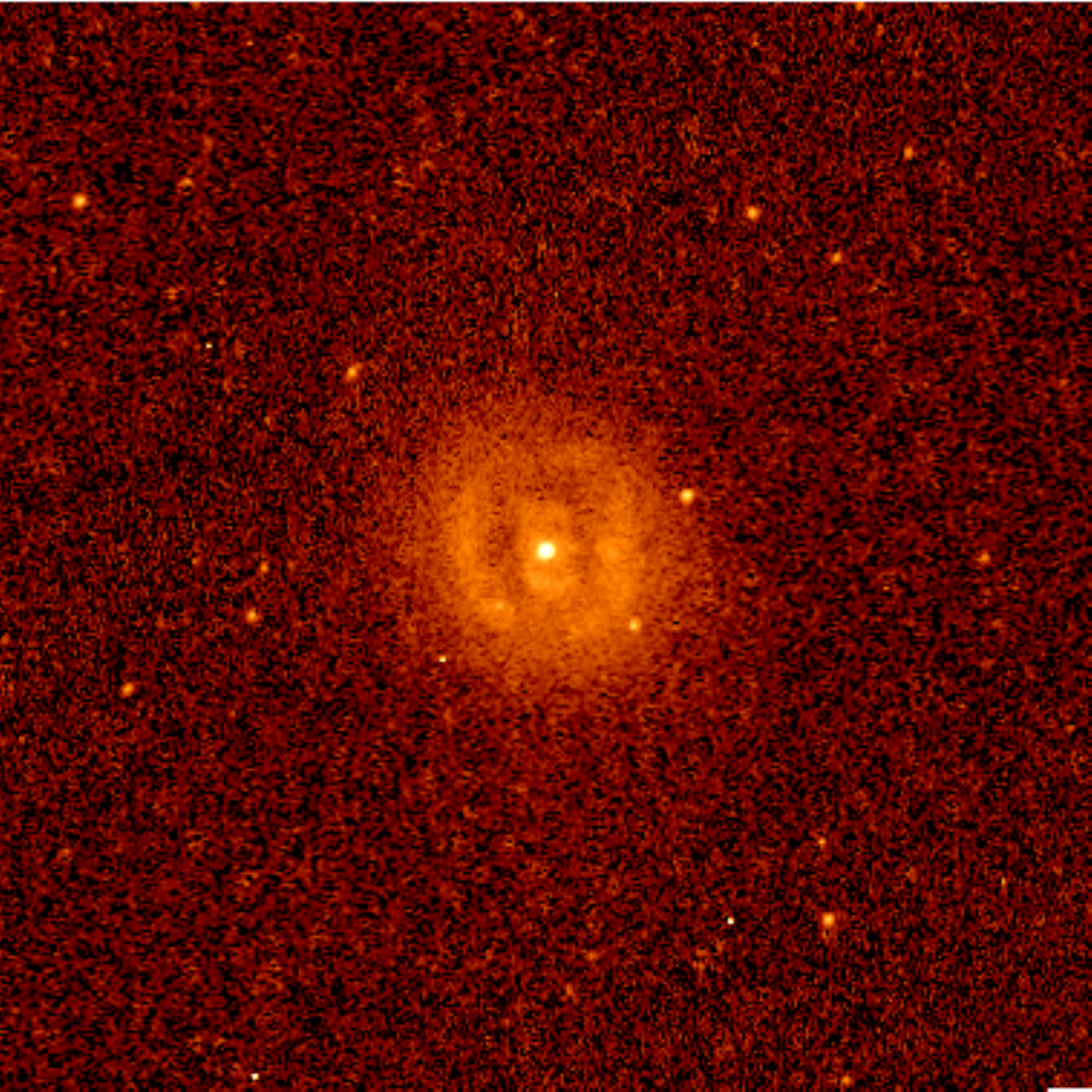}
\figsetgrpnote{False-color image in the F502N filter.}
\figsetgrpend

\figsetgrpstart
\figsetgrpnum{12.46}
\figsetgrptitle{PN~G345.0+04.3}
\figsetplot{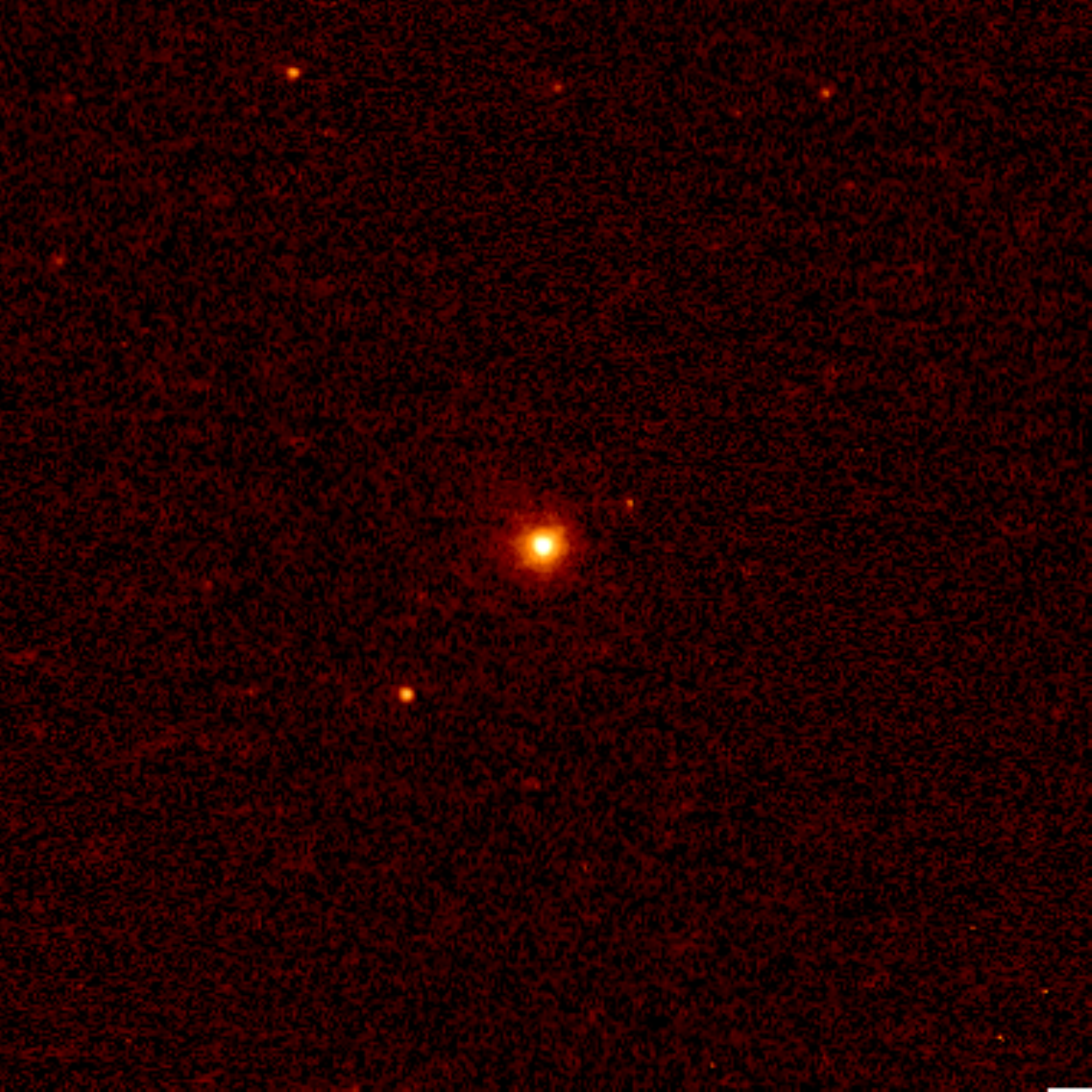}
\figsetgrpnote{False-color image in the F502N filter.}
\figsetgrpend

\figsetgrpstart
\figsetgrpnum{12.47}
\figsetgrptitle{PN~G348.4--04.1}
\figsetplot{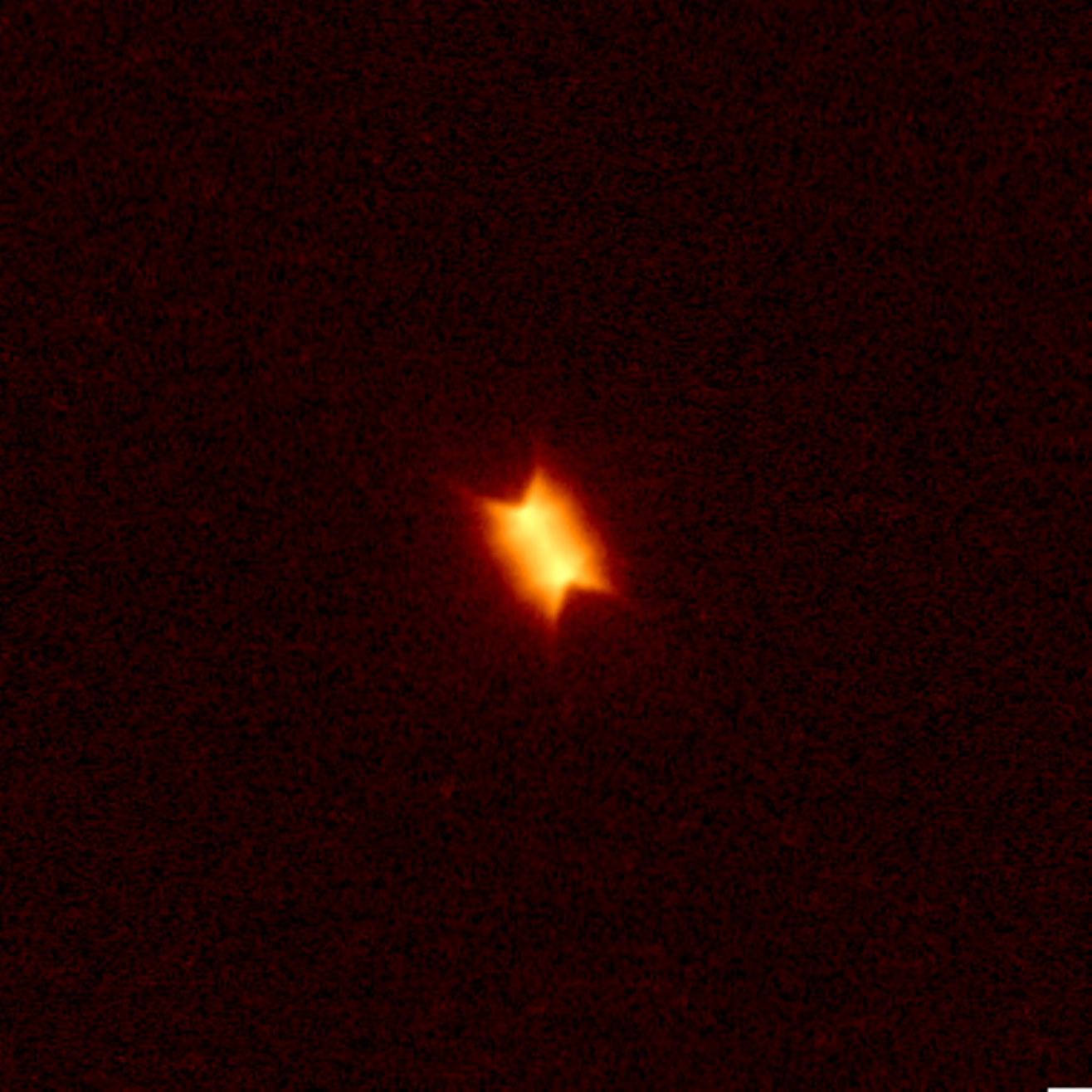}
\figsetgrpnote{False-color image in the F502N filter.}
\figsetgrpend

\figsetgrpstart
\figsetgrpnum{12.48}
\figsetgrptitle{PN~G348.8--09.0}
\figsetplot{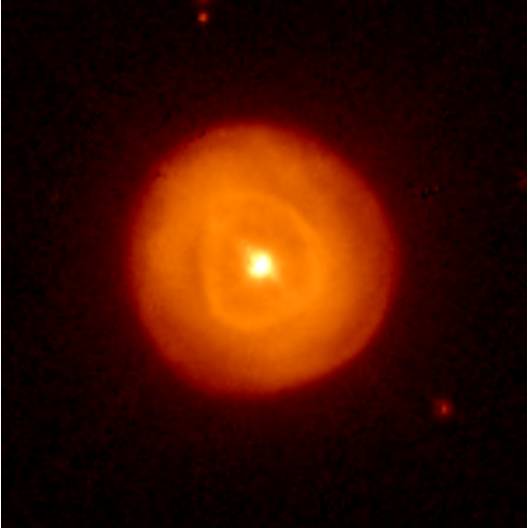}
\figsetgrpnote{False-color image in the F200LP filter.}
\figsetgrpend

\figsetgrpstart
\figsetgrpnum{12.49}
\figsetgrptitle{PN~G351.3+07.6}
\figsetplot{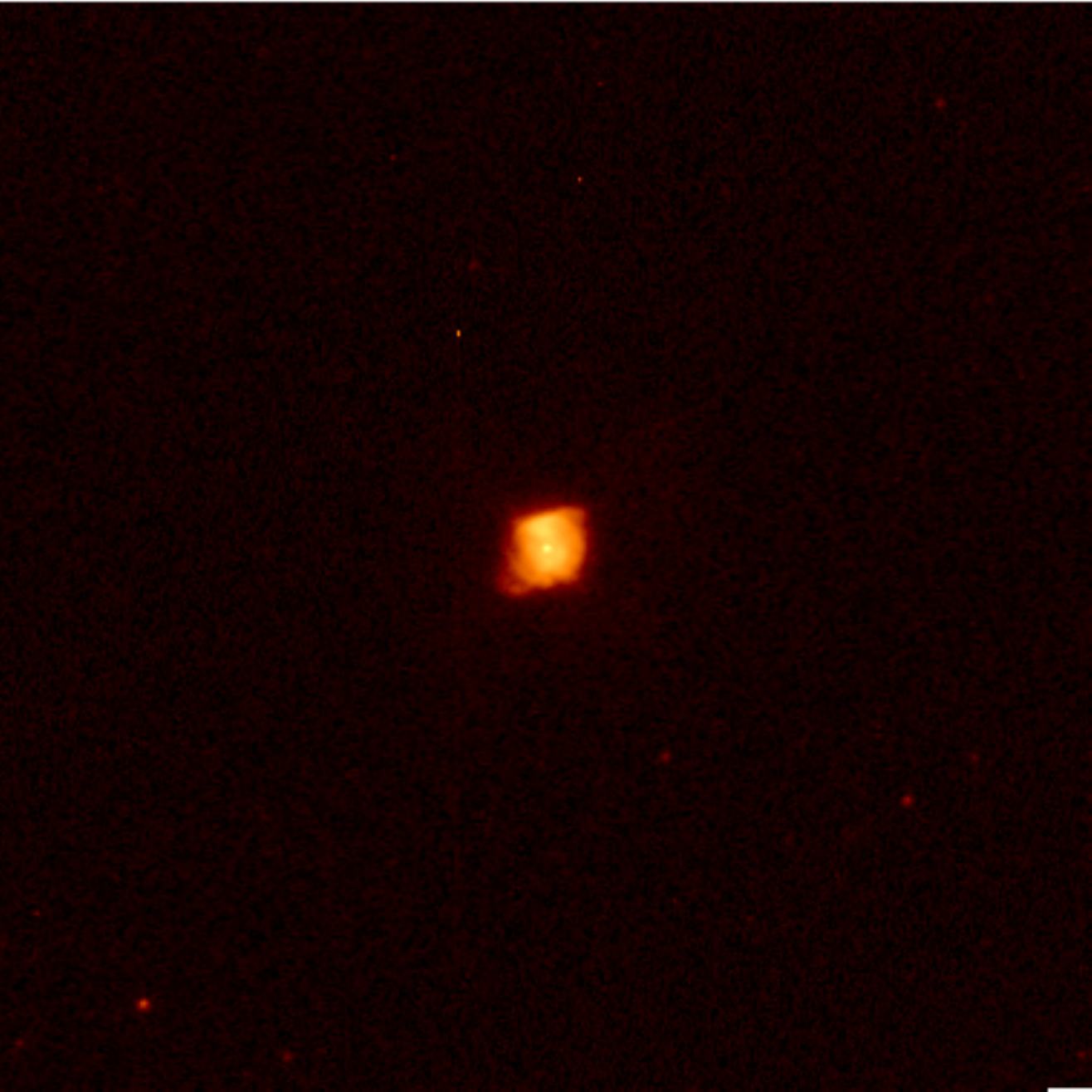}
\figsetgrpnote{False-color image in the F502N filter.}
\figsetgrpend

\figsetgrpstart
\figsetgrpnum{12.50}
\figsetgrptitle{PN~G356.5+01.5}
\figsetplot{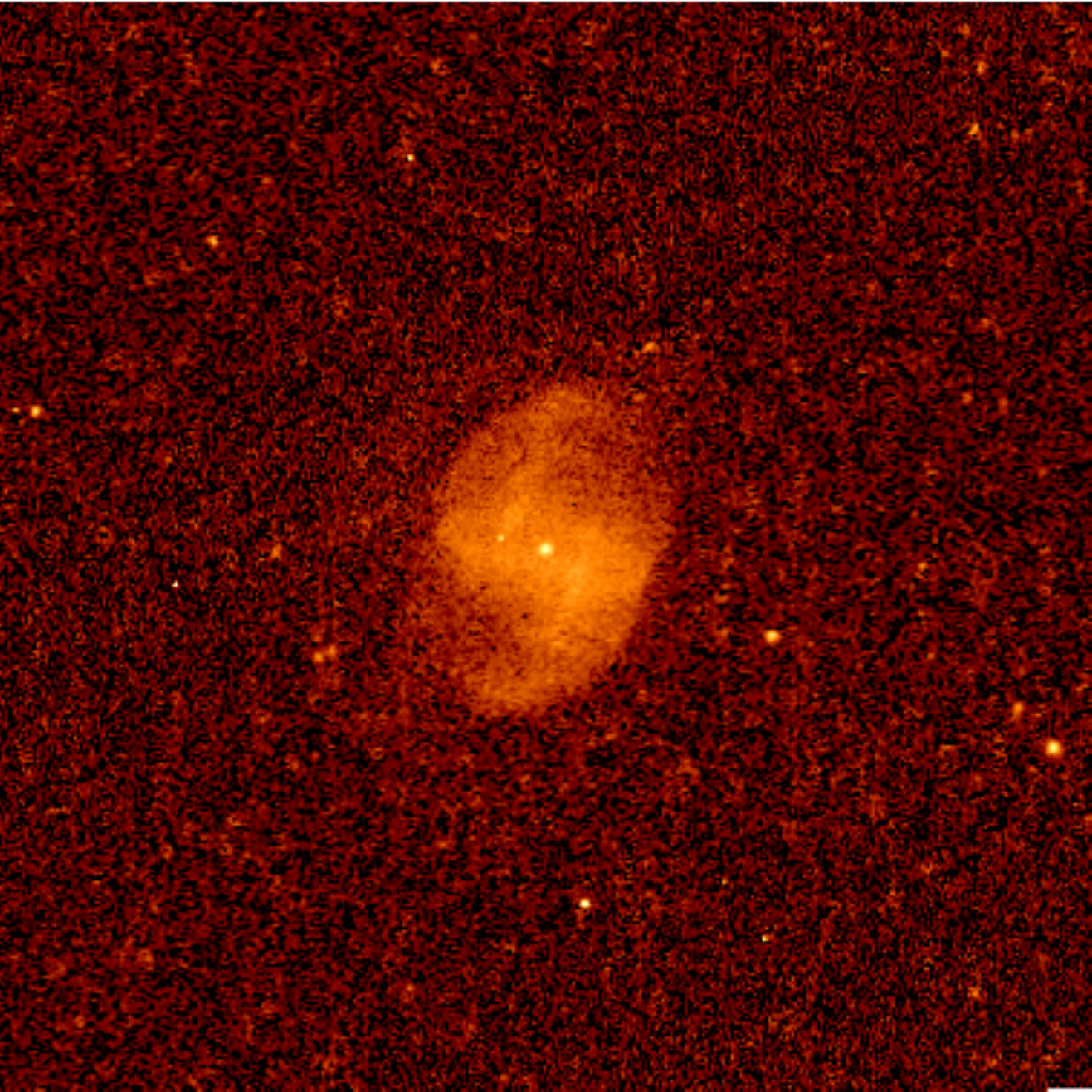}
\figsetgrpnote{False-color image in the F502N filter.}
\figsetgrpend

\figsetgrpstart
\figsetgrpnum{12.51}
\figsetgrptitle{PN~G358.6+07.8}
\figsetplot{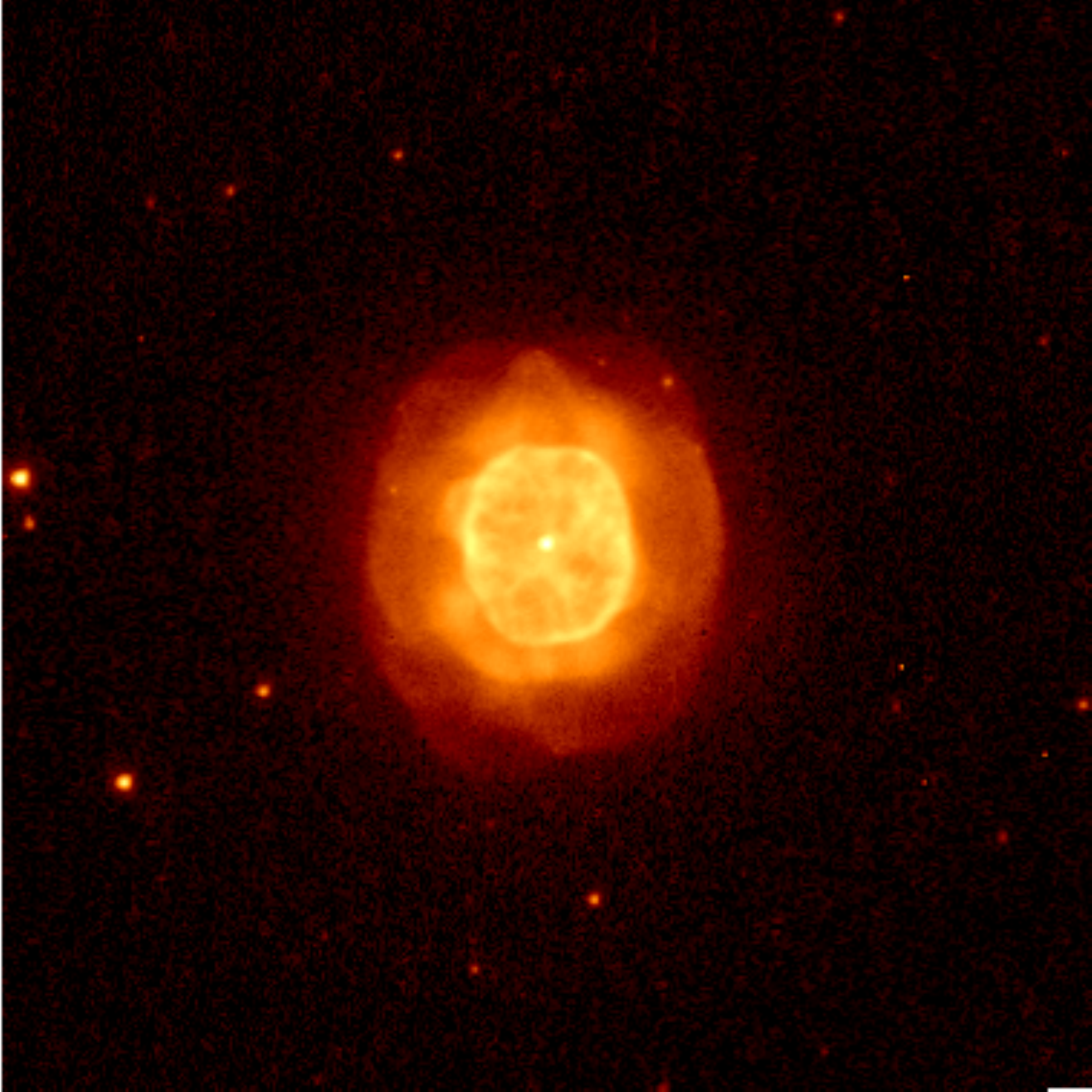}
\figsetgrpnote{False-color image in the F502N filter.}
\figsetgrpend

\figsetend

\begin{figure}   		
\epsscale{1.0}
\plotone{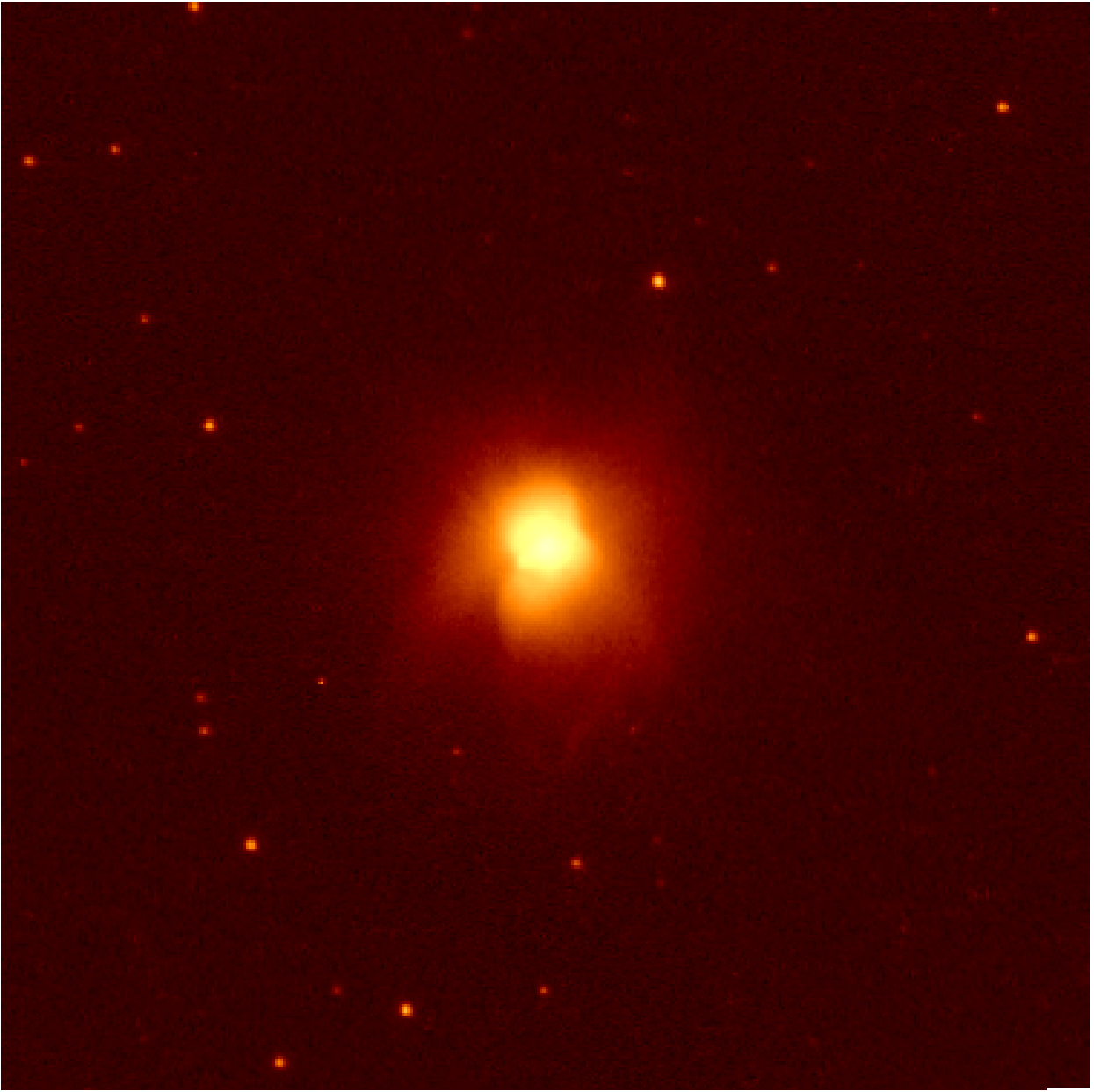}
\figcaption{
False-color image of PN~G000.8--07.6 in the F502N filter.}
\end{figure}

\clearpage

\begin{deluxetable}{c c c l l c c}
\tabletypesize{\scriptsize}
\tablecolumns{7}
\tablewidth{0pt}
\tablecaption {Observing Log \label{ObsLog}}
\tablehead {
\colhead {PN~G} & \colhead {Name} & \colhead {UT Date} & \colhead {Dataset} & 
\colhead {Filter} & \colhead {t$_{\rm exp}$} & \colhead{PA} \\
\colhead{}& \colhead{}& \colhead{}& \colhead{}& \colhead{}& \colhead{(s)}& \colhead{(deg)}\\

\colhead {(1)} & \colhead {(2)} & \colhead {(3)} & \colhead {(4)} & 
\colhead {(5)} & \colhead {(6)} &\colhead{(7)} \\
}
\startdata  

000.8--07.6 & H2--46 & 2010 Mar 07 & IB1B05011 & F502N & 375, 375 & 130.1 \\*
               &            & & IB1B05020 & F200LP & 20, 2 &\\*
               &            & & IB1B05030 & F350LP & 20, 2 &\\*
               &            & & IB1B05UNQ & F814W & 25 &\\
011.7--06.6 & M1--55 & 2009 Sep 04 & IB1B23011 & F502N & 50, 50 & -48.3\\*
               &            & & IB1B23020 & F200LP & 30, 3 &\\*
               &            & & IB1B23030 & F350LP & 30, 3 &\\*
               &            & & IB1B23RIQ & F814W & 32 &\\
014.0--05.5 & VV3--5 & 2009 Sep 22 & IB1B27011 & F502N & 50, 50& -48.3 \\*
               &            & & IB1B27020 & F200LP & 60, 6 &\\*
               &            & & IB1B27030 & F350LP & 60, 6 &\\*
               &            & & IB1B27PJQ & F814W & 40 &\\
014.3--05.5 & VV3--6 & 2009 Sep 17 & IB1B28011 & F502N & 50, 50 & -47.9\\*
               &            & & IB1B28020 & F200LP & 50, 5 &\\*
               &            & & IB1B28030 & F350LP & 50, 5 &\\*
               &            & & IB1B28WQQ & F814W & 38 &\\
021.1--05.9 & M1--63 & 2009 Oct 11 & IB1B34011 & F502N & 300, 300& -51.4 \\*
               &            & & IB1B34020 & F200LP & 20, 2 &\\*
               &            & & IB1B34030 & F350LP & 20, 2& \\*
               &            & & IB1B34F7Q & F814W & 27& \\
025.3--04.6 & K4--8 & 2010 Mar 5 & IB1B38011 & F502N & 30, 30&137.0 \\*
               &            & & IB1B38020 & F200LP & 20, 2 &\\*
               &            & & IB1B38030 & F350LP & 20, 2& \\*
               &            & & IB1B38DJQ & F814W & 30& \\
026.5--03.0 & Pe1--19 & 2010 Feb 22 & IB1B40011 & F502N & 300, 300&142.3 \\*
               &            & & IB1B40020 & F200LP & 40, 4 &\\*
               &            & & IB1B40030 & F350LP & 40, 4& \\*
               &            & & IB1B40CLQ & F814W & 35& \\
042.9--06.9 & NGC~6807 & 2009 Aug 07 & IB1B48011 & F502N & 100, 100&-1.4 \\*
               &            & & IB1B48020 & F200LP & 20, 2 &\\*
               &            & & IB1B48030 & F350LP & 20, 2 &\\*
               &            & & IB1B48PYQ & F814W & 25& \\
044.1+05.8 & CTSS--2 & 2010 Aug 30 & IB1B49011 & F502N & 300, 300&-26.5 \\*
               &            & & IB1B49020 & F200LP & 80, 8& \\*
               &            & & IB1B49030 & F350LP & 80, 8& \\*
               &            & & IB1B49ARQ & F814W & 46& \\
048.5+04.2 & K4--16 & 2010 Mar 01 & IB1B52011 & F502N & 100, 100 &153.2\\*
               &            & & IB1B52020 & F200LP & 120, 12& \\*
               &            & & IB1B52030 & F350LP & 120, 12 &\\*
               &            & & IB1B52I7Q & F814W & 52& \\
052.9--02.7 & K3--41 & 2009 Oct 03 & IB1B56011 & F502N & 300, 300&-45.2 \\*
               &            & & IB1B56020 & F200LP & 60, 6 &\\*
               &            & & IB1B56030 & F350LP & 60, 6& \\*
               &            & & IB1B56SXQ & F814W & 40& \\
053.3+24.0 & Vy1--2 & 2010 Feb 02 & IB1B57011 & F502N & 300, 300&168.0 \\*
               &            & & IB1B57020 & F200LP & 10, 1 &\\*
               &            & & IB1B57030 & F350LP & 10, 1& \\*
               &            & & IB1B57A7Q & F814W & 23 &\\
059.9+02.0 & K3--39 & 2009 Dec 29 & IB1B61011 & F502N & 300, 300 &-116.6\\*
               &            & & IB1B61020 & F200LP & 120, 75& \\*
               &            & & IB1B61030 & F350LP & 120, 75 &\\*
               &            & & IB1B61LEQ & F814W & 120& \\
063.8--03.3 & K3--54 & 2009 Nov  02 & IB1B63011 & F502N & 150, 150&-61.6 \\*
               &            & & IB1B63020 & F200LP & 20, 2 &\\*
               &            & & IB1B63030 & F350LP & 20, 2 &\\*
               &            & & IB1B63DSQ & F814W & 30& \\
068.7+01.9 & K4--41 & 2010 Jan 30 & IB1B64011 & F502N & 100, 100&-152.2 \\*
               &            & & IB1B64020 & F200LP & 110, 11& \\*
               &            & & IB1B64030 & F350LP & 110, 11& \\*
               &            & & IB1B64CMQ & F814W & 50& \\
068.7+14.8 & Sp4--1 & 2009 Jul 30 & IB1B65011 & F502N & 30, 30 &17.4\\*
               &            & & IB1B65020 & F200LP & 20, 2 &\\*
               &            & & IB1B65030 & F350LP & 20, 2& \\*
               &            & & IB1B65CPQ & F814W & 25& \\
079.9+06.4 & K3--56 & 2009 Oct 02 & IB1B67011 & F502N & 690, 690& -31.6 \\*
               &            & & IB1B67020 & F200LP & 80, 8 &\\*
               &            & & IB1B67030 & F350LP & 80, 8 &\\*
               &            & & IB1B67LTQ & F814W & 46& \\
097.6--02.4 & M2--50 & 2009 Nov  15 & IB1B69011 & F502N & 300, 300&-50.8 \\*
               &            & & IB1B69020 & F200LP & 50, 5 &\\*
               &            & & IB1B69030 & F350LP & 50, 5& \\*
               &            & & IB1B69VFQ & F814W & 36& \\
098.2+04.9 & K3--60 & 2009 Dec 04 & IB1B70011 & F502N & 300, 300&-72.9 \\*
               &            & & IB1B70020 & F200LP & 120, 80& \\*
               &            & & IB1B70030 & F350LP & 120, 80& \\*
               &            & & IB1B70WXQ & F814W & 38 &\\
104.1+01.0 & Bl2--1 & 2009 Oct 12 & IB1B71011 & F502N & 150, 150&-6.3 \\*
               &            & & IB1B71020 & F200LP & 120, 75& \\*
               &            & & IB1B71030 & F350LP & 120, 75& \\*
               &            & & IB1B71FNQ & F814W & 110& \\
%
%
107.4--02.6 & K3--87 & 2009 Nov  29 & IB1B0M011 & F502N & 710, 710 &-51.2\\*
               &            & & IB1B0M020 & F200LP & 60, 6& \\*
               &            & & IB1B0M030 & F350LP & 60, 6 &\\*
               &            & & IB1B0MLUQ & F814W & 42& \\
184.0--02.1 & M1--5 & 2009 Sep 03 & IB1B73011 & F502N & 150, 150&133.7 \\*
               &            & & IB1B73020 & F200LP & 20, 2& \\*
               &            & & IB1B73030 & F350LP & 20, 2& \\*
               &            & & IB1B73JHQ & F814W & 30 &\\
205.8--26.7 & MaC2--1 & 2009 Nov  11 & IB1B74011 & F502N & 30, 30&-170.9 \\*
               &            & & IB1B74020 & F200LP & 10, 1& \\*
               &            & & IB1B74030 & F350LP & 10, 1 &\\*
               &            & & IB1B74OXQ & F814W & 20& \\
263.0--05.5 & PB 2 & 2010 Jan 02  & IB1B75011 & F502N & 300, 300&-157.1 \\*
               &            & & IB1B75020 & F200LP & 30, 3& \\*
               &            & & IB1B75030 & F350LP & 30, 3 &\\*
               &            & & IB1B75H5Q & F814W & 33& \\
264.4--12.7 & He2--5 & 2009 Oct 04 & IB1B76011 & F502N & 150, 150&124.9 \\*
               &            & & IB1B76020 & F200LP & 20, 2& \\*
               &            & & IB1B76030 & F350LP & 20, 2& \\*
               &            & & IB1B76ZCQ & F814W & 24 &\\
274.1+02.5 & He2--34 & 2009 Dec 08 & IB1B77011 & F502N & 300, 300&164.1 \\*
               &            & & IB1B77020 & F200LP & 120, 54& \\*
               &            & & IB1B77030 & F350LP & 120, 54& \\*
               &            & & IB1B77HLQ & F814W & 100& \\
275.3--04.7 & He2--21 & 2009 Oct 17 & IB1B78011 & F502N & 300, 300&118.2 \\*
               &            & & IB1B78020 & F200LP & 120, 3& \\*
               &            & & IB1B78030 & F350LP & 120, 3& \\*
               &            & & IB1B78KCQ & F814W & 32& \\
278.6--06.7 & He2--26 & 2009 Oct 24 & IB1B79011 & F502N & 15, 15&123.1 \\*
               &            & & IB1B79020 & F200LP & 20, 2& \\*
               &            & & IB1B79030 & F350LP & 20, 2& \\*
               &            & & IB1B79BFQ & F814W & 27& \\
285.4+01.5 & Pe1--1 & 2009 Oct 05 & IB1B80011 & F502N & 300, 300&82.9 \\*
               &            & & IB1B80020 & F200LP & 120, 16& \\*
               &            & & IB1B80030 & F350LP & 120, 16 &\\*
               &            & & IB1B80GIQ & F814W & 60& \\
285.4+02.2 & Pe2--7 & 2009 Aug 17 & IB1B81011 & F502N & 50, 50&30.7 \\*
               &            & & IB1B81020 & F200LP & 70, 7& \\*
               &            & & IB1B81030 & F350LP & 70, 7& \\*
               &            & & IB1B81C8Q & F814W & 42& \\
286.0-06.5 & He2--41 & 2009 Oct 20 & IB1B82011 & F502N & 30, 30&104.9 \\*
               &            & & IB1B82020 & F200LP & 30, 3 &\\*
               &            & & IB1B82030 & F350LP & 30, 3& \\*
               &            & & IB1B82T5Q & F814W & 30& \\
289.8+07.7 & He2--63 & 2009 Dec 05 & IB1B83011 & F502N & 300, 300&141.6 \\*
               &            & & IB1B83020 & F200LP & 20, 2& \\*
               &            & & IB1B83030 & F350LP & 20, 2 &\\*
               &            & & IB1B83B9Q & F814W & 24& \\
294.9--04.3 & He2--68 & 2009 Nov  03 & IB1B84011 & F502N & 50, 50 &97.7\\*
               &            & & IB1B84020 & F200LP & 100, 10 &\\*
               &            & & IB1B84030 & F350LP & 100, 10& \\*
               &            & & IB1B84G0Q & F814W & 50 &\\
295.3--09.3 & He2--62 & 2009 Oct 29 & IB1B85011 & F502N & 30, 30 &93.9\\*
               &            & & IB1B85020 & F200LP & 20, 2 &\\*
               &            & & IB1B85030 & F350LP & 20, 2& \\*
               &            & & IB1B85RYQ & F814W & 26& \\
296.3--03.0 & He2--73 & 2009 Dec 06 & IB1B86011 & F502N & 300, 300&132.2 \\*
               &            & & IB1B86020 & F200LP & 80, 8 &\\*
               &            & & IB1B86030 & F350LP & 80, 8& \\*
               &            & & IB1B86RGQ & F814W & 45 &\\
309.0+00.8 & He2--96 & 2010 Jan 22 & IB1B87011 & F502N & 300, 300&147.7 \\*
               &            & & IB1B87020 & F200LP & 120, 20& \\*
               &            & & IB1B87030 & F350LP & 120, 20 &\\*
               &            & & IB1B87AUQ & F814W & 64& \\
309.5--02.9 & MaC1--2 & 2010 Mar 13 & IB1B88011 & F502N & 100, 100& -173.0\\*
               &            & & IB1B88020 & F200LP & 120, 20& \\*
               &            & & IB1B88030 & F350LP & 120, 20& \\*
               &            & & IB1B88ZTQ & F814W & 66& \\
311.1+03.4 & He2--101 & 2010 Jan 20 & IB1B89011 & F502N & 300, 300 &148.7\\*
               &            & & IB1B89020 & F200LP & 120, 15 &\\*
               &            & & IB1B89030 & F350LP & 120, 15& \\*
               &            & & IB1B89QJQ & F814W & 60& \\
324.8--01.1 & He2--133 & 2010 Mar 15 & IB1B91011 & F502N & 500, 500 &160.8\\*
               &            & & IB1B91021 & F200LP & 120, 120 &\\*
               &            & & IB1B91031 & F350LP & 120, 120& \\*
               &            & & IB1B91AOQ & F814W & 120& \\
327.1--01.8 & He2--140 & 2009 Aug 28 & IB1B93011 & F502N & 630, 630&-33.8 \\*
               &            & & IB1B93020 & F200LP & 120, 20& \\*
               &            & & IB1B93030 & F350LP & 120, 20& \\*
               &            & & IB1B93FHQ & F814W & 60& \\
327.8--06.1 & He2--156 & 2010 Aug 21 & IB1B94011 & F502N & 300, 300&-43.3 \\*
               &            & & IB1B94020 & F200LP & 20, 2& \\*
               &            & & IB1B94030 & F350LP & 20, 2& \\*
               &            & & IB1B94FHQ & F814W & 30& \\
334.8--07.4 & He3--1312 & 2009 Jul 28 & IB1B97011 & F502N & 30, 30 &-66.1\\*
               &            & & IB1B97020 & F200LP & 20, 2&\\*
               &            & & IB1B97030 & F350LP & 20, 2& \\*
               &            & & IB1B97HNQ & F814W & 30& \\
336.9+08.3 & He3--1312 & 2010 May 11 & IB1B98011 & F502N & 50, 50&-162.8 \\*
               &            & & IB1B98020 & F200LP & 90, 9& \\*
               &            & & IB1B98030 & F350LP & 90, 9 &\\*
               &            & & IB1B98HNQ & F814W & 50& \\
340.9--04.6 & Sa1--3 & 2010 Jan 20 & IB1B99011 & F502N & 100, 100&121.2 \\*
               &            & & IB1B99020 & F200LP & 100, 10 &\\*
               &            & & IB1B99030 & F350LP & 100, 10& \\*
               &            & & IB1B99HJQ & F814W & 50 &\\
341.5--09.1 & He2-248 & 2010 Aug 06 & IB1B0A011 & F502N & 30, 30&-54.0 \\*
               &            & & IB1B0A020 & F200LP & 20, 2 &\\*
               &            & & IB1B0A030 & F350LP & 20, 2& \\*
               &            & & IB1B0AA5Q & F814W & 25& \\
343.4+11.9 & H1--1 & 2010 May 11 & IB1B0B011 & F502N & 300, 300&-178.7 \\*
               &            & & IB1B0B020 & F200LP & 20, 2 &\\*
               &            & & IB1B0B030 & F350LP & 20, 2& \\*
               &            & & IB1B0BIFQ & F814W & 30& \\
344.2+04.7 & Vd1--1 & 2010 May 19 & IB1B0C011 & F502N & 50, 50&-173.7 \\*
               &            & & IB1B0C020 & F200LP & 60, 6& \\*
               &            & & IB1B0C030 & F350LP & 60, 6& \\*
               &            & & IB1B0CPPQ & F814W & 40& \\
344.8+03.4 & Vd1--3 & 2009 Oct 20 & IB1B0F011 & F502N & 300, 300 &-23.7\\*
               &            & & IB1B0F020 & F200LP & 20, 2& \\*
               &            & & IB1B0F030 & F350LP & 20, 2 &\\*
               &            & & IB1B0FW2Q & F814W & 30& \\
345.0+04.3 & Vd1--2 & 2010 Jun 20 & IB1B0G011 & F502N & 100, 100&-86.1 \\*
               &            & & IB1B0G020 & F200LP & 100, 10& \\*
               &            & & IB1B0G030 & F350LP & 100, 10& \\*
               &            & & IB1B0GKCQ & F814W & 50& \\
348.4--04.1 & H1--21 & 2010 Jul 25 & IB1B0H011 & F502N & 50, 50&-62.2 \\*
               &            & & IB1B0H020 & F200LP & 120, 13 &\\*
               &            & & IB1B0H030 & F350LP & 120, 13 &\\*
               &            & & IB1B0HJGQ & F814W & 54 &\\
348.8--09.0 & He2--306 & 2009 Oct 01 & IB1B0I011 & F502N & \ldots \tablenotemark{a} &-41.8\\*
               &            & & IB1B0I020 & F200LP & 20, 2 &\\*
               &            & & IB1B0I030 & F350LP & 20, 2 &\\*
               &            & & IB1B0IDZQ & F814W & 30& \\

351.3+07.6 & H1--4 & 2009 Aug 02 & IB1B0K011 & F502N & 30, 30 &-45.1\\*
               &            & & IB1B0K020 & F200LP & 30, 3& \\*
               &            & & IB1B0K030 & F350LP & 30, 3 &\\*
               &            & & IB1B0KOFQ & F814W & 30& \\
356.5+01.5 & Th3--55 & 2009 Sep 11 & IB1B0Q011 & F502N & 300, 300&-43.1 \\*
               &            & & IB1B0Q020 & F200LP & 120, 60 &\\*
               &            & & IB1B0Q030 & F350LP & 120, 60& \\*
               &            & & IB1B0QZDQ & F814W & 100& \\
358.6+07.8 & M3--36 & 2010 Mar 19 & IB1B0Z011 & F502N & 300, 300&139.8 \\*
               &            & & IB1B0Z020 & F200LP & 50, 5& \\*
               &            & & IB1B0Z030 & F350LP & 50, 5& \\*
               &            & & IB1B0ZA3Q & F814W & 40& \\
\enddata
\tablenotetext{a}{Observation aborted prior to executing the planned exposure.} 
\end{deluxetable}

%
\begin{deluxetable}{c r r c c r c c c c l}
\tabletypesize{\scriptsize}
\rotate
\tablecolumns{11}
\tablewidth{0pc}
\tablecaption {Coordinates, Fluxes, Dimensions, and Morphologies of Compact Galactic Planetary Nebulae\label{tab:Morph}}
\tablehead {
\colhead {} & \colhead {R.A.} & \colhead {Decl.} & \colhead {} & \colhead {} & \colhead {} & \colhead {R$_{phot}$} & 
\colhead {Diameter} & \colhead {Ref. Contour} & \colhead {Morph.} & \colhead {} \\
\colhead {PN~G} & \colhead {(J2000)} & \colhead {(J2000)} & \colhead {--log $F$(5007)} & 
\colhead {--log $F$($H\beta$)\tablenotemark{a}} & \colhead {$c$} & \colhead {(arcsec)} & 
\colhead {(arcsec)} & \colhead {(percent)} & \colhead {Class\tablenotemark{b}} & \colhead {Notes} \\
\colhead {(1)} & \colhead {(2)} & \colhead {(3)} & \colhead {(4)} & 
\colhead {(5)} & \colhead {(6)} & \colhead {(7)} & 
\colhead {(8)} & \colhead {(9)} & \colhead {(10)} & \colhead {(11)} 
}
\startdata
000.8--07.6 & 18 18 37.468 & $-31$ 54 45.07 & 11.69 & 12.68 & 0.39\tablenotemark{c} 
& 1.33 & $1.08\times0.99$ & 20 & B  & \\*
  &   &   &   &   &   &   & $1.82\times1.66$ & 5 &   &  \\*
  &   &   &   &   &   &   & $3.10\times3.04$ & 1 &   &  \\

011.7--06.6 & 18 36 42.475 & $-21$ 48 58.98 & \ldots & \ldots & \ldots 
& \ldots & [Unres.] & \ldots & \ldots & Not a PN \\

014.0--05.5 & 18 36 32.321 & $-19$ 19 27.62 & 11.29 & 12.32 & 1.24\tablenotemark{c}  
& 3.80 & $6.40\times5.12$ & 20 & B & Halo \\*  
  &   &   &   &   &   &   & $10.10\times8.50$ & 5 &   &  \\

014.3--05.5 & 18 37 11.096 & $-19$ 02 22.54 & 11.37 & 12.49 & 1.11\tablenotemark{c}  
& 0.50 & $0.90\times0.66$ & 20 & BC & \\*
  &   &   &   &   &   &   & $1.20\times0.80$ & 5 &   &  \\

021.1--05.9 & 18 51 30.924 & $-13$ 10 37.24 & 11.17 & 12.28 & 0.54\tablenotemark{c}  
& 2.06 & $4.24\times4.12$ & 20 & B & \\*
  &   &   &   &   &   &   & $4.76\times5.32$ & 5 &   &  \\

025.3--04.6 & 18 54 20.039 & $-08$ 47 32.93 & 11.32 & 12.44 & 0.50\tablenotemark{c}  
& 0.48 & $0.70\times0.64$ & 20 & P & Multiple axisymmetric extension \\*    
  &   &   &   &   &   &   & $0.90\times0.87$ & 5 &   &  \\*
  &   &   &   &   &   &   & $1.52\times2.48$ & 0.5 &   &  \\

026.5--03.0 & 18 49 44.685 & $-07$ 01 35.30 & 11.47 & 12.37 & 0.99\tablenotemark{c}  
& 2.93 & $4.10\times5.80$ & 20 & B & Inner lobes, irr. outer structure \\*   
  &   &   &   &   &   &   & $7.00\times7.10$ & 4 &   &  \\

042.9--06.9 & 19 34 33.541 & $+05$ 41 03.03 & 10.35 & 11.41 & 0.41\tablenotemark{c}  
& 0.48 & $0.78\times0.80$ & 20 & B &  \\* 
  &   &   &   &   &   &   & $1.00\times1.06$ & 5 &   &  \\

044.1+05.8 & 18 50 46.850 & $+12$ 37 30.10 & 12.87 & 13.51 & 1.44\tablenotemark{c} 
& 3.84 & $6.07\times4.16$ & 20 & Irr &  \\

048.5+04.2 & 19 04 51.459 & $+15$ 47 36.77 & 11.95 & 13.06 & 1.66\tablenotemark{c}  
& 1.54 & $1.50\times2.46$ & 30 & E & Inner structure \\*
  &   &   &   &   &   &   & $3.40\times4.00$ & 5 &   &  \\

052.9--02.7 & 19 39 15.885 & $+16$ 20 48.07 & 12.12 &13.30 & 1.24\tablenotemark{c} 
& 3.80 & $5.50\times6.42$ & 20 & E &  Attached halo\\*
  &   &   &   &   &   &   & $9.80\times9.60$ & 4 &   &  \\

053.3+24.0 & 17 54 23.034 & $+27$ 59 58.27 & 10.45 & 11.51 & 0.05\tablenotemark{d} 
& 2.32 & $2.46\times3.30$ & 20 & B& Halo \\*
  &   &   &   &   &   &   & $4.30\times3.04$ & 2.5 &   &  \\*
  &   &   &   &   &   &   & $10.40\times9.84$ & 0.05 &   &  \\

059.9+02.0 & 19 35 54.458 & $+24$ 54 49.96 & \ldots & 13.70\tablenotemark{e} & 2.93\tablenotemark{c}  
&1.34\tablenotemark{e} & $0.68\times0.61$\tablenotemark{e} & 20 & B &  \\* 
  &   &   &   &   &   &   & $1.04\times1.34$ & 5 &   &  \\

063.8--03.3 & 20 04 58.622 & $+25$ 26 36.82 & \ldots & 13.3 & \ldots 
& 1.06\tablenotemark{e} & $0.64\times0.46$\tablenotemark{e} & 20 & E & \\*
  &   &   &   &   &   &   & $1.06\times0.84$ & 5 &   &  \\

068.7+01.9 & 19 56 34.043 & $+32$ 22 12.72 & 12.00 & 13.06 & 1.64\tablenotemark{c}  
& 1.76 & $3.12\times2.04$ & 20 & E & Detached arc or ISM interaction \\*
  &   &   &   &   &   &   & $4.00\times3.70$ & 5 &   &  \\

068.7+14.8 & 19 00 26.665 & $+38$ 21 04.82 & 11.17 & 11.95 & 0.38\tablenotemark{c} 
& 0.59 & 1.02 & 20 & R & Inner structure\\*
  &   &   &   &   &   &   & 1.22 & 5 &   &  \\

079.9+06.4 & 20 06 55.346 & $+44$ 14 18.49 & 12.54 & 13.28 & 1.43\tablenotemark{c} 
& 2.77 & $3.20\times3.60$ & 20 & E & Attached halo\\*
  &   &   &   &   &   &   & $7.00\times6.90$ & 5 &   &  \\

097.6--02.4 & 21 57 41.773 & $+51$ 41 38.86 & 11.47 & 12.59 & 0.66\tablenotemark{d} 
& 2.36 & $4.40\times4.24$ & 20 & E & Inner structure, pole-on B?  \\*
  &   &   &   &   &   &   & $8.20\times5.70$ & 5 &   &  \\

098.2+04.9 & 21 27 26.476 & $+57$ 39 05.49 & 12.16 & 13.36 & 2.0\tablenotemark{d} 
& 1.29 & 1.00 x 1.00 & 20 & Irr & \\

104.1+01.0 & 22 20 16.608 & $+58$ 14 16.31 & 12.42 & 13.58 & 2.92\tablenotemark{c}  
& 0.97 & $1.38\times1.34$ & 20 & E & Irregular microstructure \\*
  &   &   &   &   &   &   & $1.72\times1.60$ & 5 &   &  \\

107.4--02.6 & 22 55 06.922 & $+56$ 42 30.59 & 12.50 &13.21 & 1.25\tablenotemark{d} 
& 3.54 & $5.80\times3.90$ & 20 & E & Asymmetrical inner structure \\*
  &   &   &   &   &   &   & $9.10\times8.90$ & 5 &   &  \\

184.0--02.1 & 05 46 50.023 & $+24$ 22 03.34 & 11.43 & 12.05 & 1.7\tablenotemark{d}  
& 1.03 & $2.28\times2.00$ & 20 & E & Inner structure, asymmetrical ansae \\*
  &   &   &   &   &   &   & $2.84\times2.20$ & 5 &   &  \\

205.8--26.7 & 05 03 41.877 & $-06$ 10 03.16 & 11.26 & 12.21 & $\dots$
& 1.04 & $1.54\times1.18$ & 20 & E & Inner structure, ansae \\*
  &   &   &   &   &   &   & $2.50\times2.38$ & 5 &   &  \\

263.0--05.5 & 08 20 40.310 & $-46$ 22 57.41 & 10.91 & 12.01 & 0.85\tablenotemark{c}  
& 1.35 & $2.92\times2.62$ & 10 & E & Irregular microstructure, multiple rings \\*
  &   &   &   &   &   &   & $3.20\times2.84$ & 5 &   &  \\

264.4--12.7 & 07 47 20.068 & $-51$ 15 02.99 & 10.57 & 11.35 & 0.29\tablenotemark{d} 
& 1.32 & 2.80 & 20 & R & Halo\\*
  &   &   &   &   &   &   & $3.20\times3.14$ & 5 &   &  \\*
  &   &   &   &   &   &   & $5.60\times5.46$ & 0.5 &   &  \\

274.1+02.5 & 09 41 13.978 & $-49$ 22 47.49 & \ldots & \ldots & \ldots 
& \ldots & [Unres.] & \ldots & \ldots & Not a PN \\

275.3--04.7 & 09 13 52.868 & $-55$ 28 16.62 & 11.03 & 12.15 & 0.80\tablenotemark{c}  
& 1.70 & $2.10\times1.62$ & 40 & E & Inner structure, ansae \\*
  &   &   &   &   &   &   & 3.60 & 4 &   &  \\

278.6--06.7 & 09 19 27.492 & $-59$ 11 59.97 & 10.40 & 11.55 & 0.50 
& 1.19 & $2.14\times2.02$ & 20 & E & Enclosed ring \\*  
  &   &   &   &   &   &   & $2.58\times2.32$ & 5 &   &  \\

285.4+01.5 & 10 38 27.577 & $-56$ 47 06.09 & 11.21 & 12.31 & 1.81\tablenotemark{d}  
& 1.92 & $2.35\times2.49$ & 20 & B & Detached arc \\*
  &   &   &   &   &   &   & $3.95\times3.20$ & 5 &   & \\*
  &   &   &   &   &   &   & $9.6\times5.9$ & 0.5 &   & \\

285.4+02.2 & 10 41 19.564 & $-56$ 09 16.45 & 11.60 & 12.85 & 1.28\tablenotemark{c}  
& 2.06 & $4.20\times2.80$ & 20 & E & Inner structure, ansae  \\*
  &   &   &   &   &   &   & $5.60\times3.60$ & 4 &   & size from F200LP\\

286.0--06.5 & 10 07 23.603 & $-63$ 54 29.84 & 10.90 & 11.90 & 0.70\tablenotemark{c}  
& 0.95 & $0.84\times0.74$ & 20 & B & \\*
  &   &   &   &   &   &   & $1.48\times1.23$ & 5 &   & \\*
  &   &   &   &   &   &   & $3.93\times1.68$ & 0.5 &   & \\

289.8+07.7 & 11 24 01.086 & $-52$ 51 19.37 & 11.31 & 12.36 & 0.35\tablenotemark{c}  
& 1.29 & $2.52\times2.26$ & 20 & E(s) & Ring \\*
  &   &   &   &   &   &   & $2.82\times1.72$ & 5 &   & \\

294.9--04.3 & 11 31 45.414 & $-65$ 58 13.76 & 11.56 & 11.73 & 1.53\tablenotemark{c}  
& 0.98 & $1.62\times1.46$ & 20 & E & Irregular shape \\*
  &   &   &   &   &   &   & $1.88\times1.23$ & 5 &   & \\

295.3--09.3 & 11 17 43.132 & $-70$ 49 31.78 & 10.96 & 11.94 & 0.44\tablenotemark{c}  
& 0.48 & $1.16\times1.12$ & 20 & B & \\*
  &   &   &   &   &   &   & $1.94\times1.48$ & 5 &   & \\*
  &   &   &   &   &   &   & $4.50\times2.80$ & 0.5 &   &  \\

296.3--03.0 & 11 48 38.163 & $-65$ 08 36.88 & 10.75 & 12.01 & 1.40\tablenotemark{c}  
& 1.32 & $2.47\times1.99$ & 20 & P & Halo \\*  
  &   &   &   &   &   &   & $3.46\times2.30$ & 5 &   & \\*
  &   &   &   &   &   &   & $5.04\times2.11$ & 1 &   & \\

309.0+00.8 & 13 42 36.191 & $-61$ 22 28.61 & 11.35 & 12.40 & 2.03\tablenotemark{c}  
& 1.47 & $1.32\times1.16$ & 20 & Irr &  \\*
  &   &   &   &   &   &   & $2.28\times1.71$ & 5 &   & \\*
  &   &   &   &   &   &   & $4.55\times3.68$ & 1 &   & \\

309.5--02.9 & 13 54 27.500 & $-64$ 59 32.28 & 11.90 & 13.09 & 2.08\tablenotemark{c}  
& 1.18 & $1.67\times1.49$\tablenotemark{e} & 20 & E & Irregular outer arc \\*
  &   &   &   &   &   &   & $2.22\times1.98$ & 5 &   & \\*
  &   &   &   &   &   &   & $2.76\times2.58$ & 2 &   & \\

311.1+03.4 & 13 54 55.678 & $-58$ 27 16.83 & \ldots & \ldots & \ldots 
& \ldots & [Unres.] & \ldots & \ldots & Not a PN \\

324.8--01.1 & 15 41 58.823 & $-56$ 36 25.08 & 12.15 & 13.26 & 0.43\tablenotemark{d}  
& 1.94 & $2.40\times3.46$\tablenotemark{e} & 20 & BC & Complex substructure \\*  
  &   &   &   &   &   &   & $5.16\times4.12$ & 5 &   & \\

327.1--01.8 & 15 58 08.064 & $-55$ 41 50.93 & 12.63 & 12.27 & 1.91\tablenotemark{d}  
& 1.50 & $3.64\times2.68$\tablenotemark{e} & 20 & E & Inner structure\\*
  &   &   &   &   &   &   & $4.20\times2.90$ & 5 &   & \\

327.8--06.1 & 16 23 30.599 & $-58$ 19 22.73 & 11.38 & 12.10 & 0.51\tablenotemark{c}  
& 1.05 & $0.92\times0.84$ & 20 & B & Multi-polar \\*
  &   &   &   &   &   &   & $1.74\times1.15$ & 5 &   & \\*
  &   &   &   &   &   &   & $2.98\times1.87$ & 1 &   & \\*
  &   &   &   &   &   &   & $5.0\times3.0$ & 0.2 &   & \\

334.8--07.4 & 17 03 03.125 & $-53$ 55 52.22 & 11.00\tablenotemark{f} & 11.19 & 0.55\tablenotemark{c}  
& 0.44 & $0.38\times0.24$ & 20 & B & Multi-polar \\*
  &   &   &   &   &   &   & $0.58\times0.54$ & 5 &   & faint ansae extend 6\farcs2 NW, 7\farcs6 SE \\*
  &   &   &   &   &   &   & $0.81\times0.77$ & 1 &   & \\

336.9+08.3 & 16 02 13.077 & $-41$ 33 36.09 & 11.74 & 12.73 & 1.47\tablenotemark{c}  
& 1.77 & $2.36\times2.90$ & 20 & E & Inner structure, halo\\*
  &   &   &   &   &   &   & $3.80\times4.44$ & 4 &   & \\

340.9--04.6 & 17 11 27.418 & $-47$ 25 01.66 & 11.65 & 12.80 & 1.57\tablenotemark{c}  
& 1.12 & $1.53\times1.39$\tablenotemark{e} & 20 & E(s,a) & \\*
  &   &   &   &   &   &   & $2.40\times1.99$ & 4 &   & \\

341.5-09.1 & 17 36 06.913 & $-49$ 25 45.58 & 11.28 & 12.30 & 0.40\tablenotemark{c}  
& 1.37 & $2.08\times1.56$ & 20 & E & Ansae \\*
  &   &   &   &   &   &   & $3.24\times2.38$ & 5 &   & \\*
  &   &   &   &   &   &   & $3.46\times3.44$ & 2 &   & \\

343.4+11.9 & 16 13 28.218 & $-34$ 35 39.43 & 11.25 & 12.45 & 0.45\tablenotemark{d}  
& 1.54 & $1.94\times2.64$ & 20 & BC & \\*
  &   &   &   &   &   &   & $3.15\times3.30$ & 5 &   & \\*
  &   &   &   &   &   &   & $4.96\times4.82$ & 0.5 &   & \\

344.2+04.7 & 16 42 33.470 & $-38$ 54 31.69 & 11.47 & 12.07 & 1.17\tablenotemark{c}  
& 0.55 & $0.68\times0.56$\tablenotemark{e} & 20 & B & \\*
  &   &   &   &   &   &   & $1.79\times1.29$ & 1 &   & \\*
  &   &   &   &   &   &   & $3.17\times2.04$ & 0.2 &   & \\

344.8+03.4 & 16 49 32.872 & $-39$ 21 09.36 & 13.05 & 13.04 & 0.57\tablenotemark{c}  
& 2.32 & $5.2\times4.4$\tablenotemark{d} & 20 & E & Inner structure, P? \\

345.0+04.3 & 16 46 45.080 & $-38$ 36 58.47 & \ldots &12.25 & 1.57\tablenotemark{c}  
& 2.19\tablenotemark{e} & $2.19\times0.55$\tablenotemark{e} & \ldots &B?  & Bi-lobe extension to 2\farcs45? \\

348.4--04.1 & 17 32 47.753 & $-40$ 58 29.26 & 11.55 & 12.73 & 1.72\tablenotemark{c}  
& 0.97 & $1.51\times0.78$\tablenotemark{e} & 20 & B & \\*
  &   &   &   &   &   &   & $1.90\times1.21$ & 5 &   & \\*
  &   &   &   &   &   &   & $2.57\times1.81$ & 1 &   & \\*
  &   &   &   &   &   &   & $2.97\times2.19$ & 0.5 &   & \\

348.8--09.0 & 17 56 34.343 & $-43$ 03 17.10 & \ldots & 11.80 & 0.0\tablenotemark{c} 
& 3.4 & $3.00\times2.84$\tablenotemark{e} & 20 & E & Inner structure, halo; \\*
  &   &   &   &   &   &   & $3.40\times3.30$ & 5 &   & No F502N image available \\

351.3+07.6 & 16 53 37.284 & $-31$ 40 32.88 & 11.55 & 12.35 & 0.69\tablenotemark{c}  
& 0.53 & $0.90\times0.78$ & 20 & BC & \\*  
  &   &   &   &   &   &   & $1.25\times1.09$ & 5 &   & \\

356.5+01.5 & 17 30 58.831 & $-31$ 01 05.98 & 13.07 & 13.68 & 2.77\tablenotemark{c}  
& 2.28 & $4.74\times3.04$& 20 & P & Detached arc \\*
  &   &   &   &   &   &   & $6.40\times4.50$ & 5 &   & \\

358.6+07.8 & 17 12 39.182 & $-25$ 43 37.31 & 11.60 & 12.54 & 1.05\tablenotemark{c}  
& 1.99 & $2.73\times3.52$ & 20 & E & Inner structure, halo\\*
  &   &   &   &   &   &   & $4.80\times4.40$ & 5 &   & \\*
  &   &   &   &   &   &   & $5.61\times5.10$ & 2 &   & \\
\enddata

\tablenotetext{a}{F($H\beta$) derived from F(5007) presented in this paper and the I(5007)/I($H\beta$) taken from the compilation of Acker (1992).}
\tablenotetext{b}{Primary morphological classes are E: elliptical, R: round, BC: bipolar core; B: bipolar, P: point-symmetric, Irr: irregular. Secondary morphological features  denote microstructure: inner structure, ansae, attached shell, and hallo (see text).} 
\tablenotetext{c}{Extinction constant calculated from Balmer emission line ratios from Acker et~al. (1992).}
\tablenotetext{d}{Extinction constant from Cahn et~al. (1992).}
\tablenotetext{e}{Target unresolved in F502N.} 
\tablenotetext{f}{Nebular extent obtained from image in F200LP.}

\end{deluxetable}

%
\begin{deluxetable}{crlrrrrr}
\tabletypesize{\scriptsize}
\tablecolumns{9}
\tablewidth{0pc}
\tablecaption {Distances and Sizes of Compact Galactic Planetary Nebulae\label{tab:Dist}}
\tablehead {

\colhead {PN~G Design.} & \colhead{$F_\nu$(5 GHz)\tablenotemark{a}} & \colhead{Ref. \tablenotemark{b}} & \colhead{$\tau$} & \colhead{R} & \colhead{D} & \colhead{R$_{\rm G}$} & \colhead{$|z|$}  \\
\colhead {} & \colhead {(mJy)} & \colhead {} & \colhead {} & \colhead {(pc)} & \colhead{(kpc)} & \colhead{(kpc)} & \colhead {(kpc)}  \\
\colhead{(1)}& \colhead{(2)}& \colhead{(3)}& \colhead{(4)}& \colhead{(5)}& \colhead{(6)}& \colhead{(7)}& \colhead{(8)}\\
}
\startdata

    000.8--07.6  &   1.83  &   $H\beta$  &   3.59  &   0.13 &   20.07  &  11.90  &       2.655           \\
    014.0--05.5  &  29.66  &   $H\beta$  &   3.29  &   0.11  &    5.92  &   2.69  &       0.567           \\
    014.3--05.5  &  10.00  &      A  &   2.00  &   0.05&   20.35  &  12.66  &       1.951           \\
    021.1--05.9  &  6.48  &    $H\beta$ &   3.42  &   0.12  &   11.75  &   5.12  &       1.208           \\
    025.3--04.6  &   4.07  &   $H\beta$  &   2.35  &   0.07 &   28.79  &  21.74  &       2.309           \\
    026.5--03.0  &   6.00  &    ZPB  &   3.76  &   0.14 &   10.05  &   4.58  &      0.5258           \\
    042.9--06.9  &  27.00  &    AK,IS,MAI  &   1.53  &   0.04  &   17.04  &  12.33  &       2.047           \\
    044.1+05.8  &   3.03  &   $H\beta$  &   4.29  &   0.16  &    8.77  &   6.32  &      0.8864           \\
    048.5+04.2  &   3.00  &    AK:  &   3.50  &   0.11 &   15.21  &  11.54  &       1.114           \\
    052.9--02.7  &   1.70  &    AK: &   4.53  &   0.18  &    9.91  &   8.15  &      0.4668           \\
    053.3+24.0  &  12.25  &   $H\beta$ &   0.11 &   2.32  &    9.45  &   7.48  &       3.842           \\
    059.9+02.0  &  11.00  &      AK  &   2.81  &   0.08  &   12.77  &  11.15  &      0.4456           \\
    063.8--03.3  &   7.50  &    IS,AK  &   2.80  &   0.08  &   15.97  &  14.34  &      0.9194           \\
    068.7+01.9  &  15.00  &      AK  &   2.92  &   0.09  &   10.17  &  10.41  &      0.3373           \\
    068.7+14.8  &   9.42  &   $H\beta$  &   2.17  &   0.06 &   21.51  &  19.38  &       5.495           \\
    079.9+06.4  &   5.00  &   $H\beta$  &   3.79  &   0.13  &    9.66  &  11.36  &       1.076           \\
    097.6--02.4  &   6.50  &    ZPB  &   3.53  &   0.12 &   10.08  &  13.67  &      0.4223           \\
    098.2+04.9  &  43.00  &    AK  &   2.19  &   0.06  &    9.93  &  13.58  &      0.8482           \\
    104.1+01.0  &  54.00  &    AK:  &   1.84  &   0.04  &    8.26  &  12.82  &      0.1442           \\
    107.4--02.6  &   4.50  &      ZPB  &   4.05  &   0.15  &    8.51  &  13.30  &      0.3861           \\
    184.0--02.1  &  71.00  &    AK,CR,IS &   1.78  &   0.04  &    7.01  &  15.00  &       0.257           \\
    205.8--26.7  &   2.18  &   $H\beta$  &   3.30  &   0.10 &   20.52  &  25.77  &        9.22           \\
    263.0--05.5  &  40.00  &      MAI  &   2.26  &   0.05  &    9.80  &  13.35  &      0.9396           \\
    264.4--12.7  &  29.00  &      M  &   2.38  &   0.07  &   10.60  &  13.68  &        2.33           \\
    275.3--04.7  &  16.00  &     MAI &   2.86  &   0.08 &   10.25  &  12.38  &      0.8402           \\
    278.6--06.7  &  40.00  &      M &   2.15  &   0.06  &   10.57  &  12.21  &       1.234           \\
    285.4+01.5  & 111.69  &   $H\beta$  &   2.12  &   0.06  &    5.98  &   8.62  &      0.1564           \\
    285.4+02.2  &   9.53  &   $H\beta$  &   3.25  &   0.10  &   10.14  &  11.12  &      0.3891           \\
    286.0--06.5  &  41.00  &    MAI  &   1.94  &   0.05  &   10.91  &  11.57  &       1.235           \\
    289.8+07.7  &  12.00  &    MAI &   2.74  &   0.08  &   12.82  &  12.51  &       1.717           \\
    294.9--04.3  &  34.00  &    M  &   2.05  &   0.05  &   11.36  &  10.77  &      0.8517           \\
    295.3--09.3  &  11.23  &   $H\beta$  &   1.91  &   0.05  &   21.22  &  18.96  &        3.43           \\
    296.3--03.0  &  76.00  &      M &   1.96  &   0.05  &    7.34  &   8.11  &      0.3841           \\
    309.0+00.8  & 151.81  &   $H\beta$  &   1.76  &   0.03  &    4.77  &   6.22  &     0.06661           \\
    309.5--02.9  &  34.38  &   $H\beta$  &   2.21  &   0.06  &   10.96  &   8.51  &      0.5542           \\
    324.8--01.1  & 244.60  &    MAII  &   1.79  &   0.04  &    3.82  &   5.35  &     0.07338           \\
    327.1--01.8  &  80.00  &      M&   2.05  &  0.05  &    7.43  &   4.40  &      0.2334           \\
    327.8--06.1  &   8.98  &   $H\beta$  &   2.69  &   0.08  &   15.17  &   9.35  &       1.612           \\
    334.8--07.4  &  81.47  &   $H\beta$  &   0.98  &   0.03  &   13.51  &   7.04  &       1.739           \\
    336.9+08.3  &  19.31  &   $H\beta$  &   2.81  &   0.08  &    9.64  &   3.82  &       1.392           \\
    340.9--04.6  &  21.00  &   $H\beta$  &   2.38  &   0.07  &   12.47  &   5.53  &           1           \\
    341.5--09.1  &  13.00  &    M  &   2.76  &   0.08  &   12.17  &   5.10  &       1.924           \\
    343.4+11.9  &   3.53  &   $H\beta$  &   3.43  &   0.11  &   14.72  &   7.11  &       3.035           \\
    344.2+04.7  &  44.39  &   $H\beta$  &   1.44  &   0.04  &   14.06  &   6.68  &       1.152           \\
    344.8+03.4  &   1.20  &   $H\beta$  &   4.25  &   0.16  &   14.28  &   6.87  &       0.847           \\
    345.0+04.3  &   6.65  &   $H\beta$  &   3.46  &   0.12  &   11.32  &   4.12  &      0.8491           \\
    348.4--04.1  &  34.59  &   $H\beta$  &   2.04  &   0.05  &   11.27  &   3.76  &      0.8056           \\
    351.3+07.6  &   7.67  &   $H\beta$  &   2.17  &   0.06  &   23.90  &  15.83  &       3.161           \\
    356.5+01.5  &   6.10  &      A  &   3.53  &   0.12  &   10.43  &   2.49  &      0.2729           \\
    358.6+07.8  &   3.50  &    ZPB  &   3.66  &   0.12  &   12.64  &   4.53  &       1.716           \\

\enddata
\tablenotetext{a}{Adopted 5 GHz flux}
\tablenotetext{b}{Reference for the 5 GHz flux: $H\beta$ means that the 5 GHz flux has been inferred from the relation between the 5 GHz flux and the $H\beta$ flux, as described in Cahn et~al. (1992), and with log F($H\beta$) and extinction constant from Table 2. Other references are A (Acker et~al. 1992); AK (Aaquist \& Kwok 1991); CR (Cahn \& Rubin 1974); IS (Isaacman 1984); M (Milne 1979); MAI (Milne \& Aller 1975);  MAII (Milne \& Aller 1982); ZBP (Zjilstra et~al. 1989).}
\end{deluxetable}


\begin{thebibliography}{}

\bibitem[Aaquist \& Kwok(1991)]{1991ApJ...378..599A} 
	Aaquist, O.~B., \& Kwok, S.\ 1991, 
	\apj, 378, 599 

\bibitem[Acker et~al.(1991)]{Acker91}
	Acker, A., Stenholm, B., Tylenda, R., \& Raytchev, B.\ 1991, 
	\aaps, 90, 89

\bibitem[Acker et~al.(1992)]{Acker92}
	Acker, A., Marcout, J., Ochsenbein, F., Stenholm, B., \& Tylenda, R..\ 1992, 
	Strasbourg-ESO Catalogue of Galactic Planetary Nebulae (Garching: ESO)

\bibitem[Balick et al.(2001)]{2001AJ....121..354B} Balick, B., Wilson, J., \& Hajian, A.~R.\ 2001, \aj, 121, 354 

\bibitem[Cahn \& Rubin(1974)]{1974AJ.....79..128C} 
	Cahn, J.~H., \& Rubin, R.~H.\ 1974, 
	\aj, 79, 128 

\bibitem[Cahn et~al.(1992)]{1992A&AS...94..399C} 
	Cahn, J.~H., Kaler, J.~B., \& Stanghellini, L.\ 1992, 
	\aaps, 94, 399 

\bibitem[Corradi et~al.(2003)]{Corradi03} 
	Corradi, R.~L.~M., Sch{\"o}nberner, D., Steffen, M., \& Perinotto, M.\ 2003, 
	\mnras, 340, 417 

\bibitem[Dressel et~al.(2010)]{WFC3_ihb}
	Dressel, L., Wong, M.~H.., Pavlovsky, C., \& Long, K.~S.\ (eds.) 2010,
	WFC3 Instrument Handbook (Version 2.1; Baltimore: STScI)
	
\bibitem[Dufour et~al.(2015)]{2015ApJ...803...23D} 
	Dufour, R.~J., Kwitter, K.~B., Shaw, R.~A., et~al.\ 2015, 
	\apj, 803, 23 
	
\bibitem[Garc{\'{\i}}a-Hern{\'a}ndez \& G{\'o}rny(2014)]{2014A&A...567A..12G} 
	Garc{\'{\i}}a-Hern{\'a}ndez, D.~A., \& G{\'o}rny, S.~K.\ 2014, 
	\aap, 567, A12 

\bibitem[Hsia et~al.(2014)]{2014ApJ...787...25H} 
	Hsia, C.-H., Chau, W., Zhang, Y., \& Kwok, S.\ 2014, 
	\apj, 787, 25 

\bibitem[Isaacman(1984)]{1984MNRAS.208..399I} Isaacman, R.\ 1984, \mnras, 208, 399 

\bibitem[Kerber et~al.(2003)]{Kerber03}
	Kerber, F., Mignani, R.~P., Guglielmetti, F., \& Wicenec, A.\ 2003, 
	\aap, 408, 1029

\bibitem[Kimble et~al.(2008)]{Kimble_etal08} 
	Kimble, R.~A., MacKenty, J.~W., O'Connell, R.~W., \& Townsend, J.~A.\ 2008, 
	in Proc.\ SPIE, 7010, 70101E-1 

\bibitem[Kwok(2010)]{2010PASA...27..174K} Kwok, S.\ 2010, \pasa, 27, 174 

\bibitem[Laher et~al.(2012)]{Laher12}
	Laher, R.~R., Gorjian, V., Rebull, L.~M., et~al.\ 2012, 
	\pasp, 124, 737

\bibitem[Manchado et~al.(1996)]{1996iacm.book.....M} 
	Manchado, A., Guerrero, M.~A., Stanghellini, L., \& Serra-Ricart, M.\ 1996, 
	The IAC morphological catalog of northern Galactic planetary nebulae, Publisher: La Laguna, Spain: Instituto de Astrofisica de Canarias (IAC), 1996, ISBN: 8492180609,  

\bibitem[Milne(1979)]{1979A&AS...36..227M} 
	Milne, D.~K.\ 1979, 
	\aaps, 36, 227 

\bibitem[Milne \& Aller(1975)]{1975A&A....38..183M} 
	Milne, D.~K., \& Aller, L.~H.\ 1975, 
	\aap, 38, 183 

\bibitem[Milne \& Aller(1982)]{1982A&AS...50..209M} 
	Milne, D.~K., \& Aller, L.~H.\ 1982, 
	\aaps, 50, 209 

\bibitem[Moreno-Ib{\'a}{\~n}ez et~al.(2016)]{MVSS16}
	Moreno-Ib{\'a}{\~n}ez, M., Villaver, E., Shaw, R.~A., \& Stanghellini, L.\ 2016, 
	\aap, submitted

\bibitem[Parker et~al.(2006)]{2006MNRAS.373...79P} 
	Parker, Q.~A., Acker, A., Frew, D.~J., et~al.\ 2006, 
	\mnras, 373, 79 

\bibitem[Rajan et~al.(2011)]{WFC3_dhb} 
	Rajan, A., et~al.\ 2011, 
	WFC3 Data Handbook (Version 2.1; Baltimore: STScI) 

\bibitem[Sahai et al.(2011)]{2011AJ....141..134S} Sahai, R., Morris, M.~R., \& Villar, G.~G.\ 2011, \aj, 141, 134 

\bibitem[Schwarz et~al.(1992)]{1992A&AS...96...23S} 
	Schwarz, H.~E., Corradi, R.~L.~M., \& Melnick, J.\ 1992, 
	\aaps, 96, 23 

\bibitem[Shaw et~al.(2001)]{Shaw_etal01}Shaw, R.~A., 
	Stanghellini, L., Mutchler, M., Balick, B., \& Blades, J.~C.\ 2001, 
	\apj, 548, 727 

\bibitem[Shaw et~al.(2006)]{Shaw_etal06}Shaw, R.~A., 
	Stanghellini, L., Villaver, E., \& Mutchler, M.\ 2006, 
	\apjs, 167, 201 

\bibitem[Stanghellini et~al.(2000)]{2000ApJ...534L.167S} 
	Stanghellini, L., Shaw, R.~A., Balick, B., \& Blades, J.~C.\ 2000, 
	\apjl, 534, L167 

\bibitem[Stanghellini et~al.(2003)]{Stang_etal03}Stanghellini, L., 
	Shaw, R.~A., Balick, B., Mutchler, M., Blades, J.~C., \& Villaver, E.\ 2003, 
	\apj, 596, 997 

\bibitem[Stanghellini et~al.(2006)]{Stang_etal06}
	Stanghellini, L., Guerrero, M.~A., Cunha, K., Manchado, A., \& Villaver, E.\ 2006,
	\apj, 651, 898

\bibitem[Stanghellini et~al.(2007)]{Stang_etal07}
	Stanghellini, L., Garc{\'{\i}}a-Lario, P., Garc{\'{\i}}a-Hern{\'a}ndez, D.~A., et~al.\ 2007, 
	\apj, 671, 1669 

\bibitem[Stanghellini et~al.(2008)]{Stang_etal08}Stanghellini, L., 
	Shaw, R.~A., \& Villaver, E.\ 2008, 
	\apj, 689, 194 (SSV)

\bibitem[Stanghellini et~al.(2012)]{2012ApJ...753..172S} 
	Stanghellini, L., Garc{\'{\i}}a-Hern{\'a}ndez, D.~A., Garc{\'{\i}}a-Lario, P., et~al.\ 2012, 
	\apj, 753, 172 

\bibitem[Stanghellini \& Haywood(2010)]{2010ApJ...714.1096S} 
	Stanghellini, L., \& Haywood, M.\ 2010, 
	\apj, 714, 1096 (SH10)


\bibitem[Villaver et~al.(2002)]{Vgm02}
	Villaver, E., Garc\'{\i}a-Segura, \& Manchado, A. 2002, 
	\apj, 571, 880 

\bibitem[Villaver et~al.(2003)]{Villaver03} Villaver, E., 
Garc{\'{\i}}a-Segura, G., \& Manchado, A.\ 2003, \apjl, 585, L49 

\bibitem[Villaver et~al.(2012)]{Vmg12}
	Villaver, E., Manchado, A., \& Garc{\'{\i}}a-Segura, G.\ 2012, 
	\apj, 748, 94 

\bibitem[Zijlstra et~al.(1989)]{1989A&AS...79..329Z} 
	Zijlstra, A.~A., Pottasch, S.~R., \& Bignell, C.\ 1989, 
	\aaps, 79, 329 

\end{thebibliography}
\end{document}